\documentclass{article}
\usepackage{arxiv}

\usepackage[T1]{fontenc}    
\usepackage{hyperref}       
\usepackage{url}            
\usepackage{booktabs}       

\usepackage[utf8]{inputenc} 
\usepackage[round]{natbib}
\usepackage{alltt} 
\usepackage{fancyvrb}

\usepackage{graphicx}
\usepackage{amsmath}
\usepackage{amssymb}
\usepackage{amsfonts}
\usepackage{bm}

\usepackage{epsdice}
\usepackage[linesnumbered,ruled,vlined]{algorithm2e}
\SetKwInput{KwInput}{Input}                
\SetKwInput{KwOutput}{Output}              

\usepackage{tabularx}
\usepackage{float} 
\usepackage[table,dvipsnames]{xcolor}
\usepackage{subcaption}
\usepackage{colortbl}

\usepackage[table,dvipsnames]{xcolor}
\usepackage{subcaption}
\usepackage{colortbl}

\usepackage{enumerate} 
\usepackage{mdwlist}

\usepackage{multirow,array}
\usepackage[normalem]{ulem}

\usepackage{epsdice}

\newtheorem{definition}{Definition}[section]

\newcommand{\myvec}[1]{\mathbf{#1}}
\newcommand{\tuple}[1]{\langle #1 \rangle}
\newcommand{\xtrain}{X_\text{tr}}
\newcommand{\ytrain}{\myvec{y}_\text{tr}}
\newcommand{\E}{\mathbb{E}}
\newcommand{\V}{\mathbb{V}}
\newcommand{\prob}[1]{P\left(#1\right)}
\newcommand{\est}[1]{\hat{#1}}
\newcommand{\mean}{\mu}
\newcommand{\jacobian}{\nabla}
\newcommand{\nobj}{C}
\newcommand{\objindex}{c}
\newcommand{\policyparam}{\boldsymbol{\theta}}
\newcommand{\self}{{1}}
\newcommand{\other}{{2}}
\newcommand{\payoff}{{\bf p}}
\newcommand{\nashhighlight}{\cellcolor{SpringGreen}}

\title{Opponent Learning Awareness and Modelling in Multi-Objective Normal Form Games\thanks{Some preliminary results in this article were presented at AAMAS 2020 \citep{zhang2020opponent} and the Adaptive and Learning Agents Workshop 2020 \citep{zhang2020opponentALA}.}}

\author{
  Roxana R\u{a}dulescu\\
  Vrije Universiteit Brussel\\
  Belgium \\
  \texttt{roxana.radulescu@vub.be} \\
   \And
     Timothy Verstraeten\\
  Vrije Universiteit Brussel\\
  Belgium \\
  \texttt{timothy.verstraeten@vub.be} \\
   \And    
   Yijie Zhang \\
  Universiteit van Amsterdam\\
  The Netherlands\\
  \texttt{philip.yijie.zhang@gmail.com} \\
  \And
 Patrick Mannion \\
  National University of Ireland Galway\\
  Ireland \\
  \texttt{patrick.mannion@nuigalway.ie} \\
  \And
   Diederik M. Roijers \\
        Vrije Universiteit Brussel, Belgium \\
        HU~Univ.~of~Appl.~Sci.~Utrecht\\
        The~Netherlands\\
  \texttt{diederik.yamamoto-roijers@hu.nl} \\
    \And
   Ann Now\'{e}\\
  Artificial Intelligence Lab\\
  Vrije Universiteit Brussel\\
  Belgium \\
  \texttt{ann.nowe@vub.be} \\
}

\begin{document}
\maketitle

\begin{abstract}
Many real-world multi-agent interactions consider multiple distinct criteria, i.e. the payoffs are multi-objective in nature. However, the same multi-objective payoff vector may lead to different utilities for each participant. Therefore, it is essential for an agent to learn about the behaviour of other agents in the system. In this work, we present the first study of the effects of such opponent modelling on multi-objective multi-agent interactions with non-linear utilities. Specifically, we consider two-player multi-objective normal form games with non-linear utility functions under the scalarised expected returns optimisation criterion. We contribute novel actor-critic and policy gradient formulations to allow reinforcement learning of mixed strategies in this setting, along with extensions that incorporate opponent policy reconstruction and learning with opponent learning awareness (i.e., learning while considering the impact of one's policy when anticipating the opponent's learning step). Empirical results in five different MONFGs demonstrate that opponent learning awareness and modelling can drastically alter the learning dynamics in this setting. When equilibria are present, opponent modelling can confer significant benefits on agents that implement it. When there are no Nash equilibria, opponent learning awareness and modelling allows agents to still converge to meaningful solutions that approximate equilibria.
\end{abstract}


\section{Introduction}
\label{sec:introduction}
Game theory classically studies multi-agent decision making with one-dimensional payoffs \citep{nisan2007algorithmic}. However, many real-life decision problems are much more intricate. For example, while hammering out a contract for building a new piece of software, the different agents may care about price, delivery time, functionality, and so on. In other words, many multi-agent decision problems are inherently multi-objective \citep{Radulescu2020Survey}.

In such multi-objective settings, the utility derived  from the payoffs may differ from agent to agent. For example, imagine a multi-player online game where a team of players does a quest together. The quest will lead to the same expected amount of experience points, loot and currency for each player in the team. However, depending on their level, class, and play style, different agents may care about these objectives differently, leading to different individual utilities. While the expected payoffs may be common knowledge, the utility each agent would derive from these payoffs may be private information. Furthermore, it may even be non-trivial for the individual agents to quantify these utilities themselves \citep{zintgraf2018ordered}. In such cases, it is critical to study the emergent behaviour after multiple interactions as the agents learn more about each other. In other words, it is key to look at it from a reinforcement learning perspective.

An elegant model to study agent interaction in multi-objective settings is the \emph{multi-objective normal form game (MONFG)} \citep{blackwell1956analog,shapley1959equilibrium}. To date, most papers studying MONFGs have considered different -- specifically multi-objective -- equilibria, which are often agnostic about the utility functions of the individual agents \citep{borm1988pareto,voorneveld1999axiomatizations}. Furthermore, most research implicitly assumes that the agents are interested in the expected utility of the payoff vector of a single play. This is called the \emph{expected scalarised returns (ESR)} optimality criterion \citep{roijers2018multi}. However, in many games, especially when the game is played multiple times, agents may instead be interested in the utility of the expected payoff (over multiple plays), which is called the \emph{scalarised expected returns (SER)} optimality criterion. As we study repeated interaction and long-term rewards, befitting the reinforcement learning setting, we are interested in the SER criterion. Recent work by \citet{radulescu2019equilibria} demonstrated that the difference between ESR and SER in MONFGs can drastically alter the equilibria, and that, under SER, Nash equilibria (NE) need not exist at all. The payoffs in MONFGs are common knowledge, but the utilities that the agents derive from these are not. It is therefore important to learn about the opponents, i.e., other agents, through interaction. 

In this paper, we investigate whether opponent modelling (OM) benefits agents in reinforcement learning for multi-objective multi-agent decision making problems under SER. Although opponent modelling techniques have a long history of use within the MAS community \citep{albrecht2018survey}, to date their potential applications to multi-objective multi-agent systems (MOMAS) have not been comprehensively explored. Furthermore, we build on the recent advances from the multi-agent learning literature and introduce the idea of learning with opponent learning awareness \citep{foerster2018lola} to MOMAS. In a multi-objective setting, modelling the opponents' learning step is not straightforward, since the learning direction is defined by the opponents' utility, information that is usually not available. The key idea behind our opponent learning awareness method is to train a Gaussian process \citep{rasmussen2003gaussian} as an estimator for the opponents' learning step that considers both the opponent's current policy, as well as the influence of the agent's own policy.

The contributions of this paper are:
\begin{enumerate}
    \item Using both a policy gradient and an actor-critic framework, we develop the first reinforcement learning methods that can learn stochastic best response strategies for MONFGs under SER.
    \item We contribute novel algorithms developed specifically for opponent learning awareness and modelling in MONFGs under SER with non-linear utility functions.
    \item We provide the first empirical evidence that opponent modelling can confer significant advantages in MONFGs under SER with non-linear utility functions when Nash equilibria are present. Our results demonstrate that when both agents implement opponent modelling, opponent modelling can increase the probability of converging to (better) Nash equilibria.
    \item When NE are present, we demonstrate that when only a single agent implements opponent modelling, there is an increased probability of converging to the best Nash equilibrium for that agent. 
    \item Our experimental results show that when no NE are present,
    learning with opponent learning awareness allows agents to still converge to meaningful solutions that approximate equilibria, opening the discussion for new solution concepts to be adopted for such settings.
\end{enumerate}

The next section of the paper introduces the necessary background material. In Section \ref{sec:opponent_modelling} we introduce our novel algorithms along with extensions for opponent learning awareness and modelling. Section \ref{sec:experiments} presents an experimental evaluation of our proposed algorithms in several different MONFGs. Section \ref{sec:related_work} surveys related prior work on opponent modelling. Finally, Section \ref{sec:conclusion} concludes the paper with some closing remarks and a discussion of promising directions for future research. 

\section{Background}
\label{sec:background}
In this section, we discuss the necessary background material on MONFGs, multi-objective optimisation criteria, utility functions, solution concepts, policy-based learning and opponent modelling algorithms. For a more complete overview of the field of multi-objective multi-agent decision making we refer the interested reader to a recent survey article by \citet{Radulescu2020Survey}.

\subsection{Multi-Objective Normal Form Games}
\label{sec:MONFGs}

We are interested in a setting where multiple agents, each having different preferences with respect to the objectives, are interacting and learning to optimise the utility they receive. We use the framework of multi-objective normal form games (MONFG) to model the agents' interactions.

\begin{definition}[Multi-objective normal-form game]
\label{def:monfg}
An $n$-person finite multi-objective normal-form game $G$ is a tuple $\tuple{N, \mathcal{A}, \mathbf{P}}$, with $n\ge2$ and $\nobj\ge2$ objectives, where:
\begin{itemize}
    \item $N=\{1, \ldots, n\}$ is a finite set of agents. 
    \item $\mathcal{A} = A_1 \times \dots \times A_n$, where $A_i$ is the finite action set of agent~$i$ (i.e., the pure strategies of $i$). An \textit{action (pure strategy) profile} is a vector $\mathbf{a}=\tuple{a_1, \ldots, a_n} \in \mathcal{A}$. 
    \item $\mathbf{P}=\tuple{\mathbf{p}_1, \ldots, \mathbf{p}_n}$, where $\mathbf{p}_i \colon \mathcal{A} \to \mathbb{R}^\nobj$ is the vectorial payoff of agent $i$, given an action profile. 
\end{itemize}
\end{definition}

\noindent
We adopt a utility-based perspective \citep{roijers2013survey}, by assuming that for each agent there exists a utility function that maps its vectorial payoffs to a scalar utility.

\subsubsection{Utility Functions}
In multi-objective normal-form games, the term payoff is used to denote the numeric vector received by agents after each interaction. As mentioned above, we also assume each agent $i$ has a utility function that maps this payoff to a scalar value: $u_i \colon \mathbb{R}^\nobj \to \mathbb{R}$, where $\nobj$ is the number of objectives.

In general, we only require that the utility functions $u_i$ belong to the class of monotonically increasing functions, i.e., given two joint strategies $\boldsymbol{\pi}$ and $\boldsymbol{\pi'}$: $(\forall \objindex: p^{\boldsymbol{\pi}}_{i,\objindex} \geq p^{\boldsymbol{\pi'}}_{i,\objindex}) \Rightarrow u_i({\bf p}_i^{\boldsymbol{\pi}}) \geq u_i({\bf p}_i^{\boldsymbol{\pi'}})$, where $p^{\boldsymbol{\pi}}_{i,\objindex}$ is the payoff in objective $\objindex$ for agent $i$ when the agents follow a joint strategy $\boldsymbol{\pi}$. In other words, if the value of one strategy is superior in at least one objective, we expect to maintain the same ranking after applying the utility function.

We are interested in the setting of repeated interactions, while going beyond the widely used class of linear utility functions, i.e., $u_i(\mathbf{p}_i) = \sum\limits_{\objindex=1}^{\nobj} w_{i,\objindex} p_{i,\objindex}$, and considering more general function classes. Furthermore, while the payoffs in MONFGs are known to the players, the utility that each agent derives from it remains hidden from the other agents. Learning about other agents through repeated interactions then becomes an essential component for allowing an agent to reach favourable outcomes.

\subsubsection{Optimisation Criteria}
In MONFGs each agent aims to optimise its utility. The utility of an agent can be derived by applying its utility function to its received payoffs. Contrary to single-objective games however, it matters when the utility function is applied. We distinguish between two options \citep{roijers2013survey,roijers2017multi}: (1) first computing the expectation over the payoffs obtained according to a joint strategy $\boldsymbol{\pi}$ and only then applying the utility function is called the \emph{scalarised expected returns (SER)} approach:
\begin{equation}
    u(\E[{\bf p}_i^{\boldsymbol{\pi}}]),
    \label{eqn:ser}
\end{equation}
 and (2) first applying the utility function before computing the expectation is called the \emph{expected scalarised returns (ESR)} approach:
\begin{equation}
    \E[u({\bf p}_i^{\boldsymbol{\pi}})].
    \label{eqn:esr}
\end{equation}

The distinction between these options only appears when considering non-linear utility functions \citep{roijers2017multi}. 
The choice between these criteria depends on what an agent is interested in optimising. ESR should be chosen when what matters is the utility of the payoff vector after every single interaction. Most previous research on MONFGs implicitly assumes ESR \citep{borm2003structure,lozovanu2005multiobjective}. Contrary, SER is more natural in the case of repeated interactions, as in SER the average payoff over multiple interactions determines the utility. SER is the most common  choice in the reinforcement learning (RL) literature \citep{roijers2013survey}, and has recently been analysed in MONFGs \citep{Radulescu2020Survey,radulescu2019equilibria}. As we are interested in learning over repeated interactions, we focus on SER.

\subsubsection{Solution concepts for MONFGs}

In a MONFG under SER, a Nash equilibrium (NE) \citep{Nash1951Non} is defined as a set of strategies for each agent, such that no agent can increase her SER by deviating from the equilibrium joint strategy \citep{radulescu2019equilibria}.

\begin{definition}[Nash equilibrium in a MONFG under SER]
\label{def:ne_ser}
A mixed strategy profile $\boldsymbol{\pi}^{\text{NE}}$ is a Nash equilibrium in a MONFG under SER if for all $i \in \{1,...,n\}$ and all $\pi_i \in \Pi_i$, with $\Pi_i$ the set of mixed strategies for agent $i$:
\begin{equation}
u_i \left( \E\left[ \mathbf{p}_i(\pi_{i}^{\text{NE}}, \boldsymbol{\pi}_{-i}^{\text{NE}}) \right]\right) \geq  u_i \left( \E\left[ \mathbf{p}_i(\pi_i, \boldsymbol{\pi}_{-i}^{\text{NE}}) \right]\right)
\label{eqn:ne_ser}
\end{equation}
\noindent i.e. $\boldsymbol{\pi}^{\text{NE}}$ is a Nash equilibrium under SER if no agent can increase the \emph{utility of her expected payoffs} by deviating unilaterally from $\boldsymbol{\pi}^{\text{NE}}$, where $\boldsymbol{\pi}_{-i}$ is the strategy profile without the strategy of agent $i$.
\end{definition}

Recent work \citep{radulescu2019equilibria} has demonstrated that NE need not exist in MONFGs under SER with non-linear utility functions; whether any NE exist in this setting depends on the payoff scheme of the MONFG and the utility functions of the agents. Given the lack of theoretical results for the behaviour or learning dynamics for these cases, it is interesting to experimentally determine and characterise the output in these settings.

\subsection{Policy Gradient and Actor-Critic}
Policy gradient \citep{Sutton1998,Williams:1992:SSG:139611.139614} is a family of reinforcement learning algorithms that directly learns a policy $\pi_{\policyparam}$ parameterised by $\policyparam$ instead of indirectly inferring a policy based on value functions as done in value-based methods. Policy gradient methods calculate the gradients of the objective, $J(\policyparam)$, with respect to $\policyparam$ using the agent's  experiences from interacting with the environment (i.e., observed states, actions and rewards) and update the parameters $\policyparam$ by taking a step in the direction of this gradient:

\begin{align}
\label{eq:pg}
    \policyparam \leftarrow \policyparam + \alpha \nabla J\left(\policyparam\right)
\end{align}

In addition to policy gradient methods, there is another powerful class of learning methods, called actor-critic methods. These methods learn a policy, referred to as the \emph{actor} as well as a value function, referred to as the \emph{critic} \citep{Sutton1998}. Policy gradient methods are therefore also known as actor-only methods. Compared to actor-only methods, using a critic typically reduces the variance in the gradients and thus often achieve a more stable policy update. For single-objective settings, popular state-of-the-art methods exist in both classes \citep{foerster2018lola,foerster2018dice,lowe2017multi}.

\subsection{Opponent Modelling}
\label{sec:b_om}
As the agents do not know each other's utility functions, it becomes increasingly important to explicitly learn about the other agents.
For modelling the opponent's policy, we consider here the approach of policy reconstruction using conditional action frequencies \citep{albrecht2018survey}. This implies that an agent will maintain a set of beliefs regarding the strategy of the opponent. Similar to the idea introduced for Opponent Modelling Q-learning \citep{uther1997adversarial}, joint-action learners \citep{claus1998dynamics} and fictitious play \citep{fudenberg1998theory}, we consider empirical distributions derived from observing the actions of the opponent over $w$ interactions, during which the policies of the agents remain unchanged. 

Let $\kappa_{i}(a)$ be the number of times agent $i$ observed agent $j$ take action $a \in A_{j}$. The probability that agent $j$ plays action $a$, according to agent $i$, is defined as: 
\begin{equation}
    P_{i}(a) = \frac{\kappa_{i}(a)}{\sum_{a' \in A_{j}}\kappa_{i}(a')}
    \label{eq:counter_OM}
\end{equation}

These probabilities can then be used by agent $i$ to represent the parameters $\policyparam_{j}$ of her opponent's policy $\pi_{j}$.

\subsubsection{Gaussian Processes}
\label{sec:gps}

Recent advances in multi-agent learning approaches have introduced the idea of learning with opponent learning awareness \citep{foerster2018lola}, or, in other words, an agent can learn while taking into consideration her opponent's learning step together with how this step is influenced by her own policy. This approach, thus allows one to also shape the learning process of the opponents. In a multi-objective setting, modelling the opponent's learning step is not straightforward. This is due to the fact that we also need to consider the unknown utility function of the other agent, since it is involved in the computation of the opponent's objective. To overcome this issue, we propose to use Gaussian processes as a function approximator for modelling and shaping the opponent's learning step. Gaussian processes are Bayesian regression models that are known for their sample-efficiency. One can easily define the class of functions considered by the fitting procedure and track the uncertainty over this class given the training set. Such a framework allows us to locally capture the updates of the opponents using a limited amount of samples.

Gaussian processes (GPs) \citep{rasmussen2003gaussian} are an extension of multivariate normal distributions. Specifically, a GP describes an infinite set of random variables, such that any arbitrary finite subset of variables follows a multivariate normal distribution.
In the context of regression, the outputs of the unknown function $f(\myvec{x})$ can be described as random variables,
\begin{equation}
\begin{split}
y(\myvec{x}) &= f(\myvec{x}) + \varepsilon\text{, with measurement noise}\\
\varepsilon &\sim \mathcal{N}(0, \sigma^2).
\end{split}
\end{equation}
As the number of random variables are infinite, they can be modelled using a Gaussian process. Formally, when we assume a zero-mean GP prior on the unknown function:
\begin{equation}
f(\myvec{x}) \sim \mathcal{GP}\left(0, k(\myvec{x}, \myvec{x}')\right),
\end{equation}
any finite set of measured outcomes $\myvec{y}$ can be modelled as
\begin{equation}
\myvec{y}\ |\ X \sim \mathcal{N}\left(0, K\right),
\end{equation}
where $K_{i,j} = k(x_i,x_j)$ is the correlation between the random outputs $y_i$ and $y_j$, based on their associated inputs $x_i$ and $x_j$, respectively.

When fitting a training set $\tuple{\xtrain,\ytrain}$, we can use Bayesian inference to compute the posterior's statistics for any set of outputs $\myvec{f}$,
\begin{equation}
\begin{split}
\E\left[\myvec{f}\ |\ X, \xtrain, \ytrain\right] &= K_{X,\xtrain}C^{-1}_{\xtrain,\xtrain} \ytrain\\
\V\left[\myvec{f}\ |\ X, \xtrain, \ytrain\right] &= K_{X, X} - K_{X, \xtrain} C^{-1}_{\xtrain,\xtrain} K_{\xtrain, X}\\
C_{\xtrain,\xtrain} &= K_{\xtrain, \xtrain} + \sigma^2 I,
\end{split}
\end{equation}
where $K_{X, X'}$ describes the pair-wise correlations between the outputs associated with sets $X$ and $X'$.

The choice of covariance kernel $k(\cdot, \cdot)$ defines various characteristics of the unknown function. A popular choice is the squared exponential (SE) kernel, defined as:
\begin{equation}
k_\text{SE}(\myvec{x}, \myvec{x}') = \exp\left(-0.5\sum^D_{d=1} \frac{\left(x_d - x'_d\right)^2}{l^2_d}\right),
\end{equation}
where $l_d$ is the length scale along input dimension $d$, describing the smoothness of the function. This kernel models continuous and differentiable functions, rendering it a popular choice for general modelling purposes. Evidence maximisation can be used to optimise the hyperparameters $l_d$ \citep{rasmussen2003gaussian}.

In the case of functions with multiple outputs, it is possible to define correlations between the different output variables as well. For example, one can use the following multi-task kernel:
\begin{equation}
k_\text{multi}(\tuple{\myvec{x}, e}, \tuple{\myvec{x'}, e'}) = k_\text{SE}(\myvec{x}, \myvec{x}')F_{e, e'},
\label{eq:kernel_multi}
\end{equation}
where $F_{e,e'}$ is the cross-covariance between the $e$-th and $e'$-th outputs. When the outputs are the same, the squared exponential kernel evaluates to $1$ and $F_{e,e}$ reflects the variance on a single output signal. When $F_{e,e'} = 0$ for $e \neq e'$, the resulting GP will consider all outputs to be independent. Similar to the length scales of the squared exponential kernel, the cross-covariance matrix $F$ can be optimised using evidence maximisation. For more information about the construction of this multi-task kernel, we refer the interested reader to the work by \citet{bonilla2008multi}.

\section{Opponent Modelling in MONFGs}
\label{sec:opponent_modelling}
In this paper, we investigate the effects of opponent modelling in the setting of MONFGs under SER with non-linear utility functions. We focus on understanding if opponent modelling can speed up learning or confer a significant advantage for agents who implement it in this setting. Furthermore, when considering MONFGs under SER, we also investigate whether there is a difference in the observed effect of opponent modelling in games with Nash equilibria, compared to games without Nash equilibria.

To investigate the effects of opponent modelling in MONFGs under SER, we design a series of policy-based and actor-critic algorithms specially adapted for this framework to optimise SER. 
Compared to value-based methods, policy-based and actor-critic methods allow the agents to learn an explicitly stochastic policy. This enables effective exploration and exploitation strategies that are significantly better than the often-used hard-coded epsilon-greedy exploration in value-based methods. 
More importantly however, stochastic policies are essential for the SER optimality criterion, 
as even if the opponents policy is fixed, the best response may still necessarily be stochastic. Therefore, enabling such explicitly stochastic policies is a significant improvement over recent work on reinforcement learning in MONFGs \citep{radulescu2019equilibria}, which used Q-learning with $\epsilon$-greedy, and required to be coupled with a nonlinear optimisation solver for allowing agents to determine at each step their optimal mixed strategy.

When considering opponent modelling, an intuitive approach is to model the opponent's policy 
$\pi^{\prime}$ directly; the simplest way is to represent the opponent's policy as an empirical distribution of action frequencies $\pi^{\prime}(a^{\prime})$ (Section~\ref{sec:b_om}). By using this modelling approach, the agent is able to aggregate information about the opponent's decision patterns and hence use it to improve its own policy.

Beyond just modelling the current opponent policy, we build on the idea of taking the fact that the opponent is learning into account \citep{foerster2018lola}. This is especially difficult in the multi-objective setting because, as stated above, an agent does not know the utility function of the opponents, so it does not know what the opponents are optimising for. The key idea behind our new methods is to train a Gaussian process as a function approximator to predict the opponent's learning step, while taking into consideration not only the opponent's current policy, but also the influence of the agent's own policy. We present a more in-depth explanation for this approach in Section~\ref{sec:gp_om}, followed by a detailed description of all our proposed learning algorithms.

From this section onward, for the sake of simplicity, we use two-agent MONFGs only. We therefore denote the agents as $\self$ and $\other$. Please note that our methods can be straightforwardly extended to more than two agents, by keeping an opponent model for each opposing agent, and adjusting the equations accordingly.

\subsection{Opponent Learning Awareness and Modelling using Gaussian Processes}
\label{sec:gp_om}

Let us denote the estimated policy parameters of the opponent at time $t$ as $\est{\policyparam}^t_{\other}$. The agent assumes that the opponent has the same type of policy parameters as itself.\footnote{Specifically, the parameters of a softmax policy for ACOLAM (Section~\ref{sec:acolam}) and the parameters of a sigmoid policy for LOLAM (Section~\ref{sec:lolam}).} The goal is to model the opponent's learning step, however, agents do not have access to each other's utility functions. Therefore, we must approximate this update step based on the observed interactions. The update performed by the opponent can be approximated by the change in policy parameters, divided by the learning rate, i.e.,
\begin{equation}
\begin{split}
    \est{\myvec{\Delta}}^t_{\other} = \frac{\est{\policyparam}^{t+1}_{\other} - \est{\policyparam}^{t}_{\other}}{\alpha_\text{in}},
\end{split}
\end{equation}
with $\alpha_\text{in}$ representing the supposed learning rate of the opponent under the assumption that the opponent is using a policy gradient approach for this update (Equation~\ref{eq:pg}).

Due to the central limit theorem, the uncertainty about the estimated values of ${\policyparam}_{\other}$ and $\myvec{\Delta}_{\other}$ can be described by Gaussians for large rollout batches \citep{billingsley2008probability}. Therefore, we model the Jacobian of the opponent using:
\begin{equation}
\begin{split}
    \myvec{\Delta}_{\other}\left(\policyparam_{\self}, \policyparam_{\other}\right) &= \jacobian_{\policyparam_{\other}} J_{\other}\left(\policyparam_{\self}, \policyparam_{\other}\right) + \varepsilon\\
    \jacobian_{\policyparam_{\other}} J_{\other}\left(\policyparam_{\self}, \policyparam_{\other}\right) &\sim \mathcal{GP}\left(0, k_\text{multi}(\policyparam_{\self}, \policyparam_{\other})\right)\\
    \varepsilon &\sim \mathcal{N}\left(0, \sigma^2\right),
\end{split}
\end{equation}
where $\varepsilon$ captures the approximation error. Note that we used the multi-task kernel, defined in Equation~\ref{eq:kernel_multi}, to capture correlations between the elements of the Jacobian.

At each time step, $t_\text{current}$, we define a training set:
\begin{equation}
\label{eq:trainset}
\begin{split}
    \xtrain &= \{\tuple{\policyparam^t_{\self}, \est{\policyparam}^t_{\other}}\}^{t_\text{current}}_{t=t_\text{lower}}\\
    \ytrain &= \{\est{\myvec{\Delta}}^t_{\other}\}^{t_\text{current}}_{t=t_\text{lower}},
\end{split}
\end{equation}
where $t_\text{lower} = \max(1,t_\text{current}-H+1)$ and $H$ is the maximum number of samples in the training set.
Using the posterior statistics described in Section~\ref{sec:gps}, we can compute the mean function:
\begin{equation}
\begin{split}
    \mean^\jacobian_{\other}\left(\policyparam_{\self}, \policyparam_{\other}\right) &= \E\left[\jacobian_{\policyparam_{\other}} J_{\other}\left(\policyparam_{\self}, \policyparam_{\other}\right)\ |\ \xtrain, \ytrain\right].
\end{split}
\end{equation}

\subsection{Actor-Critic for MONFGs} 
\label{sec:acmonfgs}

In this section we propose a line of algorithms of increasing complexity in the actor-critic family for SER. AC (Section~\ref{sec:ac-sec}) introduces the actor-critic framework for reinforcement learning in MONFGs, modelling the opponent as part of the environment. ACOM (Section~\ref{sec:acom}) improves upon AC by adding a learned model of the opponent's current policy. ACOLAM (Section~\ref{sec:acolam}) also considers the opponent's learning by predicting the opponent's learning updates using a Gaussian process. 

\subsubsection{Actor-Critic without Opponent Modelling (AC)} 
\label{sec:ac-sec}

When maximising its SER, we optimise the inner product of the action-values $\boldsymbol{Q}(a) \in \mathbb{R}^{\nobj}$ and the stochastic policy $\pi(a|\policyparam)$, parameterised by $\policyparam$. We define the SER objective of an agent as:

\begin{align}
    J(\policyparam)=u\left(\sum_{a \in A} \pi(a | \policyparam) \boldsymbol{Q}(a)\right)
\end{align}

\noindent where $u$ is the agent's (non-linear) utility function. Specifically, $\sum_{a} \pi(a | \policyparam) \boldsymbol{Q}(a)$ is an estimation of the expected multi-objective return vector, $\E[\payoff^{\boldsymbol{\pi}}_\self]$ (Equation~\ref{eqn:ser}).
We propose a base algorithm without opponent modelling as well as algorithms with opponent learning awareness and modelling within the actor-critic framework that optimise the SER objective,  $J(\policyparam)$.

To optimise SER, we have to take the gradients of $J(\policyparam)$ w.r.t $\policyparam$. We divide this into 2 iterative steps. 
First, the multi-objective action value vector $\boldsymbol{Q}(a)$ needs to be learned. After an action $a$ is chosen using  $\pi(a | \policyparam)$, the agent observes a vectorial payoff $\payoff$, and applies a stateless Q-learning update rule (as per \citep{radulescu2019equilibria}):

\begin{equation}
    \label{eqn:mo_qlearning}
    \boldsymbol{Q}(a) \leftarrow \boldsymbol{Q}(a)+\alpha_{\boldsymbol{Q}} \left(\payoff-\boldsymbol{Q}(a)\right),
\end{equation}

\noindent where $\alpha_{\boldsymbol{Q}}$ is the learning rate for Q-learning. After the action values have been updated, the objective $J$ is calculated. 

Second, the agent updates $\policyparam$ using the computed gradient of $J(\policyparam)$, by performing a gradient ascent step:
\begin{equation}
    \policyparam \leftarrow \policyparam + \alpha_{\policyparam} \jacobian J(\policyparam),
\end{equation}

\noindent where $\alpha_{\policyparam}$ is the learning rate. We detail in Algorithm~\ref{alg:ac} the update under AC for agent $\self$. 

\begin{algorithm}[h]
\DontPrintSemicolon
  \KwInput{experience $\tuple{a_{\self}, {\bf p}}$, learning rates $\alpha_{\boldsymbol{Q}}$, $\alpha_{\policyparam}$, policy parameters $\policyparam_{\self}$, utility function $u_{\self}$}
  
  \KwOutput{$\pi_{\self}$, $\boldsymbol{Q}$}
        
        Update Q-function: $\boldsymbol{Q}\left(a_{\self}\right) \leftarrow \boldsymbol{Q}\left(a_{\self}\right)+\alpha_{Q}\left[{\bf p}-\boldsymbol{Q}\left(a_{\self}\right)\right]$\\
        
        Calculate the gradient of the objective: $\nabla_{\policyparam_{\self}}J(\policyparam_{\self})=\nabla_{\policyparam_{\self}} u_{\self}\left(\sum_{a \in A_{\self}} \pi_{\self}(a|\policyparam_{\self}) \boldsymbol{Q}\left(a_{\self}\right)\right)$\\
        
        Update policy parameters: 
        $\policyparam_{\self} \leftarrow \policyparam_{\self} + \alpha_{\policyparam} \nabla_{\policyparam_{\self}} J(\policyparam_{\self})$

\caption{AC update for agent $\self$}
\label{alg:ac}
\end{algorithm}

\subsubsection{Actor-Critic with Opponent Modelling (ACOM)} 
\label{sec:acom}

We combine opponent modelling with our actor-critic algorithm, and propose the \emph{Actor-Critic with Opponent Modelling (ACOM)} algorithm. To do so, we make the following modifications. Firstly, instead of learning $\boldsymbol{Q}(a)$, a joint action value $\boldsymbol{Q}(a_{\self}, a_{\other})$ is learned to estimate the expected vectorial payoff for each possible joint action. Then, after each episode, the agent combines the updated $\boldsymbol{Q}(a_{\self}, a_{\other})$ and estimate of the opponent's policy $\pi_{\other}$ to evaluate the expected utility of its next action. Note that as stochastic policies are used by both the agent and its opponent, this requires marginalising out both policies: 

\begin{align}
    J(\policyparam_{\self})=u\left(\sum_{a_{\self} \in A_{\self}} \pi(a_{\self} | \policyparam_{\self}) \sum_{a_{\other} \in A_{\other}} \pi_{\other}\left(a_{\other}\ |\ \policyparam_{\other} \right) \boldsymbol{Q}\left(a_{\self}, a_{\other}\right)\right).
\end{align}
This results in the update step for agent $\self$ under ACOM (Algorithm \ref{alg:acom}).

\begin{algorithm}[h]
\DontPrintSemicolon
  \KwInput{experience $\tuple{a_{\self}, a_{\other}, {\bf p}}$, learning rates $\alpha_{\boldsymbol{Q}}$, $\alpha_{\policyparam}$, policy parameters $\policyparam_{\self}$, utility function $u_{\self}$, estimated opponent policy parameters $\est{\policyparam}_{\other}$}
  
  \KwOutput{$\pi_{\self}$, $\boldsymbol{Q}$}
        
        Update joint Q-function: $\boldsymbol{Q}\left(a_{\self},a_{\other}\right) \leftarrow \boldsymbol{Q}\left(a_{\self},a_{\other}\right)+\alpha_{Q}\left[{\bf p}-\boldsymbol{Q}\left(a_{\self},a_{\other}\right)\right]$\\
        
        Calculate the gradient of the objective: $\nabla_{\policyparam_{\self}}J(\policyparam_{\self})=\nabla_{\policyparam_{\self}} u_{\self}\left(\sum_{a \in A_{\self}} \pi_{\self}(a|\policyparam_{\self})\sum_{a^{\prime}\in A_{\other}}\pi_{\other}(a^{\prime}|\est{\policyparam}_{\other}) \boldsymbol{Q}\left(a,a^{\prime}\right)\right)$\\
        
        Update policy parameters: 
        $\policyparam_{\self} \leftarrow \policyparam_{\self} + \alpha_{\policyparam} \nabla_{\policyparam_{\self}} J(\policyparam_{\self})$

\caption{ACOM update for agent $\self$}
\label{alg:acom}
\end{algorithm}

\subsubsection{Actor-Critic with Opponent Learning Awareness and Modelling (ACOLAM)} 
\label{sec:acolam}

Finally, we propose \emph{Actor-Critic with Opponent Learning Awareness and Modelling (ACOLAM)}, which incorporates the use of a Gaussian process to model and shape the opponent's learning step. This requires the agent to also maintain a history of estimated opponent's policies, together with her own policies for the last $H$ steps, to create the training set $\xtrain$ and $\ytrain$ (Equation~\ref{eq:trainset}) for training the GP.

We detail in Algorithm~\ref{alg:acomgp} the update under ACOLAM for agent $\self$. 

\begin{algorithm}[h]
\DontPrintSemicolon
  \KwInput{experience $\tuple{a_{\self}, a_{\other}, {\bf p}}$, learning rates $\alpha_{\boldsymbol{Q}}$, $\alpha_{\policyparam}$,  $\alpha_{\text{in}}$ policy parameters $\policyparam_{\self}$, utility function $u_{\self}$, estimated opponent policy parameters $\est{\policyparam}_{\other}$, history of policies $\xtrain$ and opponent policy differences $\ytrain$,  lookahead $L$}
  
  \KwOutput{$\pi_{\self}$, $\boldsymbol{Q}$}
  
    Update joint Q-function: $\boldsymbol{Q}\left(a_{\self},a_{\other}\right) \leftarrow \boldsymbol{Q}\left(a_{\self},a_{\other}\right)+\alpha_{\boldsymbol{Q}}\left[{\bf p}-\boldsymbol{Q}\left(a_{\self},a_{\other}\right)\right]$\\
            
     Initialise opponent's $\policyparam^{\prime}_{\other} = \est{\policyparam}_{\other}$\\
  Train GP on $\tuple{\xtrain, \ytrain}$\\
    \For{l $\in$ \{1...L\}}{ 
        Predict posterior mean $\mean^\jacobian_{\other}\left(\policyparam_{\self}, \policyparam^{\prime}_{\other}\right)$ using GP inference\\
        Update $\policyparam'_{\other} \leftarrow \policyparam^{\prime}_{\other} + \alpha_{\text{in}} \mean^\jacobian_{\other}\left(\policyparam_{\self}, \policyparam^{\prime}_{\other}\right)$
    }
    
    Calculate the gradient of the objective: $\nabla_{\policyparam_{\self}}J(\policyparam_{\self})=\nabla_{\policyparam_{\self}} u_{\self}\left(\sum_{a \in A_{\self}} \pi_{\self}(a|\policyparam_{\self})\sum_{a^{\prime}\in A_{\other}}\pi_{\other}(a^{\prime}|\policyparam^{\prime}_{\other}) \boldsymbol{Q}\left(a,a^{\prime}\right)\right)$\\
        
    Update policy parameters: 
    $\policyparam_{\self} \leftarrow \policyparam_{\self} + \alpha_{\policyparam} \nabla_{\policyparam_{\self}} J(\policyparam_{\self})$
\caption{ACOLAM update for agent $\self$}
\label{alg:acomgp}
\end{algorithm}

\subsection{Policy Gradient for MONFGs}
\label{sec:pg}

In this section, we propose an algorithm in the policy gradient family for SER. Firstly, as a baseline algorithm, we extend the single-objective LOLA-DiCE (Learning with Opponent-Learning Awareness using the Infinitely Differentiable Monte-Carlo Estimator) algorithm \citep{foerster2018dice} to reinforcement learning in MONFGs by making some unrealistic assumptions, leading to Multi-Objective LOLA. Specifically, Multi-Objective LOLA requires the utility function of the opponent as well as the opponent's policy parameters to be known. This is of course unrealistic, as this information is not in fact accessible in an MONFG. Therefore, we propose LOLAM (Section~\ref{sec:lolam}) which predicts the opponent's learning updates using a Gaussian process, removing the need for the assumptions in Multi-Objective LOLA. 

Multi-Objective LOLA is a policy gradient method that allows agents to learn by optimising the following objective w.r.t. $\policyparam_{\self}$:

\begin{equation}
\label{eq:lolaobj}
    J_{\self}(\policyparam_{\self}, \policyparam_{\other}) = u_{\self} \left( \mathbb{E}_{\pi_{\policyparam_{\self}},\pi_{\policyparam_{\other} + \alpha_\text{in}\widetilde{\mathrm{\Delta}}_{\other}(\policyparam_{\self}, \policyparam_{\other})}} \left[ \bm{\mathcal{L}}_{\self}\right] \right),
\end{equation}
with $\bm{\mathcal{L}}_{\self} = \sum\limits^K_{k=1} \gamma^k \mathbf{p}^k_{\self}$ representing the return and
\begin{equation}
    \widetilde{\mathrm{\Delta}}_{\other}(\policyparam_{\self}, \policyparam_{\other}) = \nabla_{\policyparam_{\other}} u_{\other} \left( \mathbb{E}_{\pi_{\policyparam_{\self}},\pi_{\policyparam_{\other}}}  \left[ \bm{\mathcal{L}}_{\other}\right] \right)
\end{equation}
representing the opponent's anticipated learning step.

DiCE \citep{foerster2018dice} provides an estimator for the original objective, that can also be differentiated repeatedly, thus supporting higher-order gradient estimation. The implementation of this method relies on the $\epsdice{2}$ operator, defined as follows\footnote{For a detailed explanation of this approach, please refer to \citet{foerster2018dice}}:
 
 \begin{equation}
 \begin{aligned}\epsdice{2}(\mathcal{W}) &=\exp (\tau-\perp(\tau)) \\ \tau &=\sum_{w \in \mathcal{W}} \log (\prob{w\ |\ \policyparam}),
 \end{aligned}
 \end{equation}
 where $\mathcal{W}$ is the set of stochastic nodes that influence the original objective, and $\perp$ sets the gradient of the operand to zero, i.e., $\nabla_x\perp(x) = 0$. 

Equation~\ref{eq:lolaobj} can be defined as a DiCE-objective in the following manner: 

\begin{equation}
\label{eqn:diceobj}
    J_{\self,\epsdice{2} \left(\policyparam_{\self},\policyparam_{\other}\right)} = u_{\self} \left( \sum\limits_{k} \epsdice{2} \left( \left\{ a^{k' \le k}_{j \in \{1, 2\}} \right\} \right) 
    \gamma^k \mathbf{p}^k_{\self} \right), 
\end{equation}
 with $\left\{ a^{k' \le k}_{j \in \{\self, \other\}} \right\}$ being the set of actions taken by the agents up until time $k$, when performing a rollout of length $K$. The notation is kept consistent with the original work \citep{foerster2018dice}. For full implementation details, please refer to our Github repository: \url{https://github.com/rradules/opponent_modelling_monfg}.
 
 
 We now introduce the Multi-Objective LOLA update in Algorithm~\ref{alg:molola}, for the policy parameters of agent $\self$.

\begin{algorithm}[h]
\DontPrintSemicolon
  \KwInput{lookahead $L$, learning rates $\alpha_{\policyparam}$, $\alpha_{\text{in}}$, utility functions $u_{\self}$ and $u_{\other}$ and policy parameters $\policyparam_{\self}$ and $\policyparam_{\other}$ of each player.}
  
  \KwOutput{$\policyparam'_{\self}$}

  Initialise opponent's $\policyparam'_{\other} = \policyparam_{\other}$
  
    \For{l $\in$ \{1...L\}}{ 
        Rollout trajectories $\tau_{l}$ under $(\pi_{\policyparam_{\self}}, \pi_{\policyparam_{\other}})$
        
        Update $\policyparam'_{\other} \leftarrow \policyparam'_{\other} + \alpha_{\text{in}} \jacobian_{\policyparam'_{\other}}  J_{\other,\epsdice{2} \left(\policyparam'_{\other},\policyparam_{\self}\right)}$
    }
    Rollout trajectories $\tau$ under $(\pi_{\policyparam_{\self}},\pi_{\policyparam_{\other}})$
    
    Update $\policyparam'_{\self} \leftarrow \policyparam_{\self} + \alpha_{\policyparam} \jacobian_{\policyparam_{\self}}  J_{\self,\epsdice{2} \left(\policyparam_{\self},\policyparam'_{\other}\right)}$
    
\caption{Multi-Objective LOLA update for agent ${\self}$}
\label{alg:molola}
\end{algorithm}

Note that this initial version of the Multi-Objective LOLA approach requires full information regarding the opponent's policy parameters and utility function. In MONFGs, this information is not available, so we  need adapt the algorithm to account for this.

\subsubsection{Multi-Objective LOLA-DiCE with Opponent Modelling (LOLAM)}
\label{sec:lolam}
Since in most cases agents do not have access to their opponents' policy parameters and utility functions, these elements need to be modelled. LOLAM uses the same approach as described in ACOLAM for modelling the policy parameters of the opponent (Section~\ref{sec:b_om}), together with the opponent's update step (Section~\ref{sec:gp_om}). 

We now introduce the multi-objective LOLA-DiCE with opponent modelling update in Algorithm~\ref{alg:mololaom}, for the policy parameters of agent $\self$.

\begin{algorithm}[h]
\DontPrintSemicolon
  \KwInput{lookahead $L$, learning rates $\alpha_{\policyparam}$, $\alpha_\text{in}$, utility function $u_{\self}$, policy parameters $\policyparam_{\self}$, estimated opponent policy parameters $\est{\policyparam}_{\other}$, history of policies $\xtrain$ and opponent policy differences $\ytrain$}
  
  \KwOutput{$\policyparam'_{\self}$}

  Initialise opponent's $\policyparam'_{\other} = \est{\policyparam}_{\other}$\\
  Train GP on $\tuple{\xtrain, \ytrain}$\\
    \For{l $\in$ \{1...L\}}{ 
        Predict posterior mean $\mean^\jacobian_{\other}\left(\policyparam_{\self}, \policyparam'_{\other}\right)$ using GP inference\\
        Update $\policyparam'_{\other} \leftarrow \policyparam'_{\other} + \alpha_\text{in} \mean^\jacobian_{\other}\left(\policyparam_{\self}, \policyparam'_{\other}\right)$
    }
    
    Rollout trajectories $\tau$ under $(\pi_{\policyparam_{\self}},\pi_{\policyparam'_{\other}})$\\
    
    Update $\policyparam'_{\self} \leftarrow \policyparam_{\self} + \alpha_{\policyparam} \nabla_{\policyparam_{\self}}  J_{\self,\epsdice{2} \left(\policyparam_{\self},\policyparam'_{\other}\right)}$
    
\caption{LOLAM update for agent ${\self}$}
\label{alg:mololaom}
\end{algorithm}

As you can see, instead of the rollouts in Multi-Objective LOLA (Algorithm~\ref{alg:molola} -- Line 3), in LOLAM the rollouts are replaced with posterior mean Jacobian predictions using the GP. In that way we \emph{directly} estimate the learning of the opponent. LOLAM uses these estimates directly to predict the updates to $\policyparam_{\other}$, rather than calculating the opponent's objective $J_{\other,\epsdice{2}}$ and computing its gradient. This is because the GP estimates the learning step of the opponent from data, i.e., the GP includes estimates of how the opponent is learning. This is of course necessary, as due to us not knowing the other agent's utility function, we cannot infer the direction of the learning steps of the opponent otherwise, even if we would assume that the opponent follows the same learning algorithm. Moreover, we argue that this can be beneficial if the other agent does not in fact follow the same learning algorithm; there are no assumptions in the estimates of the learning of the other agent, it is all learned from interaction data. 

\section{Experimental Setup and Results}
\label{sec:experiments}
To evaluate the impact of opponent learning awareness and modelling, we use five 2-player 2-objective MONFGs with different properties. In all these MONFGs, we consider the utility functions as defined in \citep{radulescu2019equilibria}; the row player's utility function is: \begin{equation}
\label{eq:u1}
    u_{\self}( \myvec{p}_{\self} ) = \left(p_{\self,1}\right)^2 + \left(p_{\self,2}\right)^2,
\end{equation} while the column player's utility function is:
\begin{equation}
\label{eq:u2}
    u_{\other}(\myvec{p}_{\other}) = p_{\other,1} \cdot p_{\other,2}.
\end{equation}

We first introduce Game~\ref{table:balance_minus_R} (Table~\ref{table:balance_minus_R}) that has one NE in pure strategies under SER: (L,M). Secondly, we create a MONFG with multiple NE, inspired from \citep{radulescu2019equilibria},  referred to as Game~\ref{table:MONFG_2action_with_NE} (Table~\ref{table:MONFG_2action_with_NE}). There are two equilibria in this case: (L,L) and (M,M). (L,L) offers the highest utility for the row player, while (M,M) is the preferred outcome for the column player.  This allows us to focus closely on the competition between the agents for reaching their preferred equilibrium. For the third MONFG with NE, we use Game~\ref{table:MONFG_3action_with_NE}
(Table~\ref{table:MONFG_3action_with_NE}) \citep{radulescu2019equilibria}, having 3 pure Nash equilibria (i.e., (L,L), (M,M), (R,R)) under SER with the specified utility functions. The (R,R) NE is Pareto-dominated by the other equilibria. Again, (L,L) is the best outcome for the row player in terms of utility, while (M,M) is preferred by the column player. 

\begin{table}[ht!]
  \centering
      \caption{Game 1 -- A MONFG which has one pure strategy NE in (L,M) under SER, with expected payoffs of 10 and 3.}
    \setlength{\extrarowheight}{2pt}
    \begin{tabular}{cc|c|c|}
      & \multicolumn{1}{c}{} & \multicolumn{1}{c}{$L$}  & \multicolumn{1}{c}{$M$}  \\\cline{3-4}
      \multirow{2}*{}  & $L$ & $(4,0)$ & \nashhighlight $(3,1)$ \\\cline{3-4}
      & $M$ & $(3,1)$ & $(2,2)$ \\\cline{3-4}
    \end{tabular}
    \label{table:balance_minus_R}
\end{table}
 \begin{table}[ht!]
  \centering
      \caption{Game 2 -- A MONFG which has pure strategy NE in (L,L) -- payoffs (17, 4), and (M,M) -- payoffs (13, 6), under SER. Note that (L,L) offers the highest utility for the row player, whereas (M,M) offers the highest utility for the column player.}
    \setlength{\extrarowheight}{2pt}
    \begin{tabular}{cc|c|c|}
      & \multicolumn{1}{c}{} & \multicolumn{1}{c}{$L$}  & \multicolumn{1}{c}{$M$}  \\\cline{3-4}
      \multirow{2}*{}  & $L$ & \nashhighlight $(4,1)$ & $(1,2)$ \\\cline{3-4}
      & $M$ & $(3,1)$ & \nashhighlight $(3,2)$ \\\cline{3-4}
    \end{tabular}
    \label{table:MONFG_2action_with_NE}
\end{table}
\begin{table}[h!]
  \centering
      \caption{Game 3 -- A MONFG which has pure strategy NE in (L,L) -- payoffs (17, 4), (M,M) -- (13, 6), and (R,R) -- (10, 3), under SER \citep{radulescu2019equilibria}. Note that (L,L) and (M,M) Pareto-dominate (R,R), and that (L,L) offers the highest utility for the row player, whereas (M,M) offers the highest utility for the column player.}
    \setlength{\extrarowheight}{2pt}
    \begin{tabular}{cc|c|c|c|}
      & \multicolumn{1}{c}{} & \multicolumn{1}{c}{$L$}  & \multicolumn{1}{c}{$M$} & \multicolumn{1}{c}{$R$} \\\cline{3-5}
      \multirow{2}*{}  & $L$ & \nashhighlight $(4,1)$ & $(1,2)$ & $(2,1)$ \\\cline{3-5}
      & $M$ & $(3,1)$ & \nashhighlight $(3,2)$ & $(1,2)$ \\\cline{3-5}
      & $R$ & $(1,2)$ & $(2,1)$ & \nashhighlight $(1,3)$ \\\cline{3-5}
    \end{tabular}
    \label{table:MONFG_3action_with_NE}
 \end{table}
We also conduct experiments using two MONFGs without any NE under SER. For this setting, \citet{radulescu2019equilibria} have shown that NE need not exist. 

We use Game \ref{table:balance_2action} (Table \ref{table:balance_2action}), and the (Im)balancing Act MONFG, which we refer to as Game \ref{table:balance_original} (Table \ref{table:balance_original}), originally introduced in \citep{radulescu2019equilibria}. Both of these games exhibit similar dynamics when the players use the utility functions in Equations~\ref{eq:u1} and \ref{eq:u2}. To get the highest utility, agent 1 (row) wishes to make the objectives as imbalanced as possible, whereas agent 2 (column) prefers balanced objectives. Because of the structure of the payoffs, it is never possible to reach a stable equilibrium in pure or mixed strategies, as one of the agents always has an incentive to deviate towards its preferred pure strategy to gain extra utility \citep{radulescu2019equilibria}.

\begin{table}[ht!]
  \centering
      \caption{Game 4 -- with no NE under SER \citep{radulescu2019equilibria}}
    \setlength{\extrarowheight}{2pt}
    \begin{tabular}{cc|c|c|}
      & \multicolumn{1}{c}{} & \multicolumn{1}{c}{$L$}  & \multicolumn{1}{c}{$M$}  \\\cline{3-4}
      \multirow{2}*{}  & $L$ & $(4,0)$ & $(2,2)$ \\\cline{3-4}
      & $M$ & $(2,2)$ & $(0,4)$ \\\cline{3-4}
    \end{tabular}
    \label{table:balance_2action}
\end{table}
 \begin{table}[ht!]
  \centering
   \caption{Game 5 -- The (Im)balancing act MONFG \citep{radulescu2019equilibria}, with no NE under SER.}
    \setlength{\extrarowheight}{2pt}
    \begin{tabular}{cc|c|c|c|}
      & \multicolumn{1}{c}{} & \multicolumn{1}{c}{$L$}  & \multicolumn{1}{c}{$M$} & \multicolumn{1}{c}{$R$} \\\cline{3-5}
      \multirow{2}*{}  & $L$ & $(4,0)$ & $(3,1)$ & $(2,2)$ \\\cline{3-5}
      & $M$ & $(3,1)$ & $(2,2)$ & $(1,3)$ \\\cline{3-5}
      & $R$ & $(2,2)$ & $(1,3)$ & $(0,4)$ \\\cline{3-5}
    \end{tabular}
    \label{table:balance_original}
\end{table}

For each setting, agents interact for 3000 episodes, averaged over 30 trials. Most often, we present our results in the form of empirical outcome distributions for the last 10\% of the interactions, as it allows us to analyse the relative frequency of different joint actions being played upon convergence. In all the actor-critic settings, the gradient $\jacobian_{\policyparam}J(\boldsymbol{\theta})$ is computed analytically and the agents' policy $\pi(a|\boldsymbol{\theta})$ is represented using a softmax function: $\pi_{\self}(a = k\ |\ \policyparam_{\self})=\frac{e^{\theta_{\self,k}}}{\sum_{j=1}^{|A_{\self}|} e^{\theta_{\self,j}}}$. For the implementation of the policy gradient algorithms we use Pytorch \citep{Paszke2019pytorch} and we rely on the provided automatic-differentiation functionality to compute the required gradients.

We distinguish between two main case studies in our experiments, i.e., (1) the \textbf{full information setting}, which is a theoretical situation in which, despite the competitive games' settings, agents have access to each other's policy parameters and utility functions, and (2) the \textbf{no information setting}, which is more realistic and is the situation which we are interested in studying. In the no information setting the agents can only observe each other's actions and payoffs after each interaction.

\begin{figure}[h!]
\centering
\includegraphics[scale=0.65]{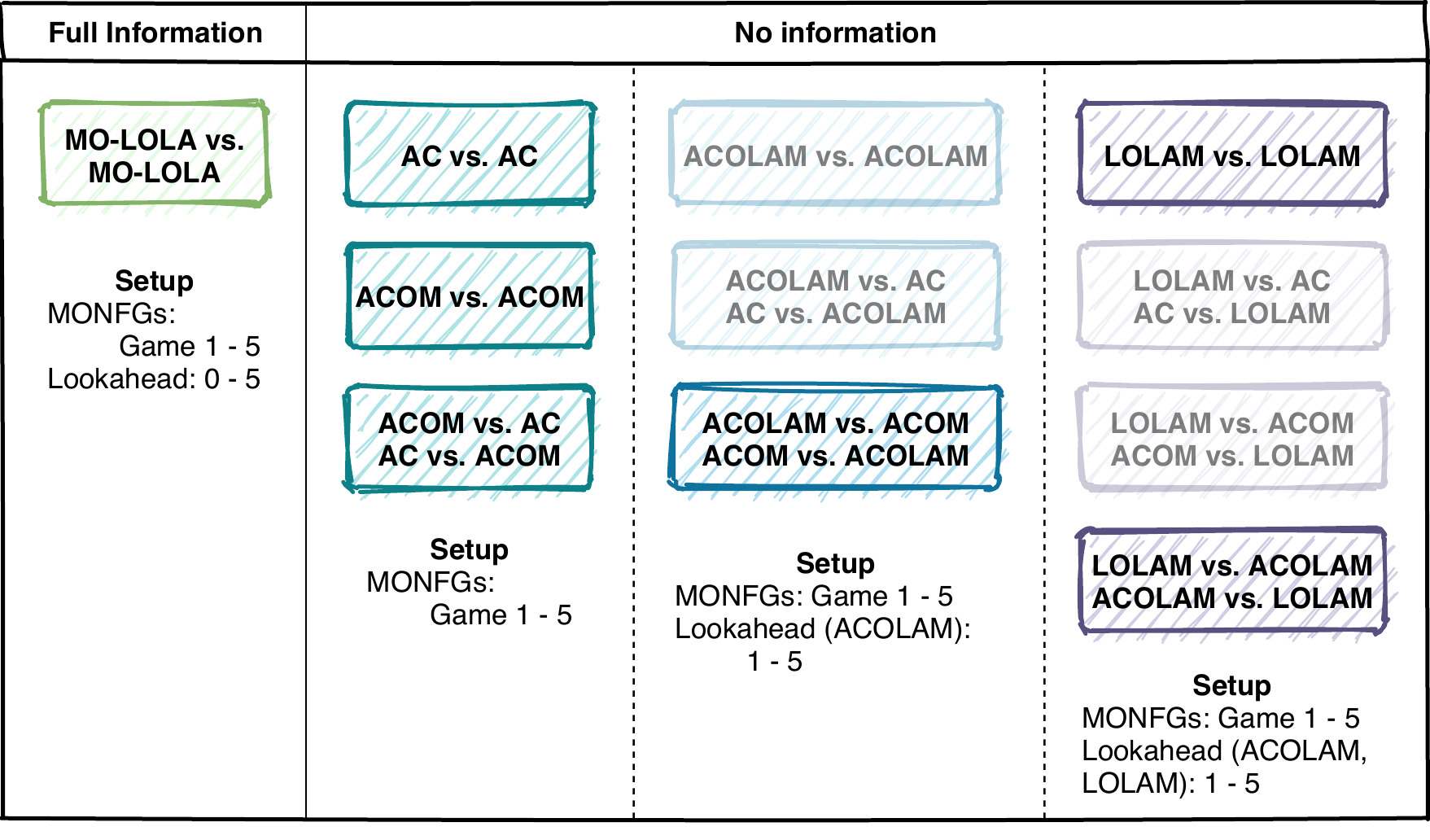}
\caption{Experimental overview, highlighting the settings we analyse in this work.}
\label{fig:algos}
\end{figure}

We present an overview of all the experiments we conducted in Figure~\ref{fig:algos}. Due to the asymmetry of our MONFGs, we analyse both variants for agent 1 and agent 2 when comparing two distinct approaches. We only discuss here the subset of results that exhibit distinct behaviours, i.e., the settings highlighted in Figure~\ref{fig:algos}. The results for all the settings are provided as supplementary material.\footnote{\url{https://github.com/rradules/opponent_modelling_monfg_results}}
\paragraph{Statement of reproducibility} The source code to reproduce all experiments is publicly available.\footnote{\url{https://github.com/rradules/opponent_modelling_monfg}}

Table~\ref{tab:params} shows an overview for the parameter values used throughout all experiments.
\begin{table}[ht!]
\centering
\caption{Overview of parameter values}
\begin{tabular}{|l|l|l|}
\hline
Algorithm           & Parameter                                         & Value \\ \hline
AC                  & $\alpha_{\boldsymbol{Q}}$                         & 0.05  \\ \hline
ACOM, ACOLAM        & $\alpha_{\boldsymbol{Q}}$                         & 1     \\ \hline
AC, ACOM, ACOLAM    & $\alpha_{\policyparam}$                           & 0.05  \\ \hline
ACOLAM        & $\alpha_{\text{in}}$                         & 0.05     \\ \hline
MO-LOLA, LOLAM         & $\alpha_{\policyparam}$                    & 0.1   \\ \hline
MO-LOLA, LOLAM         & $\alpha_{\text{in}}$                              & 0.2   \\ \hline
MO-LOLA, LOLAM         & $\gamma$                              & 1  \\ \hline
ACOLAM, LOLAM       & $H$ (GP training set size)                        & 50    \\ \hline
ACOM, ACOLAM, LOLAM & $w$ (opponent policy estimation window) & 100   \\ \hline
\end{tabular}
\label{tab:params}
\end{table}

\subsection{Full information setting - MO-LOLA vs. MO-LOLA}
\label{sec:fullinfo}

\begin{figure*}[h!]
\centering
\begin{subfigure}{0.3\textwidth}
    \centering
  \includegraphics[width=4cm]{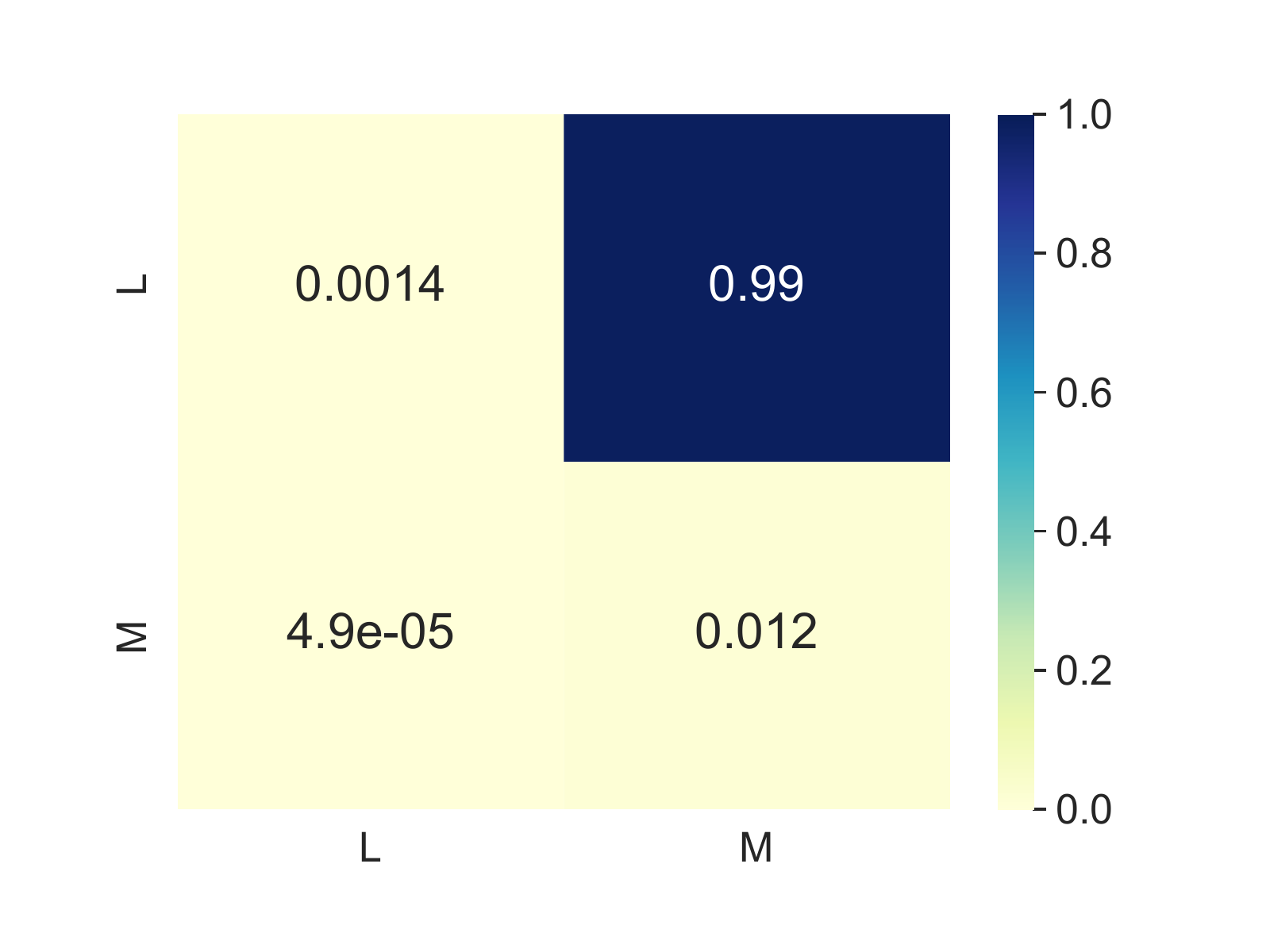}
  \caption{Game 1}
  \label{fig:lolag1}
\end{subfigure}
\begin{subfigure}{0.3\textwidth}
\centering
\includegraphics[width=4cm]{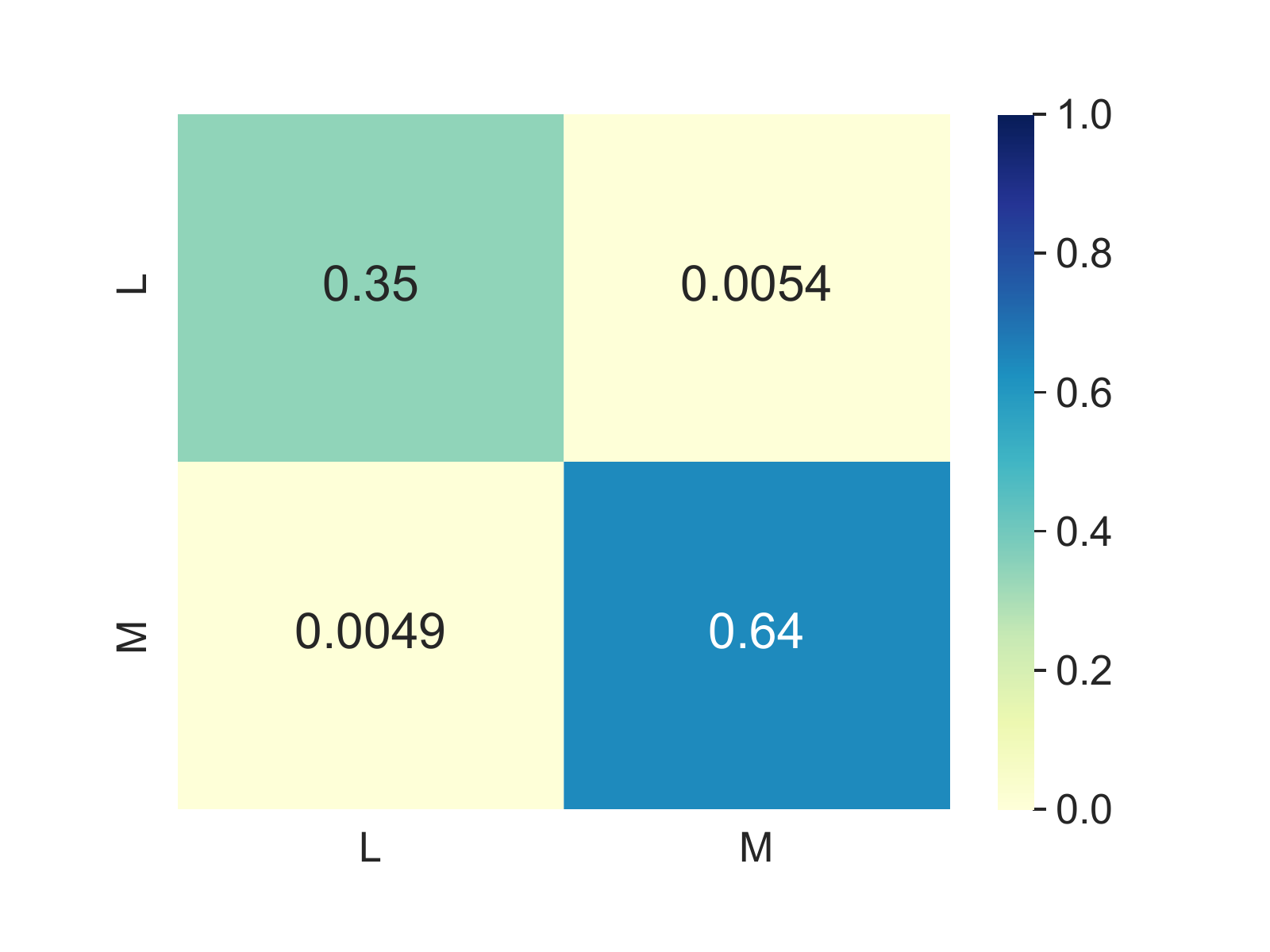}
\caption{Game 2}
\label{fig:lolag2}
\end{subfigure}
\begin{subfigure}{0.3\textwidth}
\centering
\includegraphics[width=4cm]{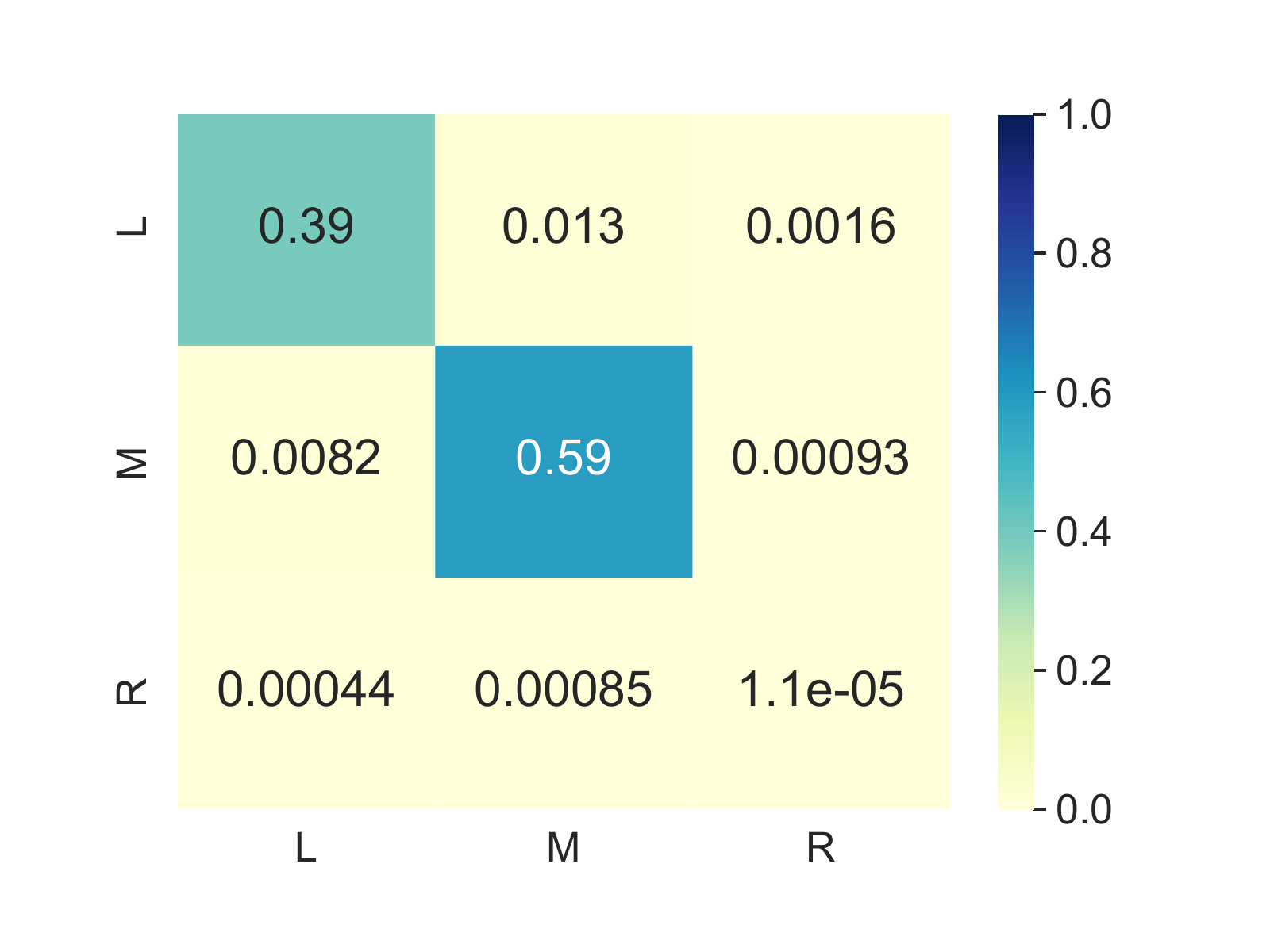}
\caption{Game 3}
\label{fig:lolag3}
\end{subfigure}
\caption{Empirical outcome distributions for MO-LOLA vs. MO-LOLA. Lookahead 1 for both agents.}
\label{fig:lolagms}
\end{figure*}

\paragraph{Game 1} Multi-Objective LOLA agents manage to reach the pure NE (L,M) with a probability of at least $\approx98\%$ under all the possible lookahead value combinations (Figure~\ref{fig:lolag1}). 

\begin{figure*}[h!]
\centering
\begin{subfigure}{0.3\textwidth}
    \centering
  \includegraphics[width=4cm]{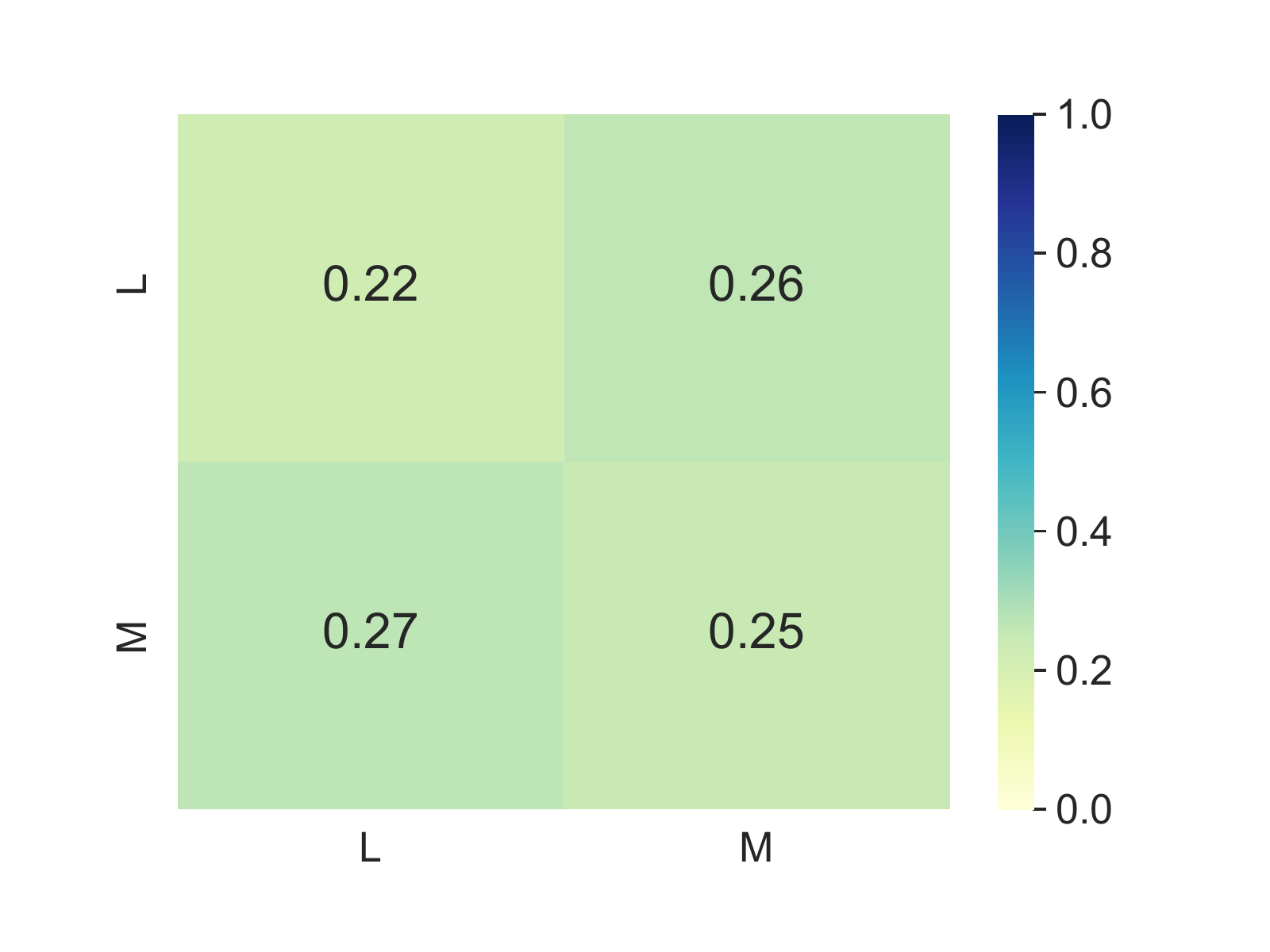}
  \caption{Empirical outcome distribution}
\end{subfigure}
\begin{subfigure}{0.3\textwidth}
\centering
\includegraphics[width=4cm]{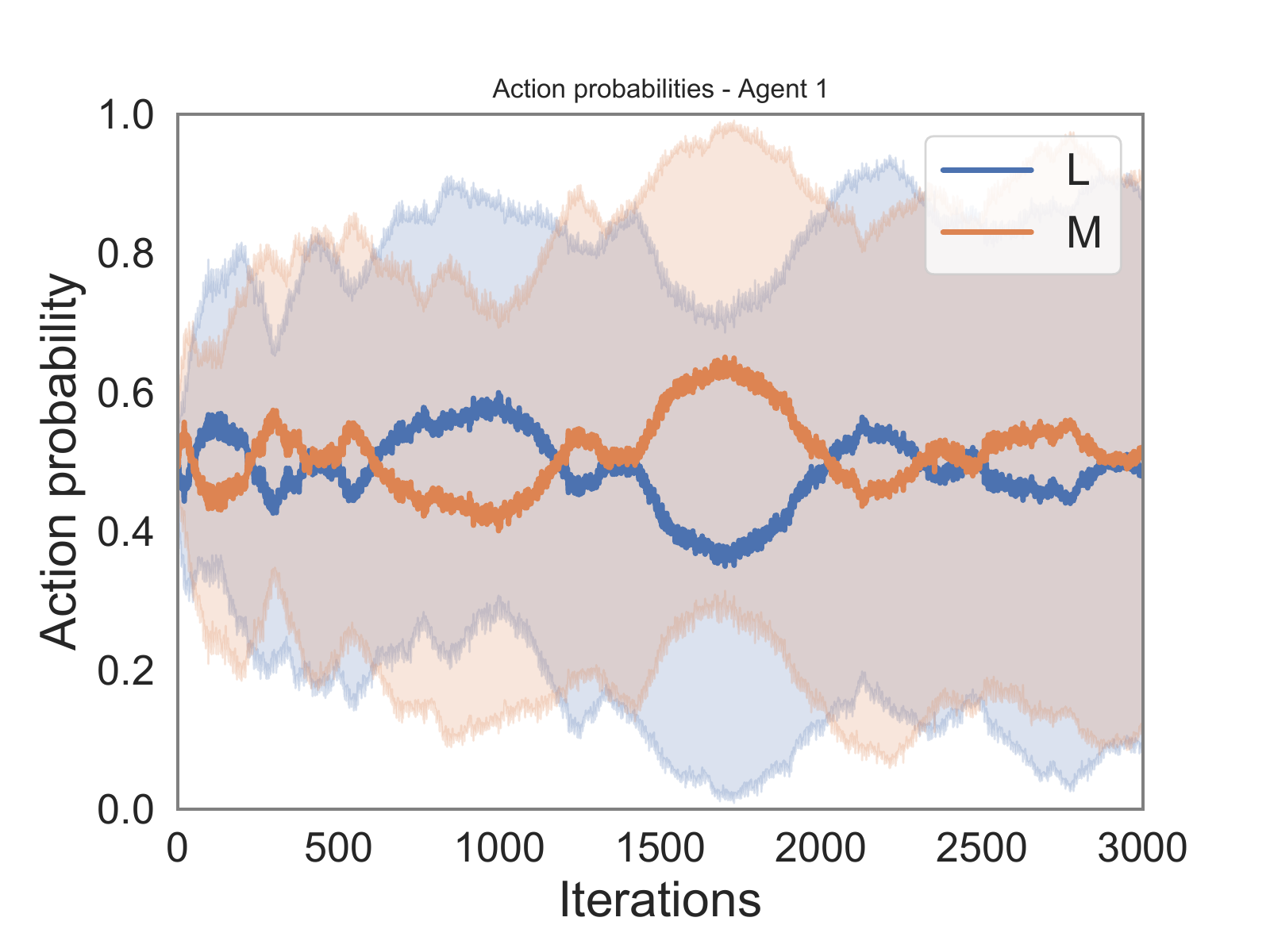}
\caption{Action probabilities of agent 1}
\end{subfigure}
\begin{subfigure}{0.3\textwidth}
\centering
\includegraphics[width=4cm]{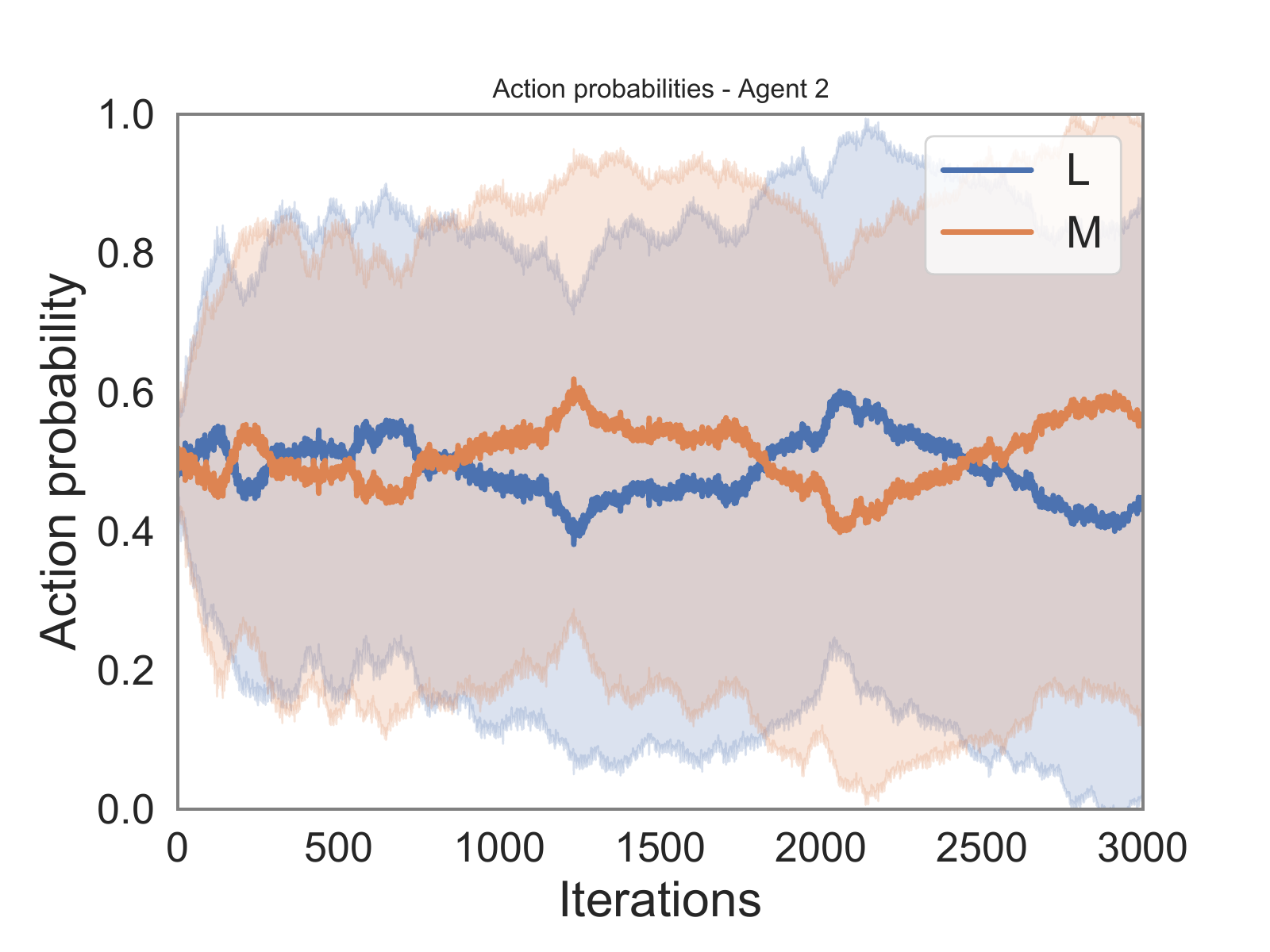}
\caption{Action probabilities of agent 2}
\end{subfigure}
\caption{Game 4 (Table~\ref{table:balance_2action}) -- MO-LOLA vs. MO-LOLA. The lookahead value for these instances is 2 for both agents.}
\label{fig:lolag4}
\end{figure*}

\begin{figure*}[h!]
\centering
\begin{subfigure}{0.3\textwidth}
    \centering
  \includegraphics[width=4cm]{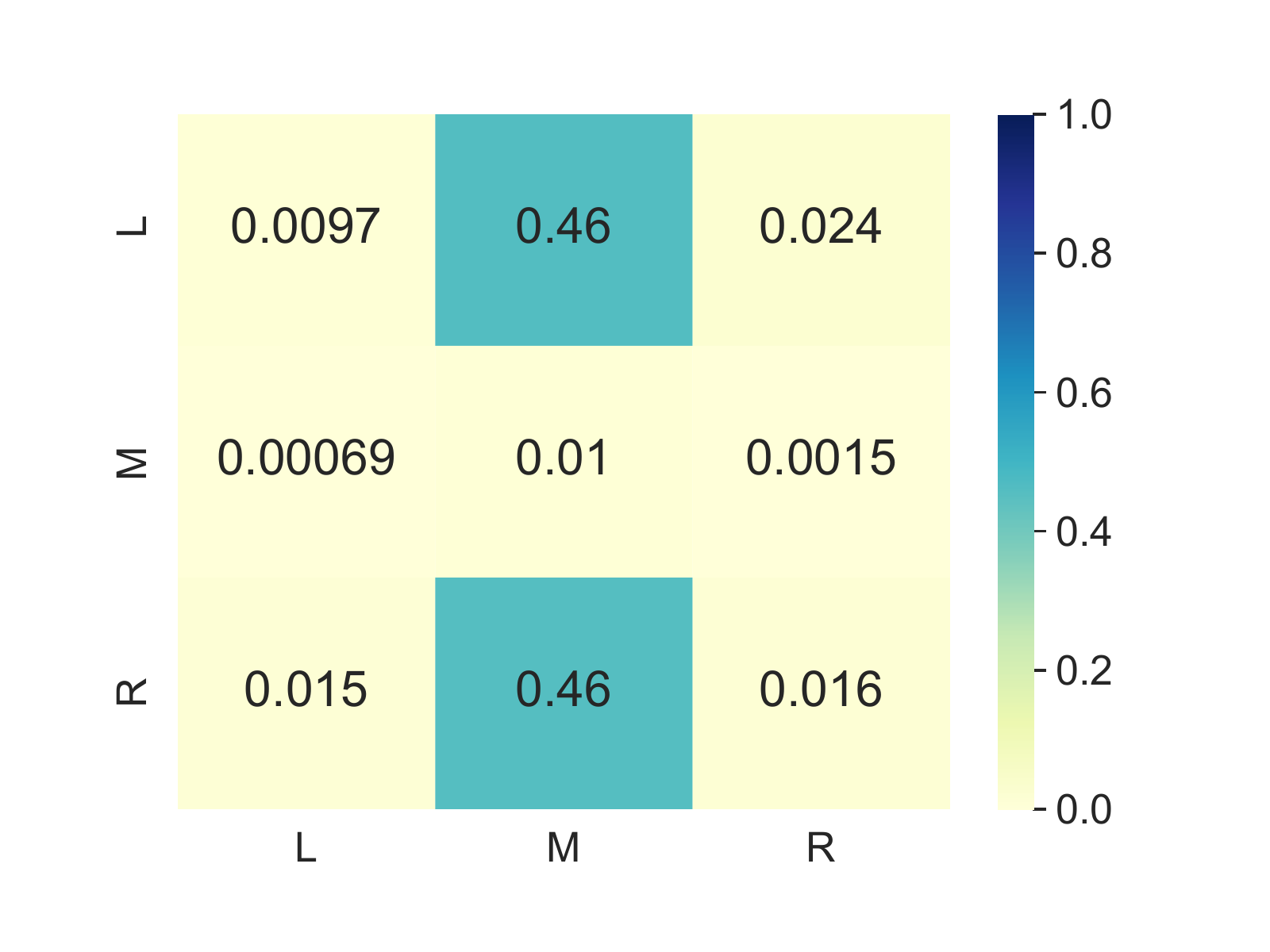}
  \caption{Empirical distribution outcome}
\end{subfigure}
\begin{subfigure}{0.3\textwidth}
\centering
\includegraphics[width=4cm]{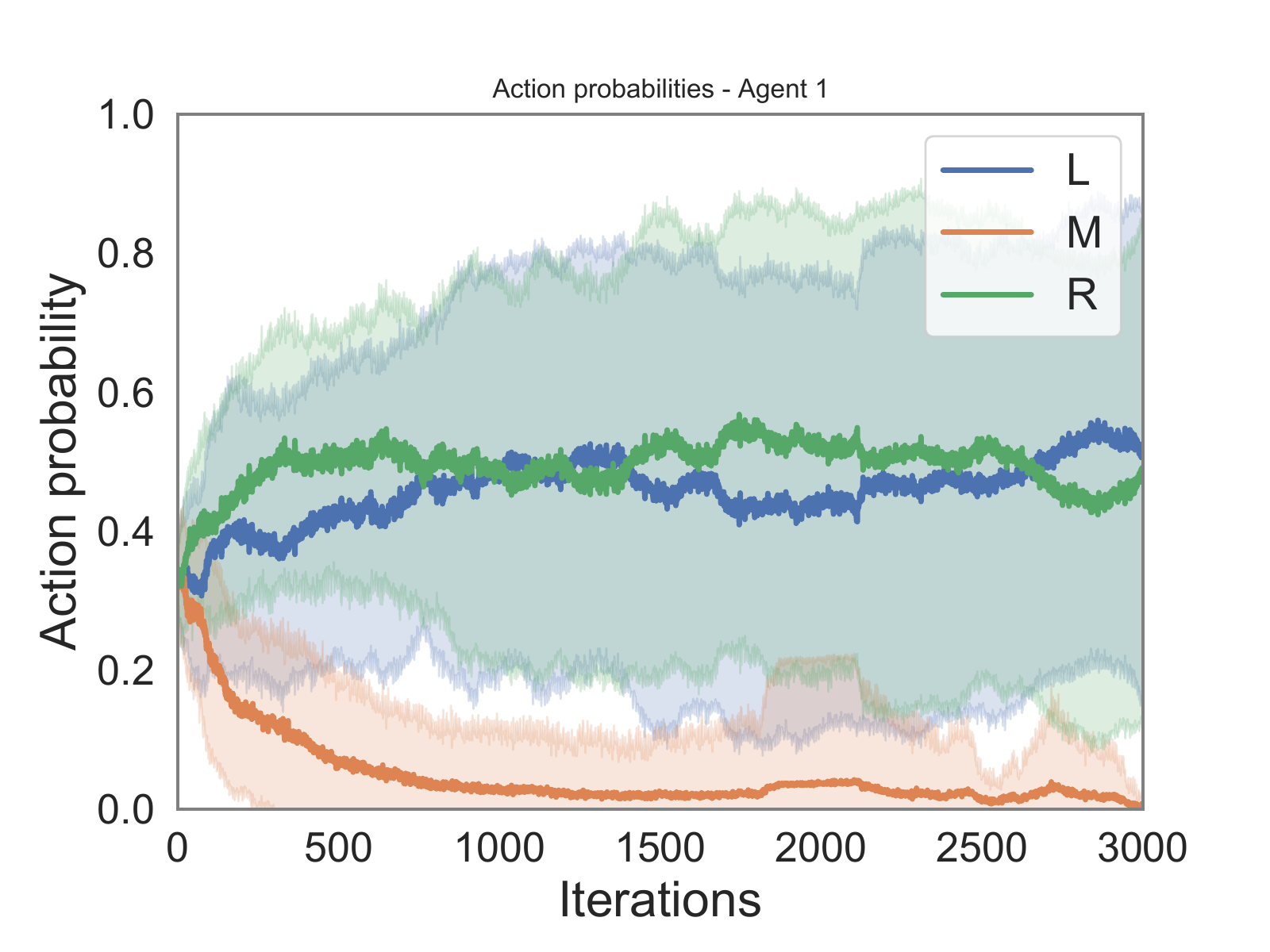}
\caption{Action probabilities of agent 1}
\end{subfigure}
\begin{subfigure}{0.3\textwidth}
\centering
\includegraphics[width=4cm]{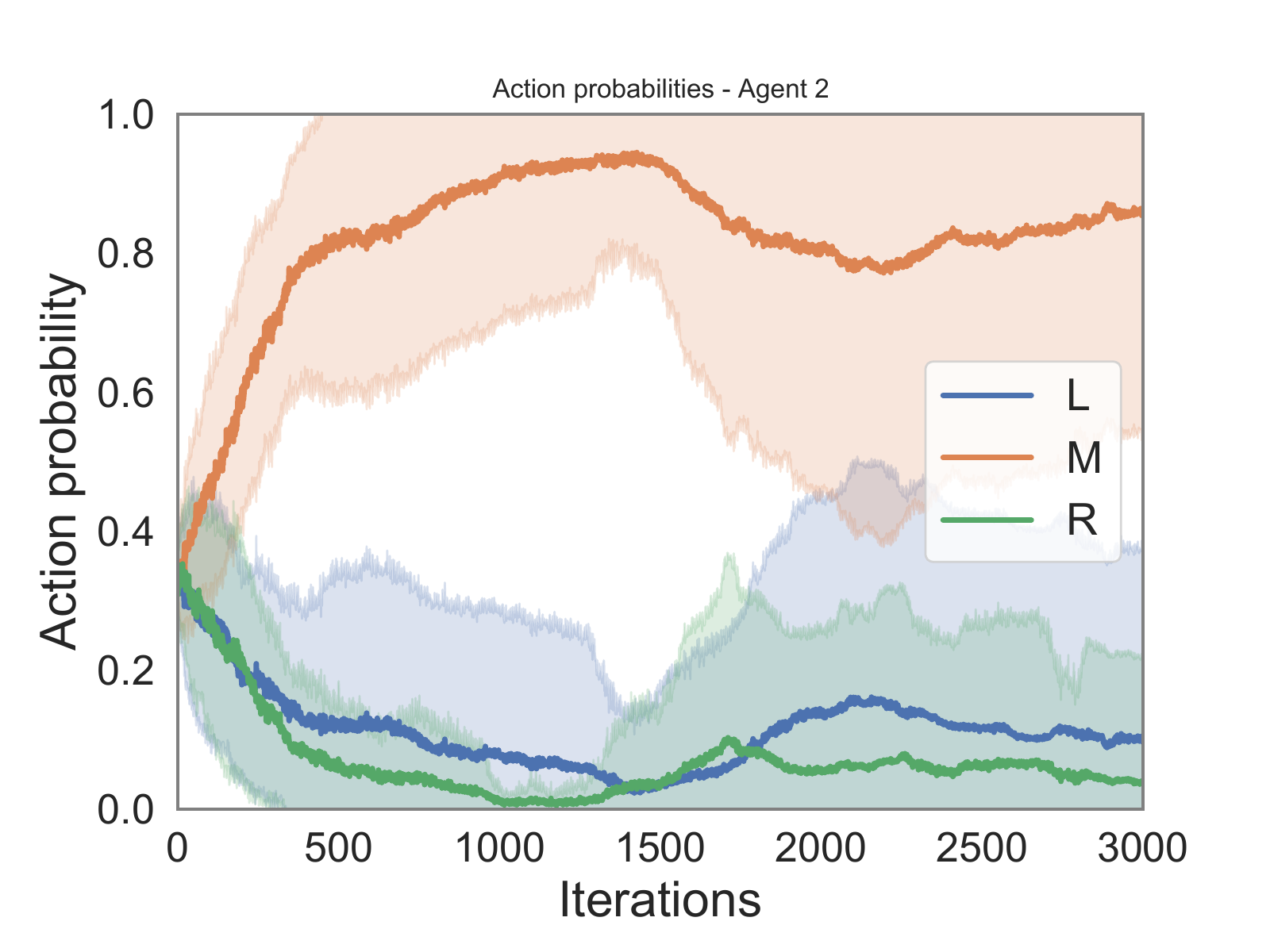}
\caption{Action probabilities of agent 2}
\end{subfigure}
\caption{Game 5 (Table~\ref{table:balance_original}) -- MO-LOLA vs. MO-LOLA. The lookahead value is 2 for both agents.}
\label{fig:lolag5}
\end{figure*}

\paragraph{Game 2} The agents reach the two pure NE (L,L) and (M,M) with a probability of $\approx99\%$ (Figure~\ref{fig:lolag2}). 

\paragraph{Game 3}
Multi-Objective LOLA agents manage to reach the two preferred pure NE (L,L) and (M,M) with a probability of $\approx99\%$. Furthermore, the agents' balance between their preferred NE outcomes and avoid the dominated NE (R,R)  (Figure~\ref{fig:lolag3}). 

\paragraph{Game 4}
Here, the agents present interesting learning dynamics, i.e., they cycle between all their possible joint actions, since there is no NE for this setting (Figure~\ref{fig:lolag4}). We notice that when the lookahead value increases, it becomes easier for the agents to converge to an equal probability distribution over their 2 actions, i.e., they presented less variance in their action probability evolution throughout the learning iterations. We note that this outcome was characterised in \citep{radulescu2019equilibria} as a correlated equilibrium for the game.

\begin{figure*}[h!]
\centering
\begin{subfigure}{0.3\textwidth}
    \centering
    \includegraphics[width=4cm]{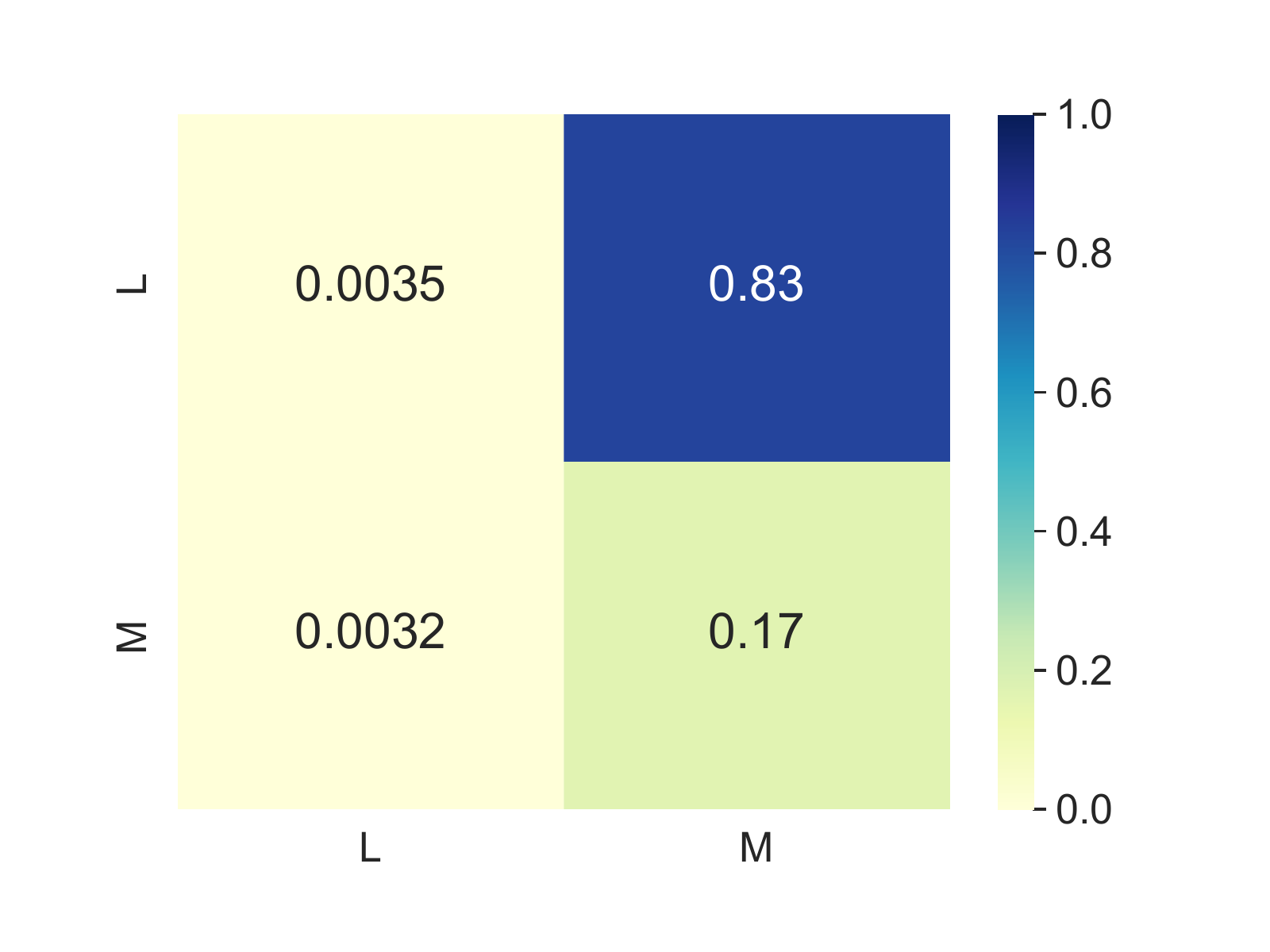}
    \caption{Game 1}
    \label{fig:acac_g1}
\end{subfigure}
\begin{subfigure}{0.3\textwidth}
    \centering
    \includegraphics[width=4cm]{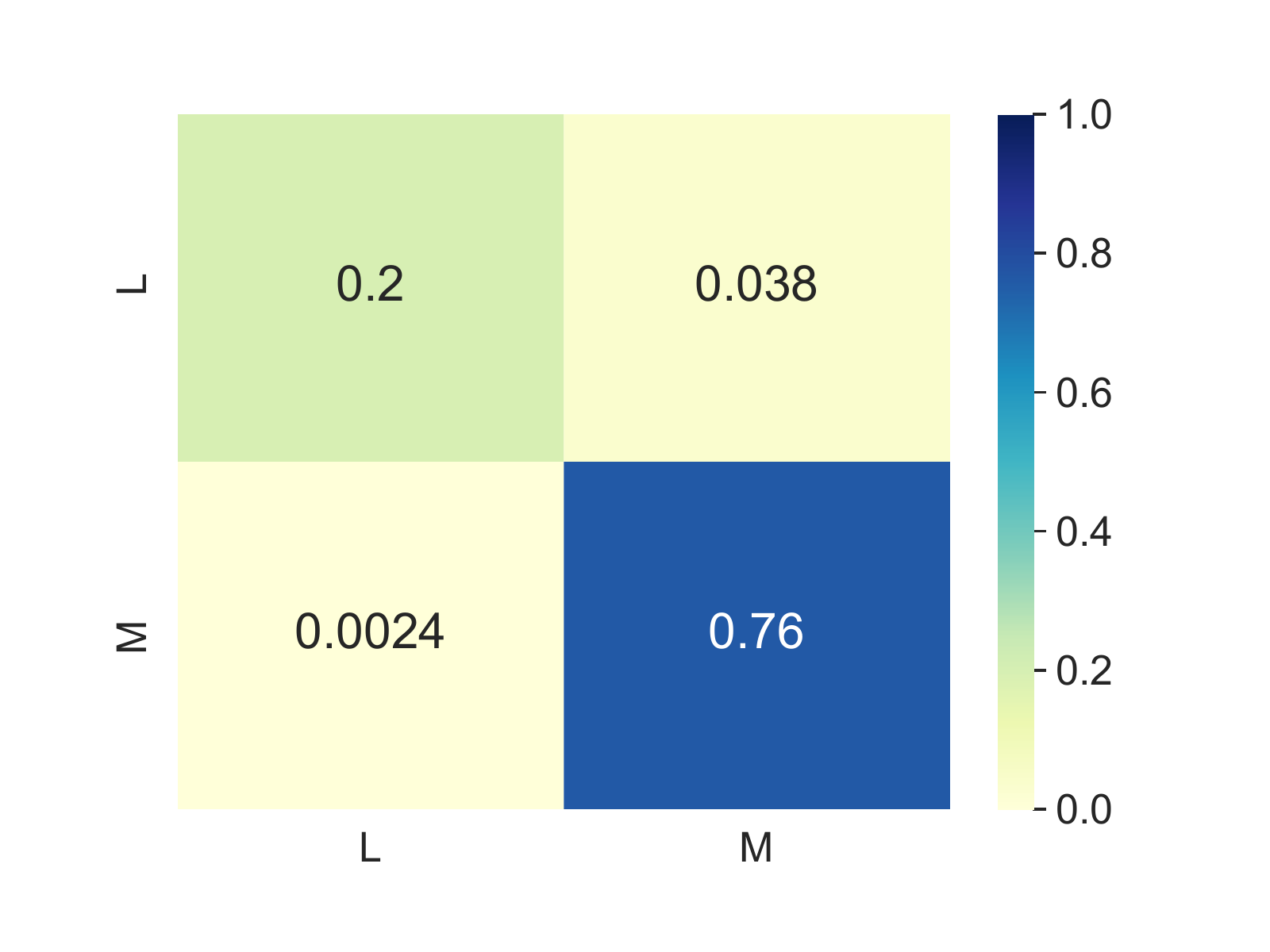}
    \caption{Game 2}
    \label{fig:acac_g2}
\end{subfigure}
\begin{subfigure}{0.3\textwidth}
    \centering
    \includegraphics[width=4cm]{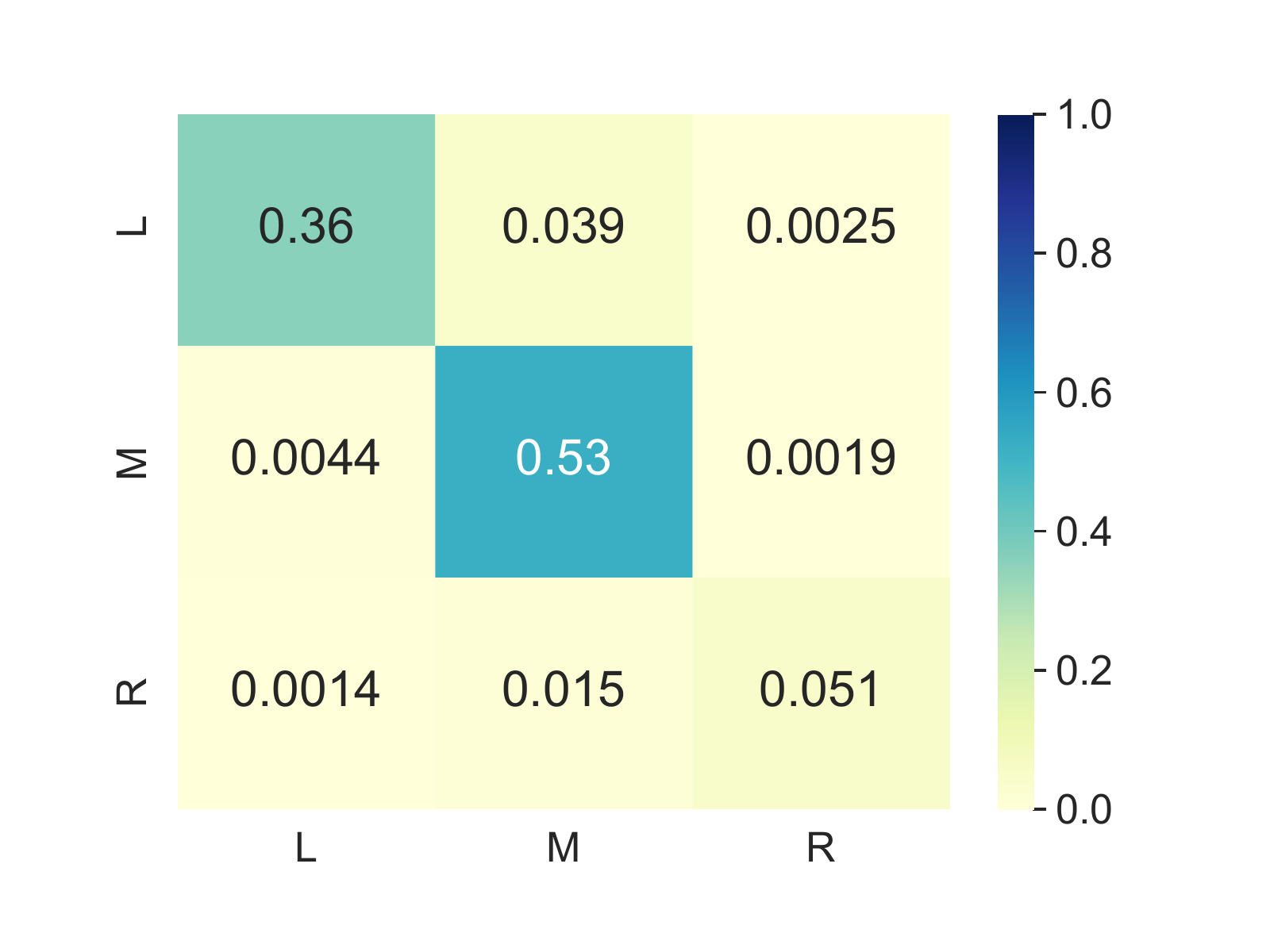}
    \caption{Game 3}
    \label{fig:acac_g3}
\end{subfigure}\\
\begin{subfigure}{0.3\textwidth}
    \centering
    \includegraphics[width=4cm]{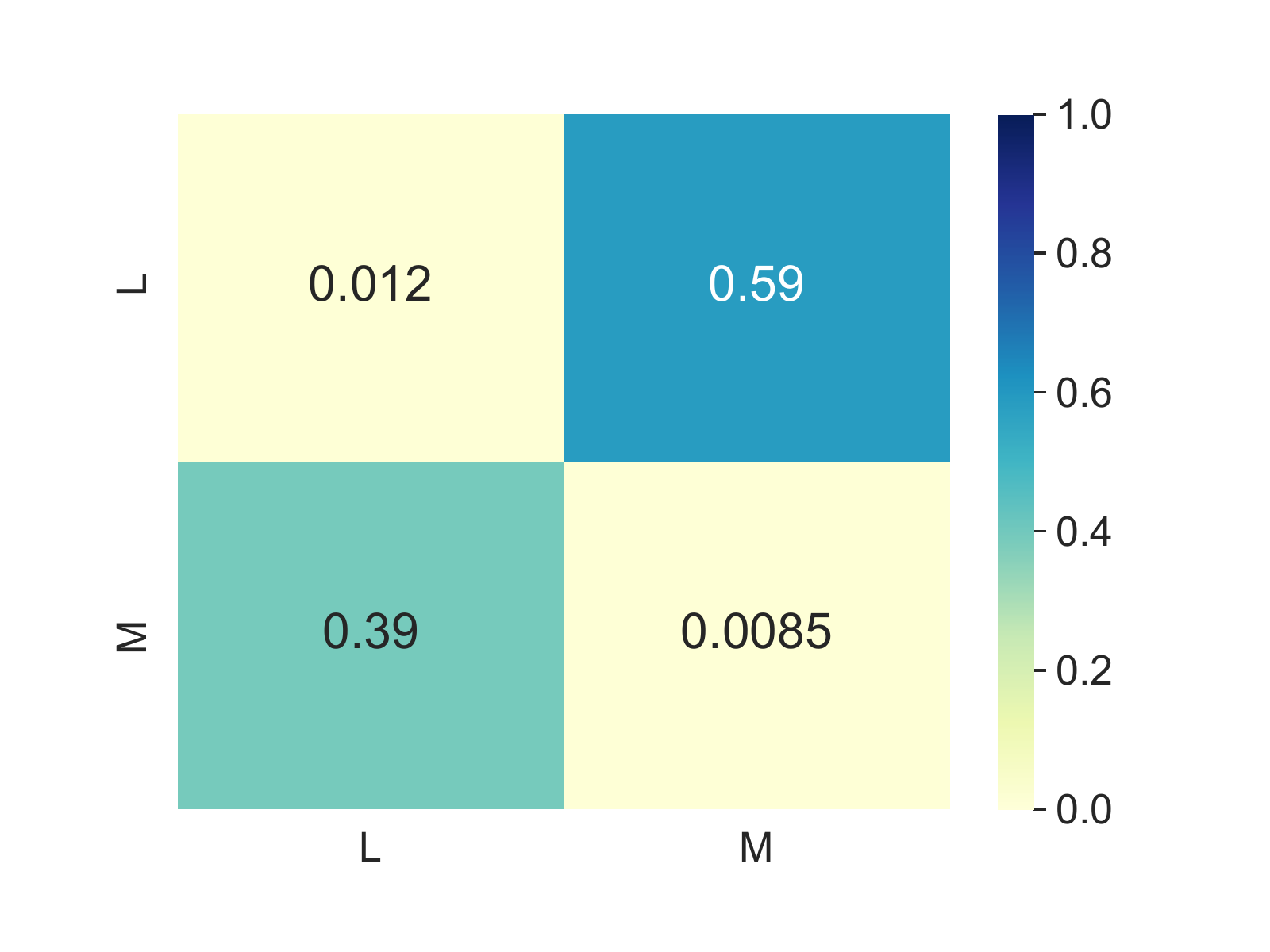}
    \caption{Game 4}
    \label{fig:acac_g4}
\end{subfigure}
\begin{subfigure}{0.3\textwidth}
    \centering
    \includegraphics[width=4cm]{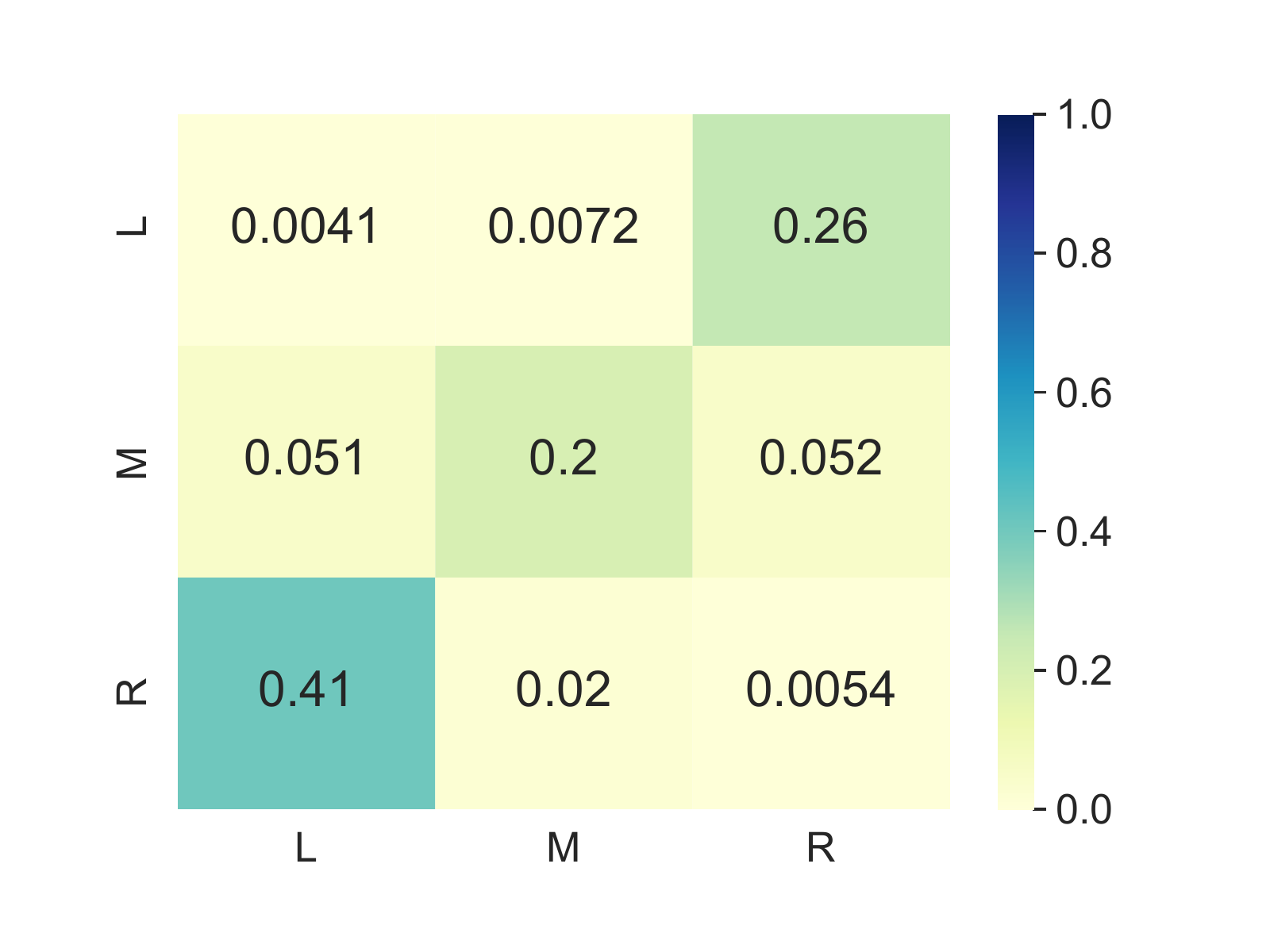}
    \caption{Game 5}
    \label{fig:acac_g5}
\end{subfigure}
\caption{Empirical outcome distributions for AC vs. AC.}
\label{fig:acac}
\end{figure*}

\paragraph{Game 5}
Despite the lack of a NE, Multi-Objective LOLA agents settle for a middle ground outcome, with agent 1 oscillating almost equally between actions L and R and agent 2 converging to action M (Figure~\ref{fig:lolag5}). This is again an interesting outcome, since it very closely matches one of the possible correlated equilibria for this game, according to the analysis of \citep{radulescu2019equilibria}. A possible correlated equilibrium in this game consists of the agents alternating between playing the joint actions (L,M) and (R,M), and our results demonstrate that the MO-LOLA agents manage to find this outcome without any external correlation signal. 

This is a significant result because there is no prior published example of a learning algorithm that converges to an equilibrium in the absence of a correlation signal in settings such as Game 4 and Game 5 (i.e., under SER with non-linear utility functions here NE do not exist). Furthermore, this provides a hopeful result for future analysis of MONFGs, since in MONFGs under SER, NE need not exist, thus requiring a different solution concept as the golden standard for MONFGs under SER. MO-LOLA (and the no-information version LOLAM) can potentially allow agents to learn to reach (approximate) equilibria in such settings, where other learning algorithms will fail entirely to reach meaningful outcomes.

\subsection{No information setting}
\label{sec:noinfo}

We now move to evaluating the no-information setting. This is the realistic MONFG setting. 

\subsubsection{AC and ACOM}

\begin{figure*}[h!]
\centering
\begin{subfigure}{0.3\textwidth}
    \centering
    \includegraphics[width=4cm]{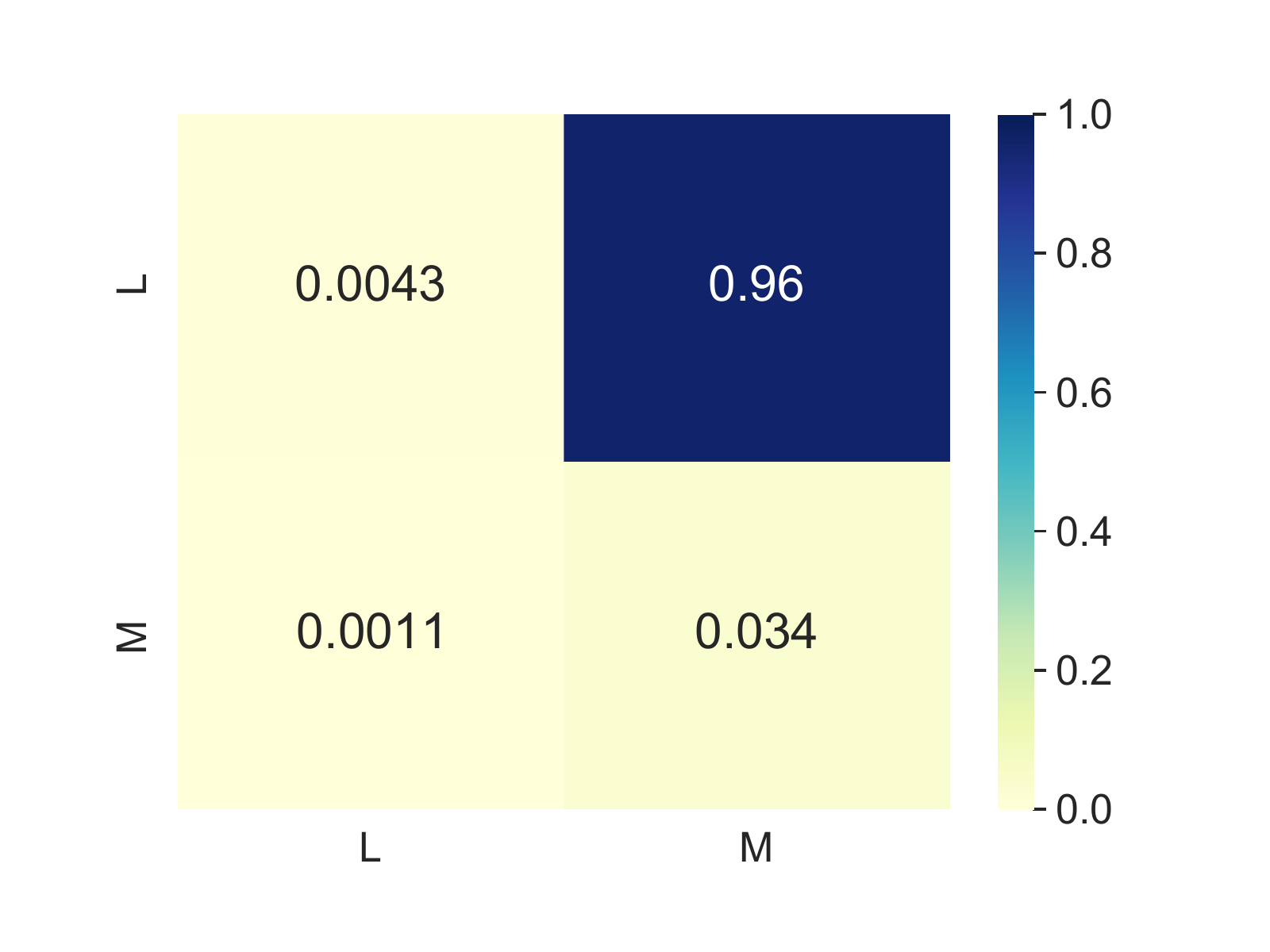}
    \caption{Game 1}
    \label{fig:acomacom_g1}
\end{subfigure}
\begin{subfigure}{0.3\textwidth}
    \centering
    \includegraphics[width=4cm]{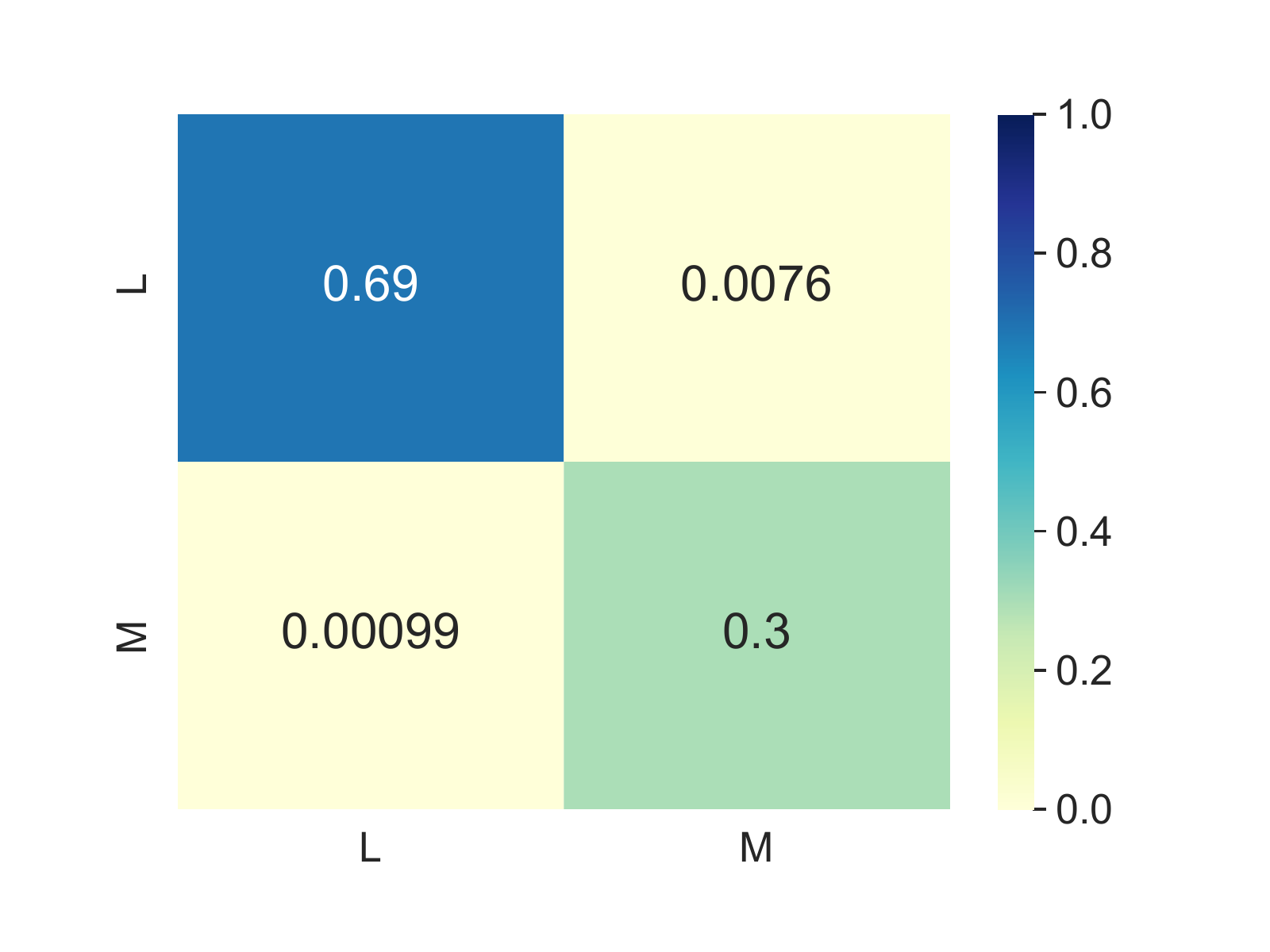}
    \caption{Game 2}
    \label{fig:acomacom_g2}
\end{subfigure}
\begin{subfigure}{0.3\textwidth}
    \centering
    \includegraphics[width=4cm]{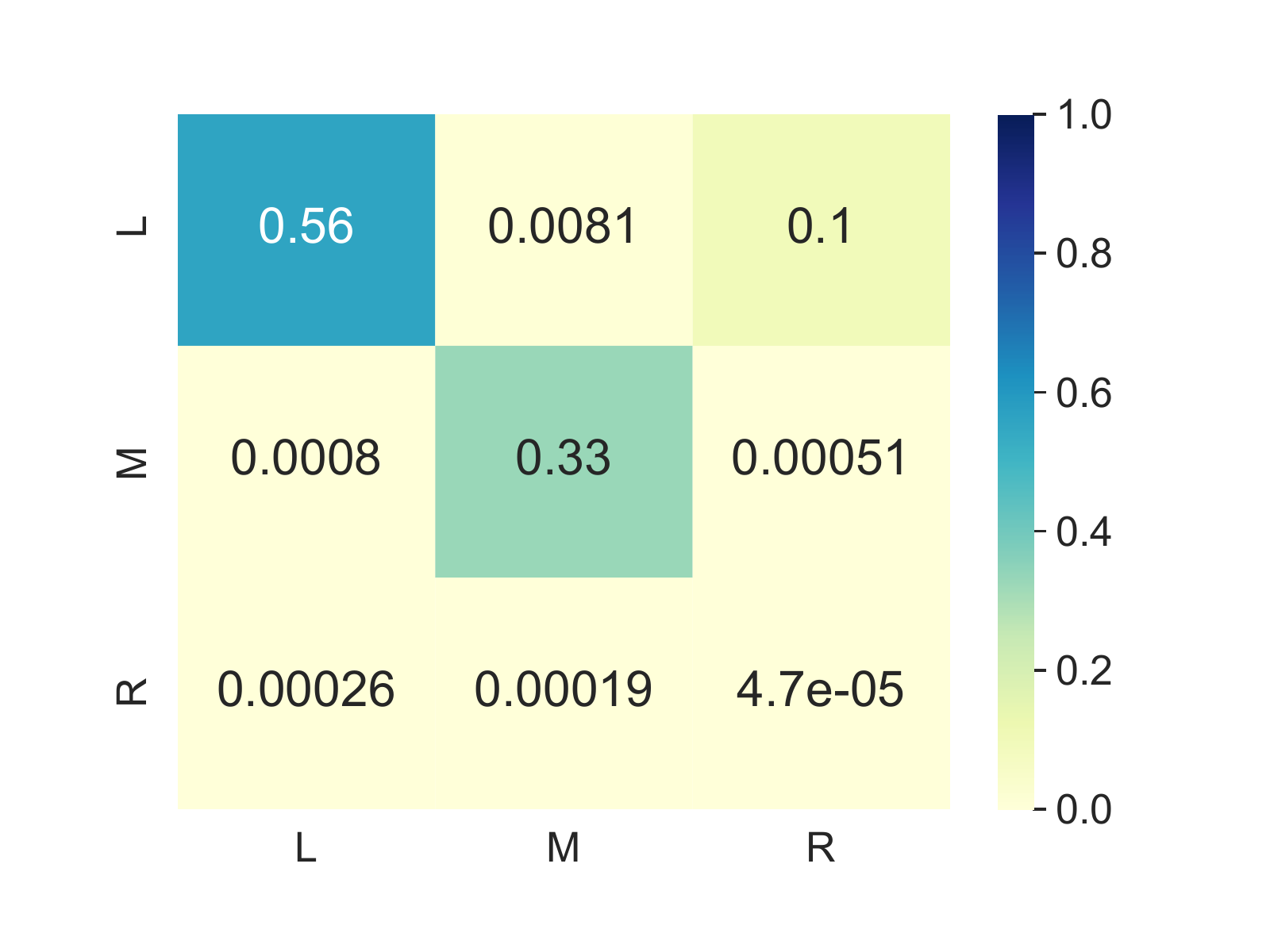}
    \caption{Game 3}
    \label{fig:acomacom_g3}
\end{subfigure}\\
\begin{subfigure}{0.3\textwidth}
    \centering
    \includegraphics[width=4cm]{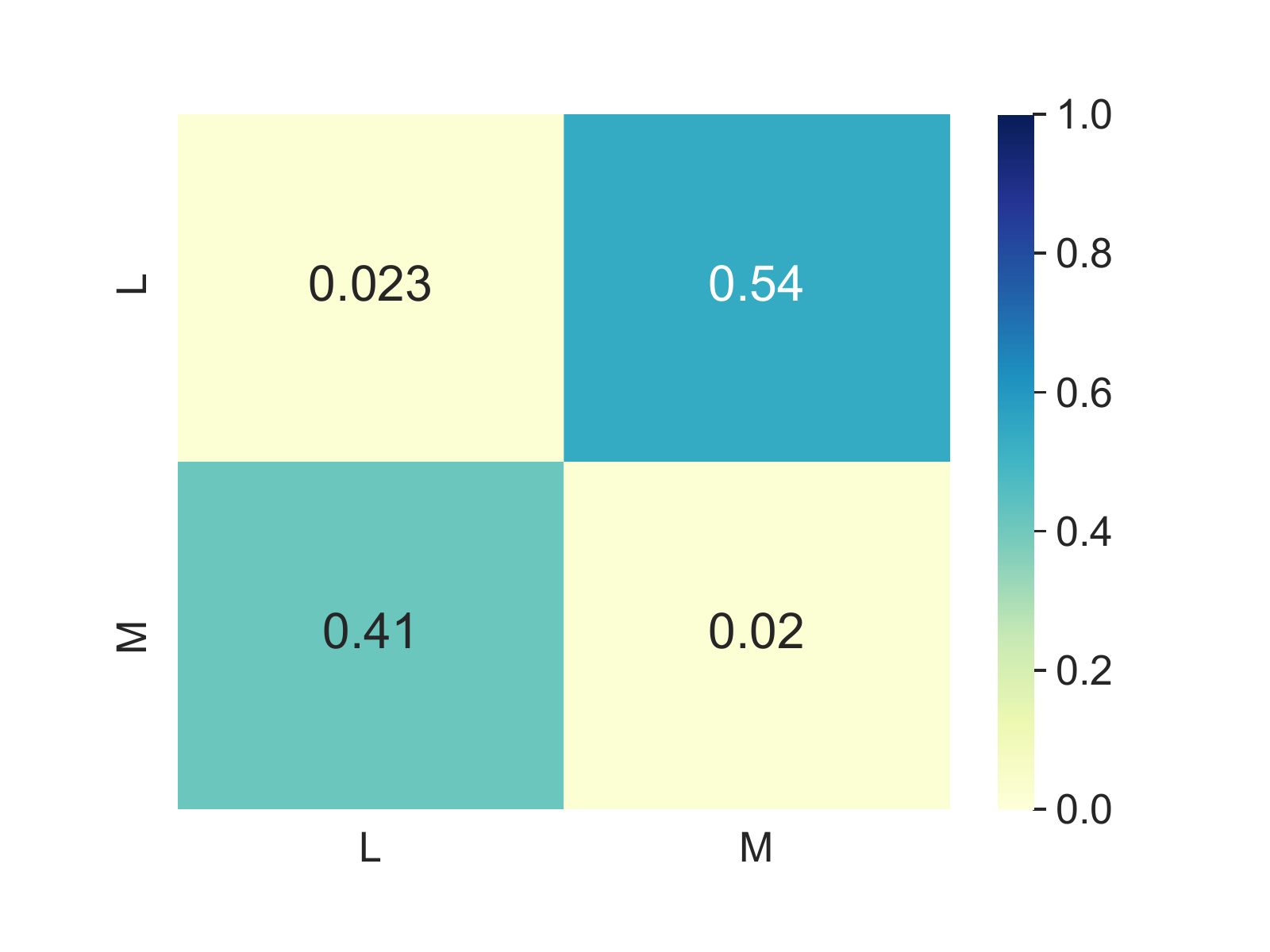}
    \caption{Game 4}
    \label{fig:acomacom_g4}
\end{subfigure}
\begin{subfigure}{0.3\textwidth}
    \centering
    \includegraphics[width=4cm]{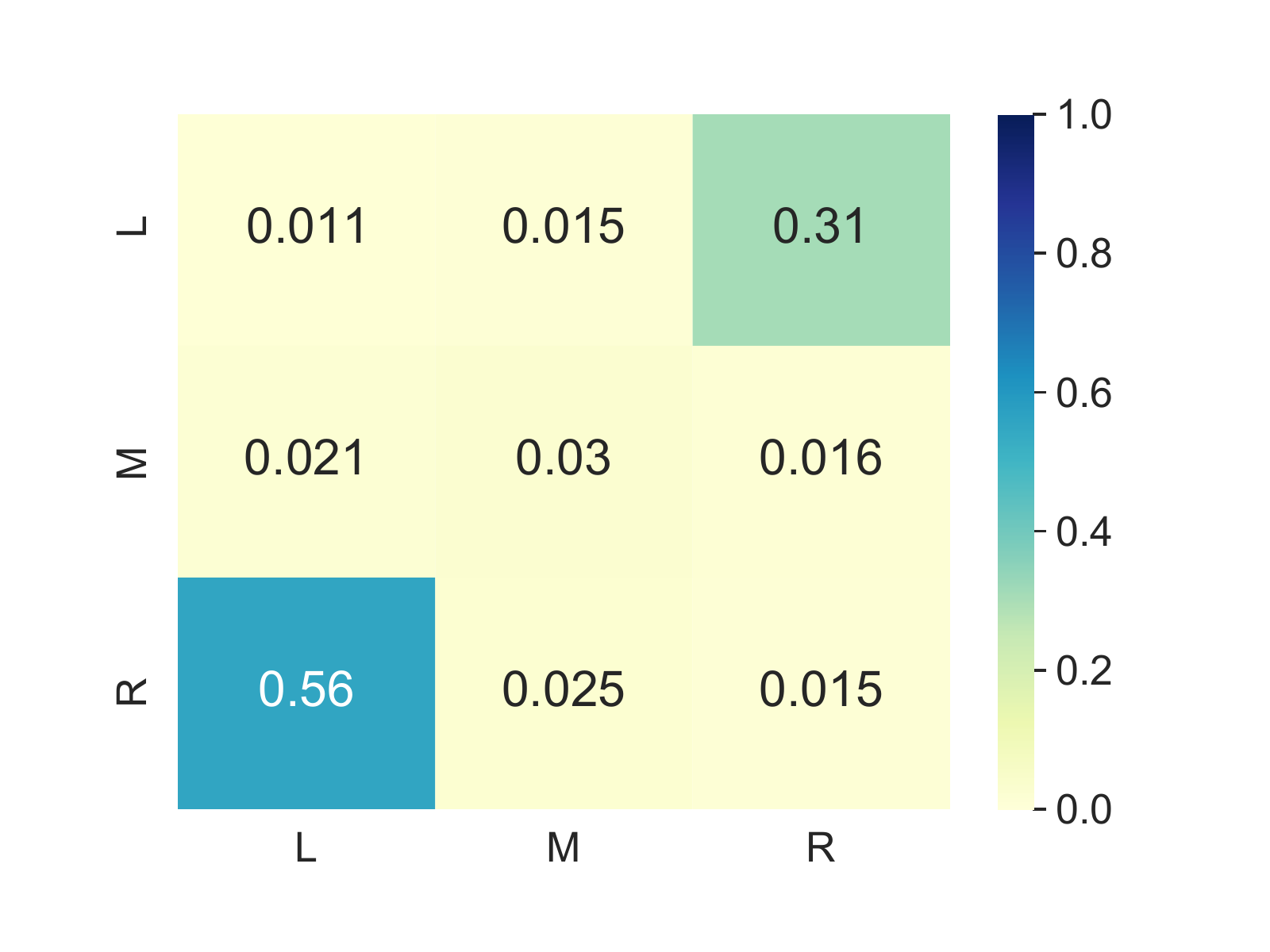}
    \caption{Game 5}
    \label{fig:acomacom_g5}
\end{subfigure}
\caption{Empirical outcome distributions for ACOM vs. ACOM.}
\label{fig:acomacom}
\end{figure*}

\begin{figure*}[h!]
\centering
\begin{subfigure}{0.3\textwidth}
    \centering
    \includegraphics[width=4cm]{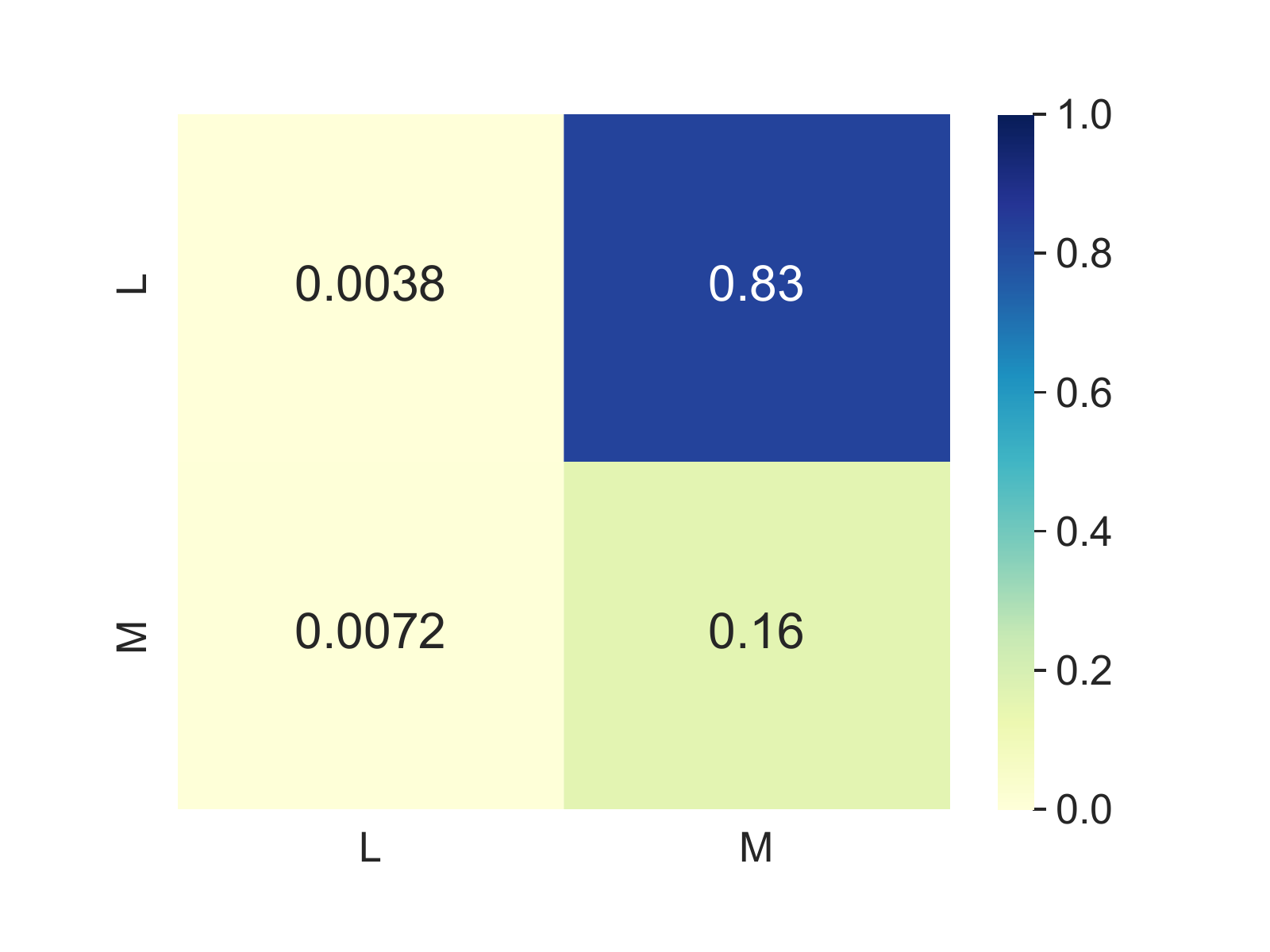}
    \caption{Game 1}
    \label{fig:acacom_g1}
\end{subfigure}
\begin{subfigure}{0.3\textwidth}
    \centering
    \includegraphics[width=4cm]{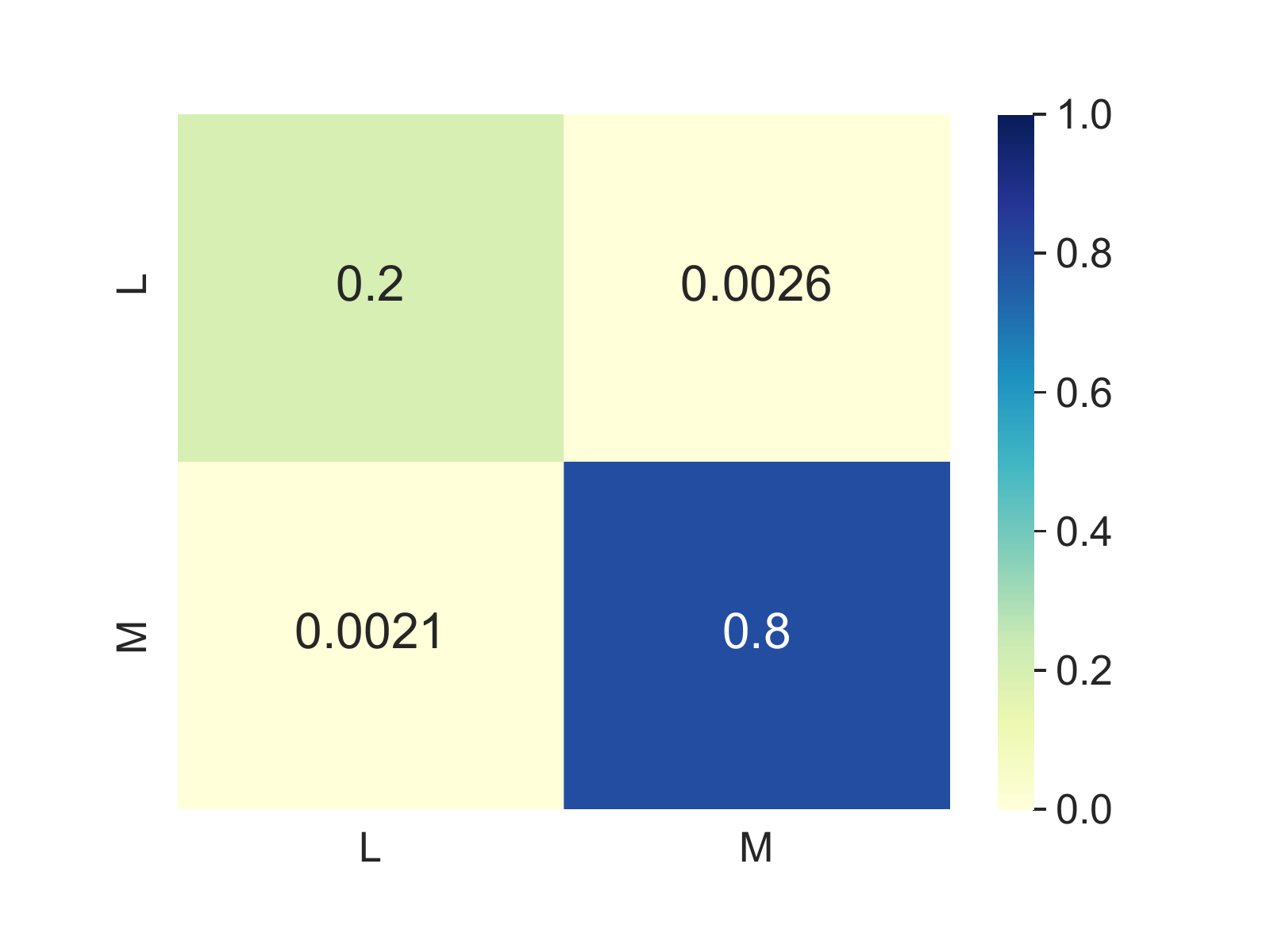}
    \caption{Game 2}
     \label{fig:acacom_g2}
\end{subfigure}
\begin{subfigure}{0.3\textwidth}
    \centering
    \includegraphics[width=4cm]{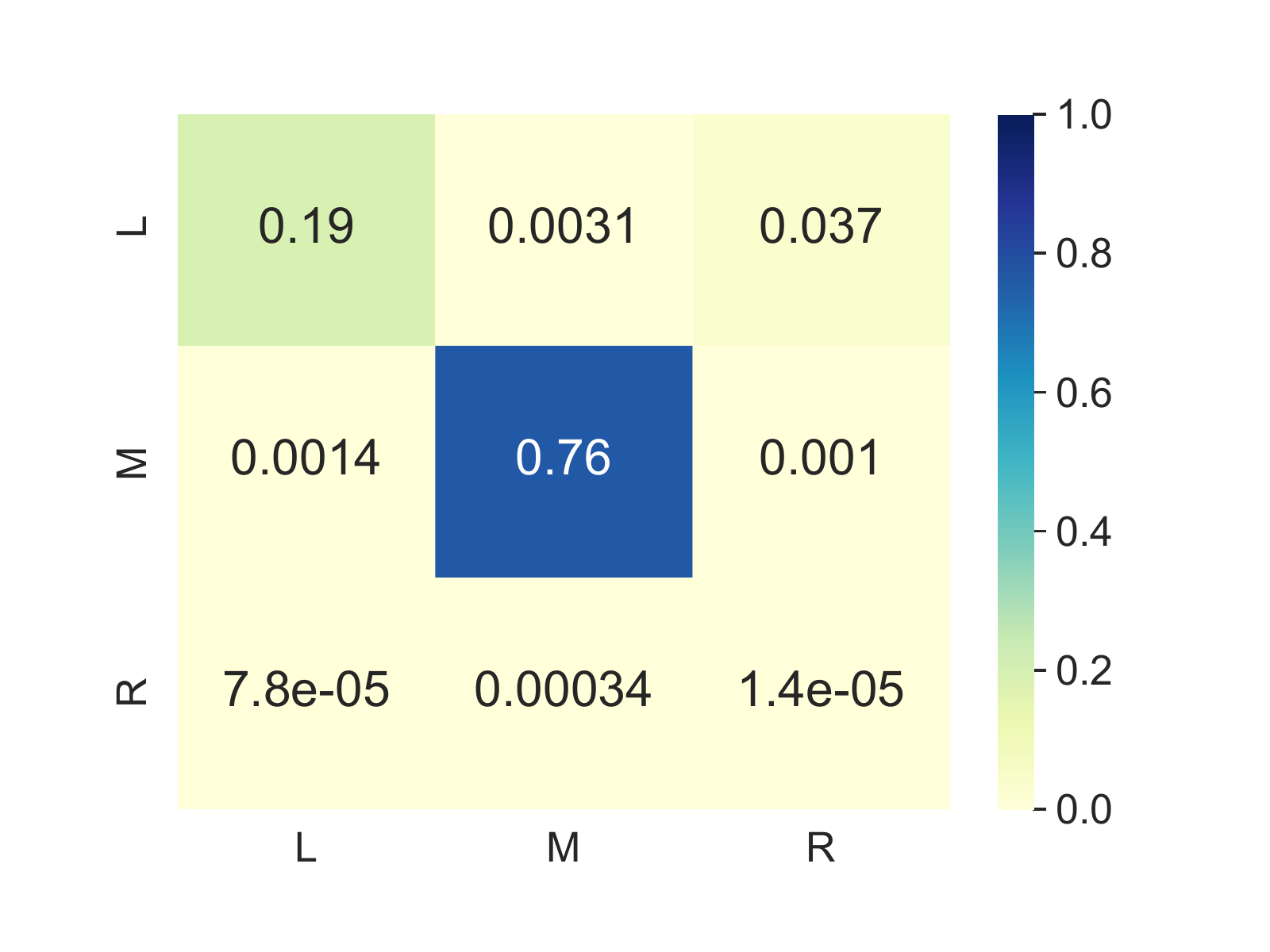}
    \caption{Game 3}
     \label{fig:acacom_g3}
\end{subfigure}\\
\begin{subfigure}{0.3\textwidth}
    \centering
    \includegraphics[width=4cm]{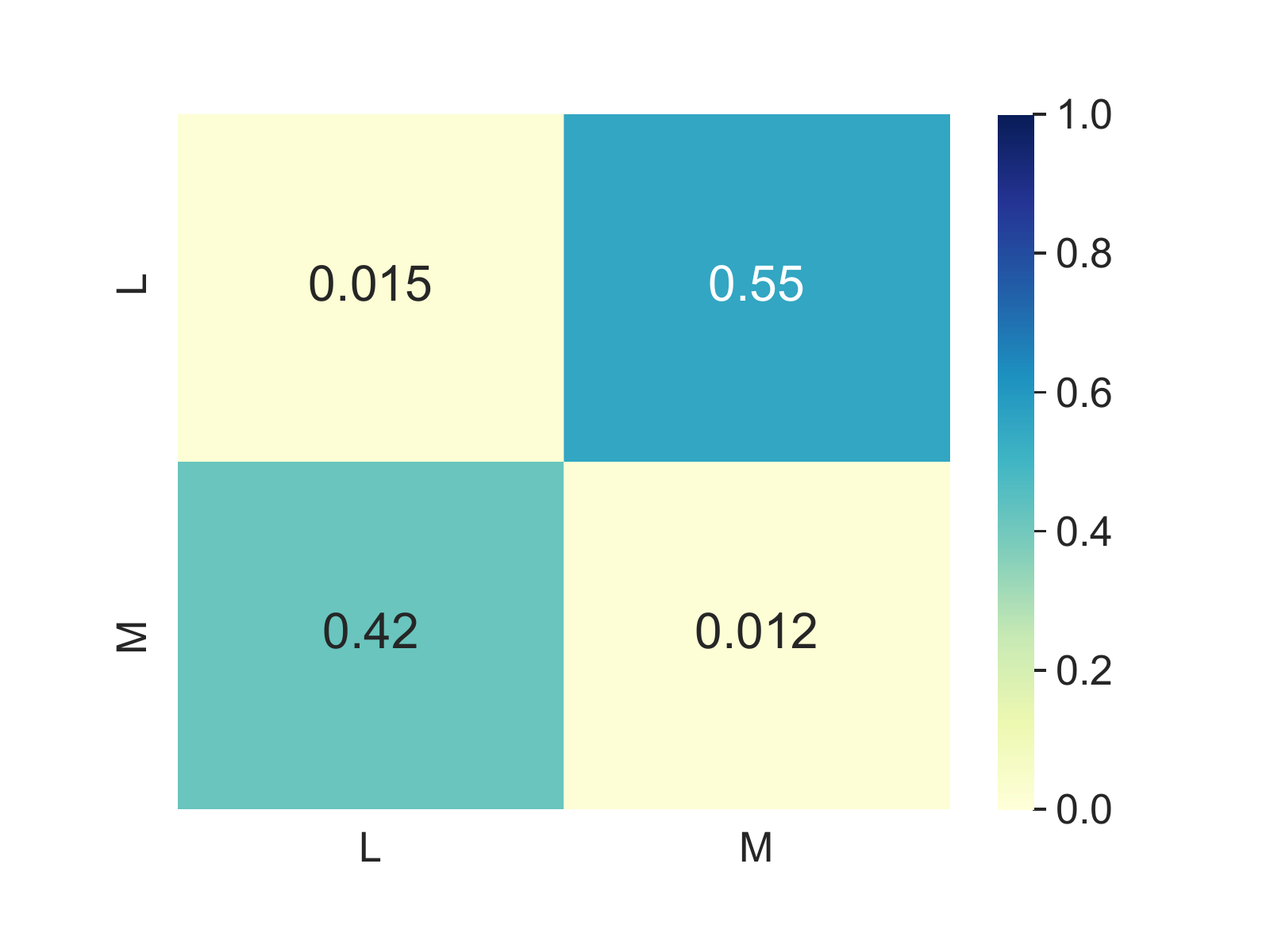}
    \caption{Game 4}
     \label{fig:acacom_g4}
\end{subfigure}
\begin{subfigure}{0.3\textwidth}
    \centering
    \includegraphics[width=4cm]{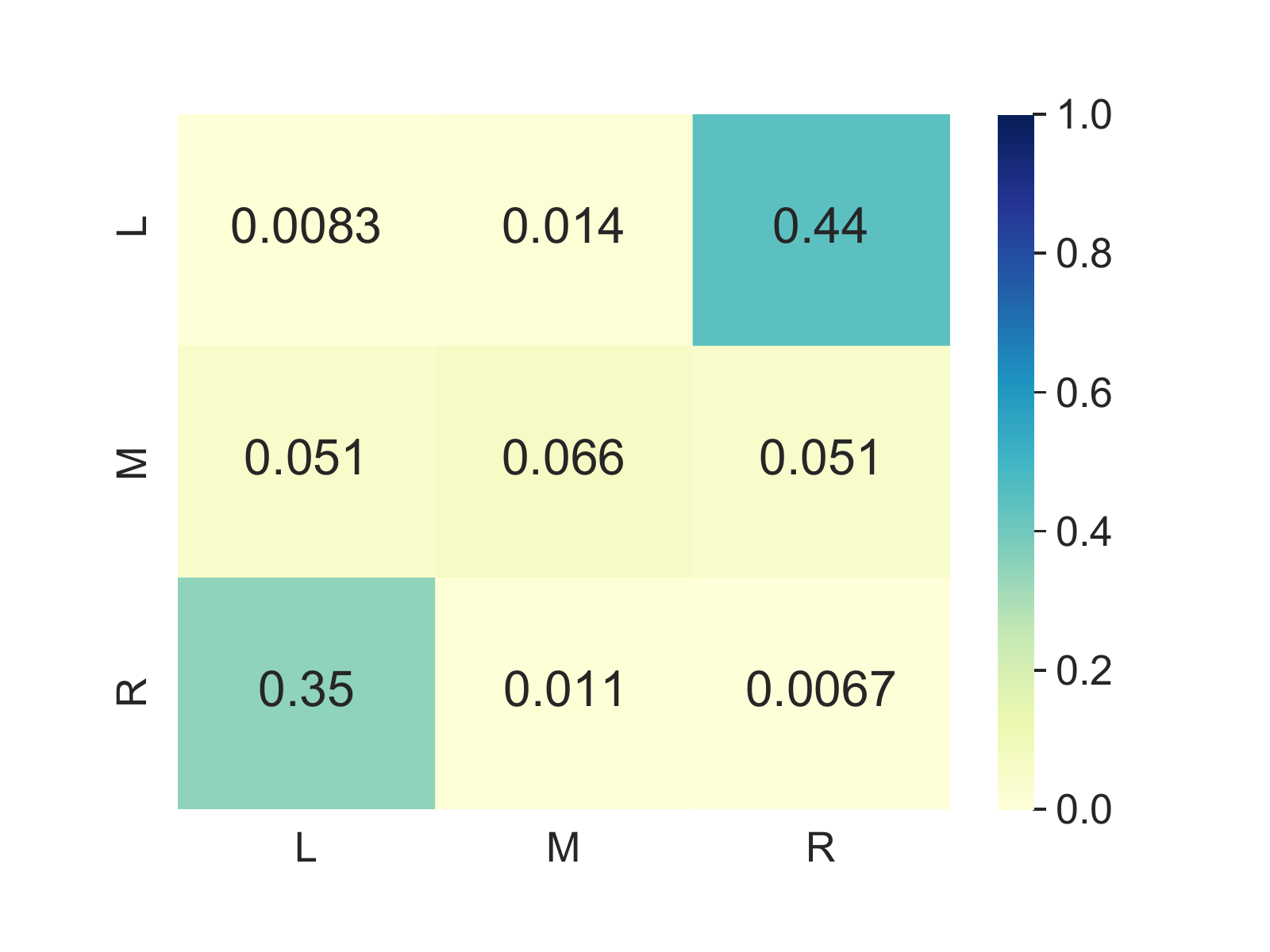}
    \caption{Game 5}
     \label{fig:acacom_g5}
\end{subfigure}
\caption{Empirical outcome distributions for AC vs. ACOM.}
\label{fig:acacom}
\end{figure*}

\begin{figure*}[h!]
\centering
\begin{subfigure}{0.3\textwidth}
    \centering
    \includegraphics[width=4cm]{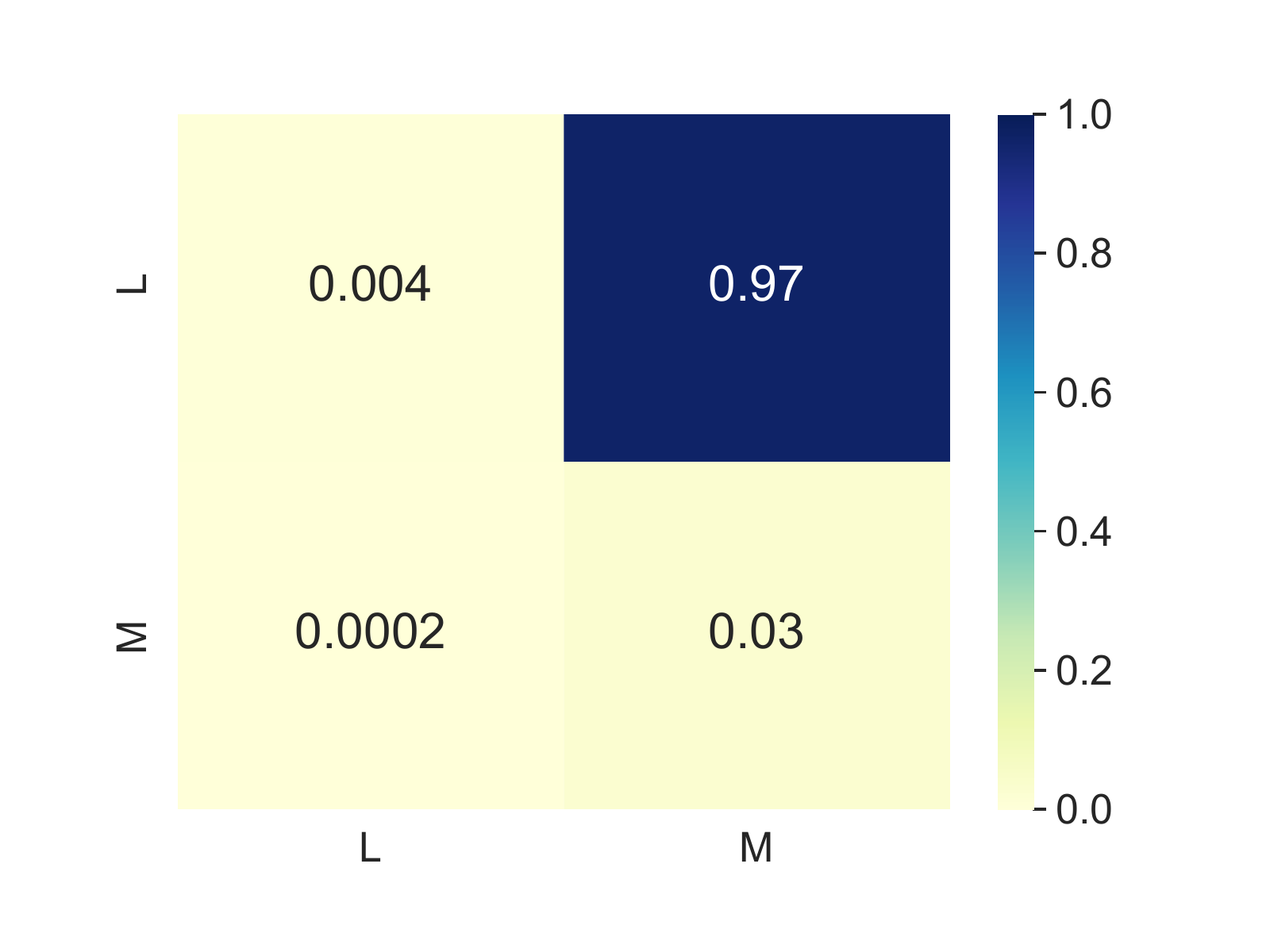}
    \caption{Game 1}
    \label{fig:acomac_g1}
\end{subfigure}
\begin{subfigure}{0.3\textwidth}
    \centering
    \includegraphics[width=4cm]{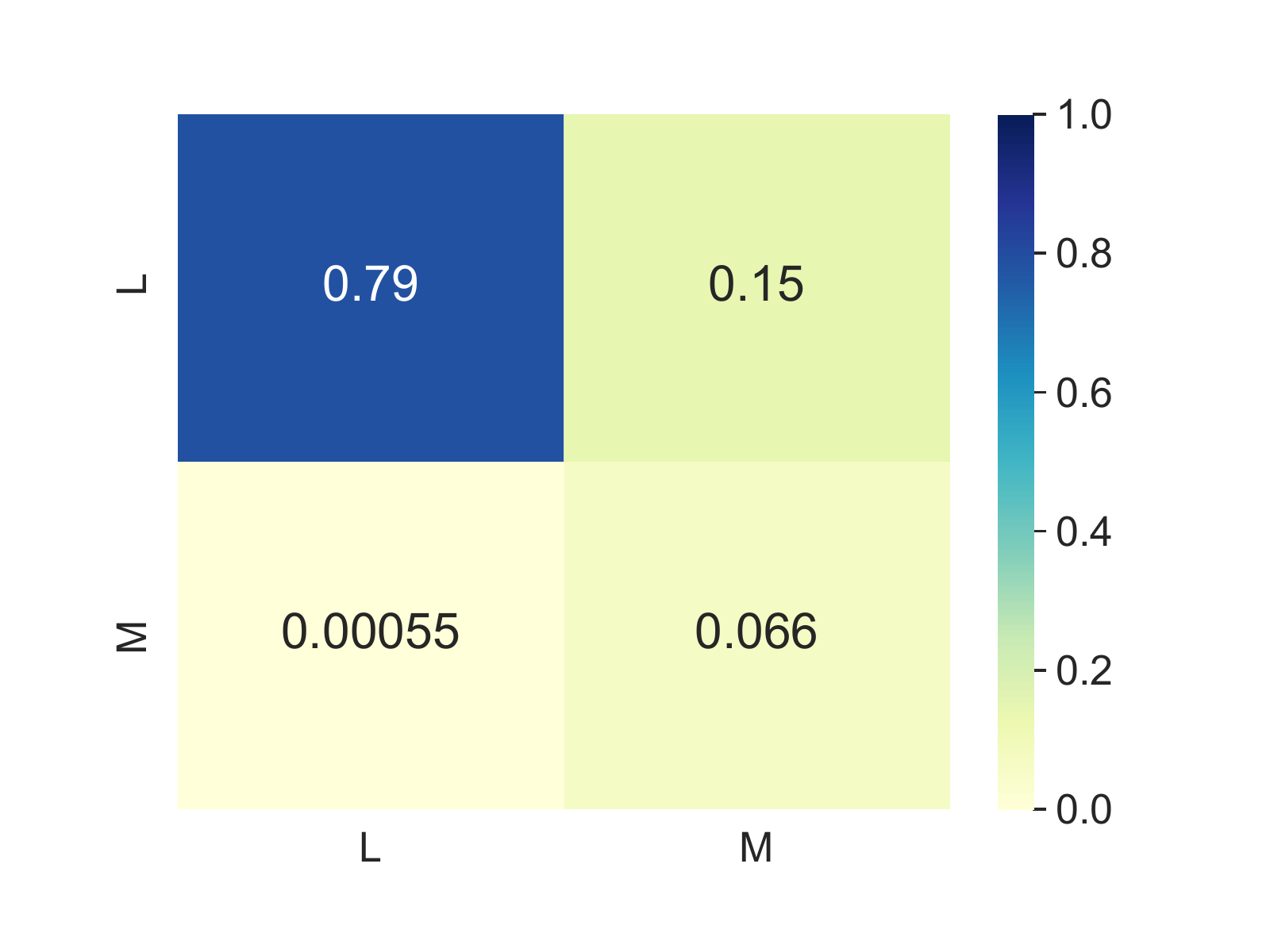}
    \caption{Game 2}
    \label{fig:acomac_g2}
\end{subfigure}
\begin{subfigure}{0.3\textwidth}
    \centering
    \includegraphics[width=4cm]{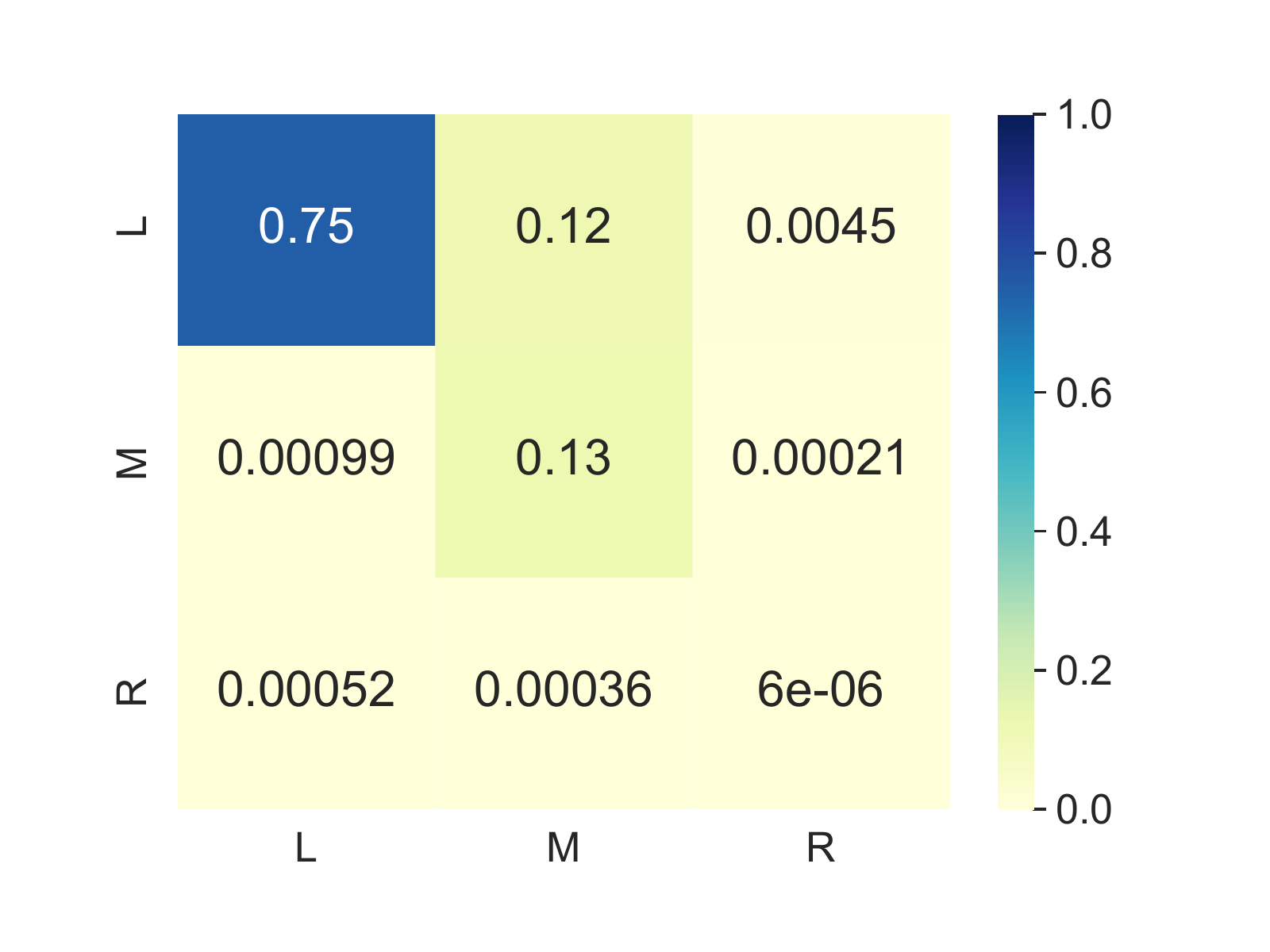}
    \caption{Game 3}
    \label{fig:acomac_g3}
\end{subfigure}\\
\begin{subfigure}{0.3\textwidth}
    \centering
    \includegraphics[width=4cm]{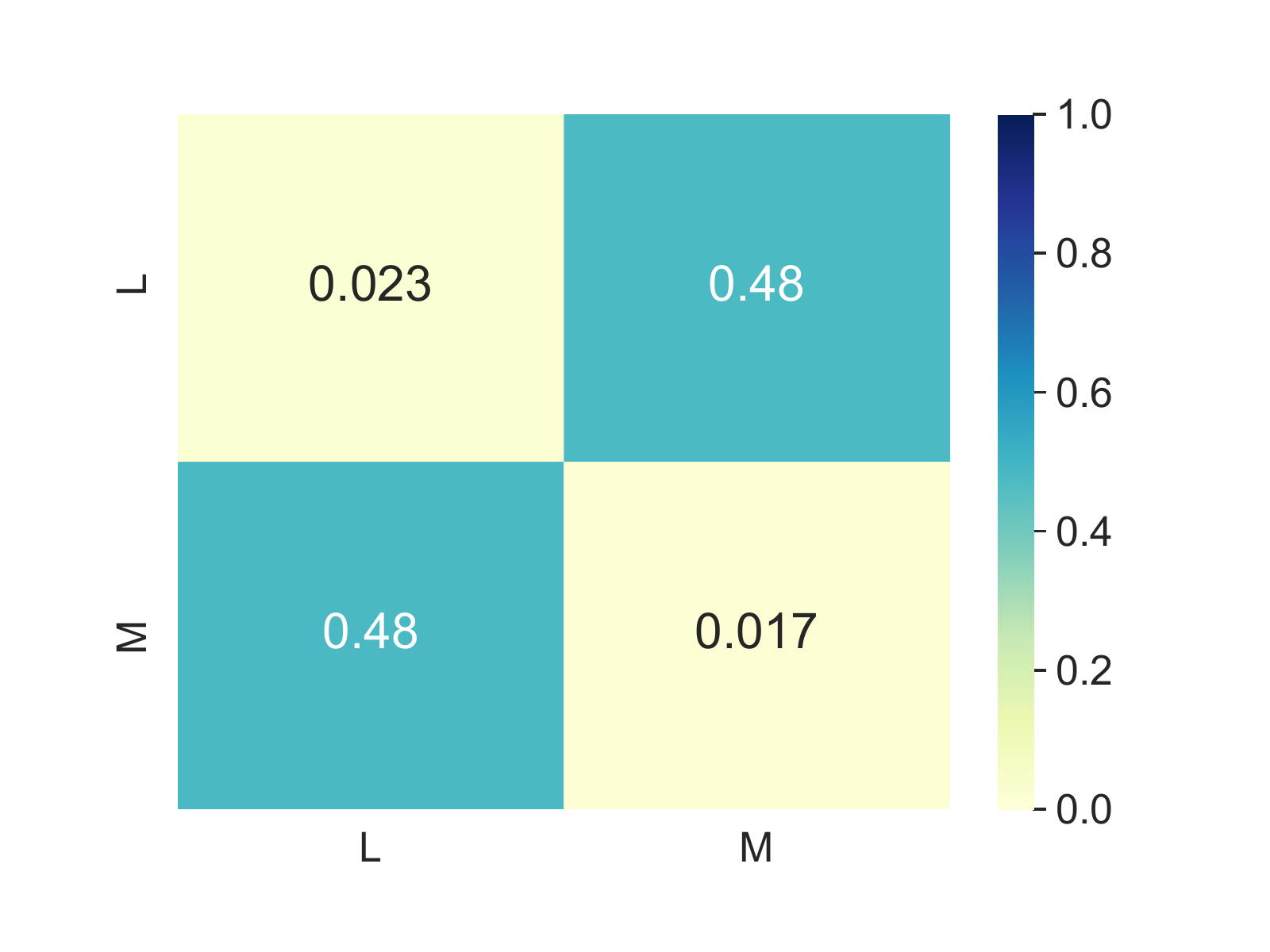}
    \caption{Game 4}
    \label{fig:acomac_g4}
\end{subfigure}
\begin{subfigure}{0.3\textwidth}
    \centering
    \includegraphics[width=4cm]{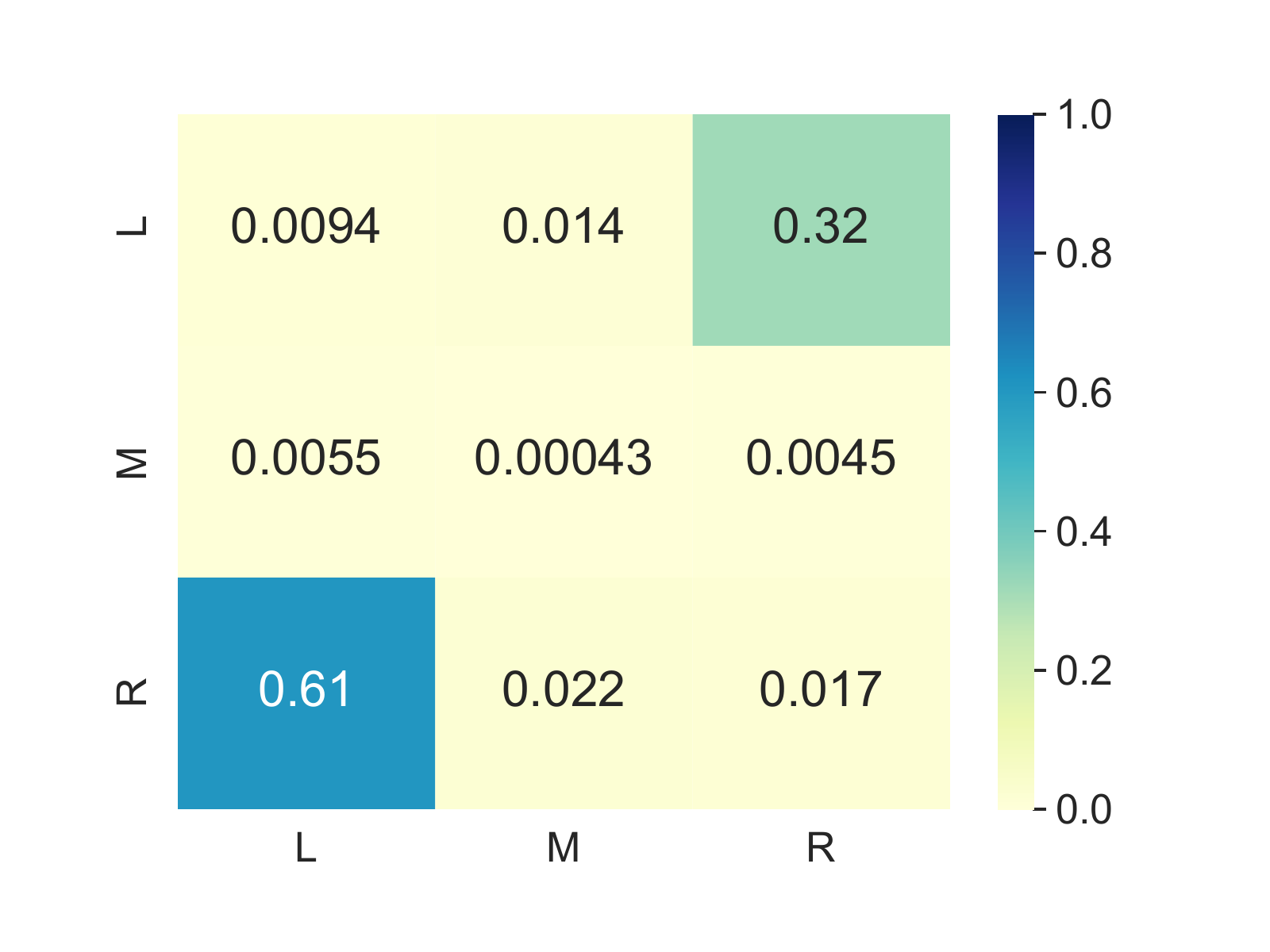}
    \caption{Game 5}
    \label{fig:acomac_g5}
\end{subfigure}
\caption{Empirical outcome distributions for ACOM vs. AC.}
\label{fig:acomac}
\end{figure*}

Let us start exploring the results for no-information setting by looking at the AC and ACOM approaches. For Games 4 and 5, where no NEs are present, agent 2 seems to have an advantage in all cases, in contrast to the behaviour of multi-objective LOLA, where a middle ground point is found (Figures~\ref{fig:acac_g4}, \ref{fig:acac_g5} -- \ref{fig:acomac_g4}, \ref{fig:acomac_g5}).

In Games 1--3, all the combinations between AC and ACOM generally manage to converge to the NE and also avoid the dominated point (R,R) from Game 3. A notable exception is presented in Game 1 in which agent 1 uses the AC approach (Figures~\ref{fig:acac_g1} and \ref{fig:acacom_g1}). Instead of converging to action L, agent 1 seems to also allot a small probability to action M, despite the fact that outcome (M,M) is less preferred for that agent compared to outcome (L,M).

\paragraph{ACOM vs. AC and AC vs. ACOM}
Games 2 and 3 also allow us to draw the first conclusions regarding the single-sided use of opponent modelling. By contrasting Figures~\ref{fig:acacom_g2}, \ref{fig:acacom_g3} with Figures~\ref{fig:acomac_g2}, \ref{fig:acomac_g3}, we notice how the ACOM approach confers the agent with a significant advantage in terms of shifting the outcome distribution towards its preferred NE. We remind the reader that agent 1 prefers (L,L), while agent 2 prefers (M,M) in both Games 2 and 3. 

\subsubsection{ACOLAM}

We continue our analysis by looking at the ACOLAM method. We note here that ACOLAM vs. ACOLAM yields the same behaviour as ACOM vs. ACOM. Moreover, the interactions between AC and ACOLAM also show the same results as for AC and ACOM. This observation points to the fact that using a Gaussian Process as an estimator for the opponent's local learning step is a valid approach and it is not detrimental for the learning process of the agents. 

For these reasons, we mainly focus here on the comparison between ACOLAM and ACOM and try to evaluate whether the extra step of learning with opponent learning awareness can bring any benefits for the studied settings.

\paragraph{ACOM vs ACOLAM and ACOLAM vs. ACOM}
In the games without a NE under SER, i.e., Games 4 and 5, the behaviour of ACOM and ACOLAM when going against each other is consistent with the previously observed situations between AC and ACOM, where agent 2 seems to have an advantage and the final joint-action they converge to (i.e., second diagonal of the games) does not represent any meaningful outcome.

\begin{figure*}[h!]
\centering
\begin{subfigure}{0.3\textwidth}
    \centering
    \includegraphics[width=4cm]{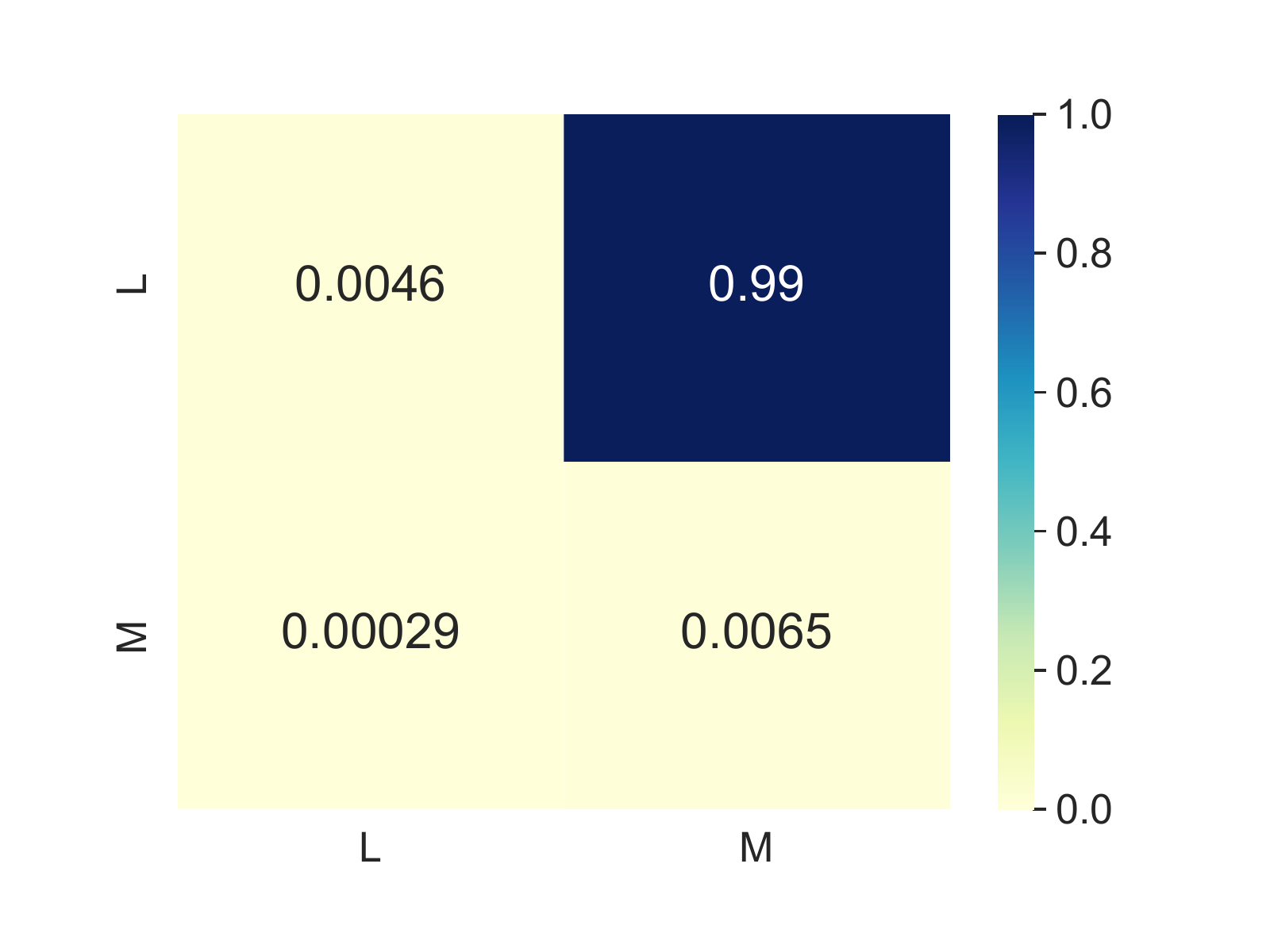}
    \caption{Game 1}
    \label{fig:acomacolam_g1}
\end{subfigure}
\begin{subfigure}{0.3\textwidth}
    \centering
    \includegraphics[width=4cm]{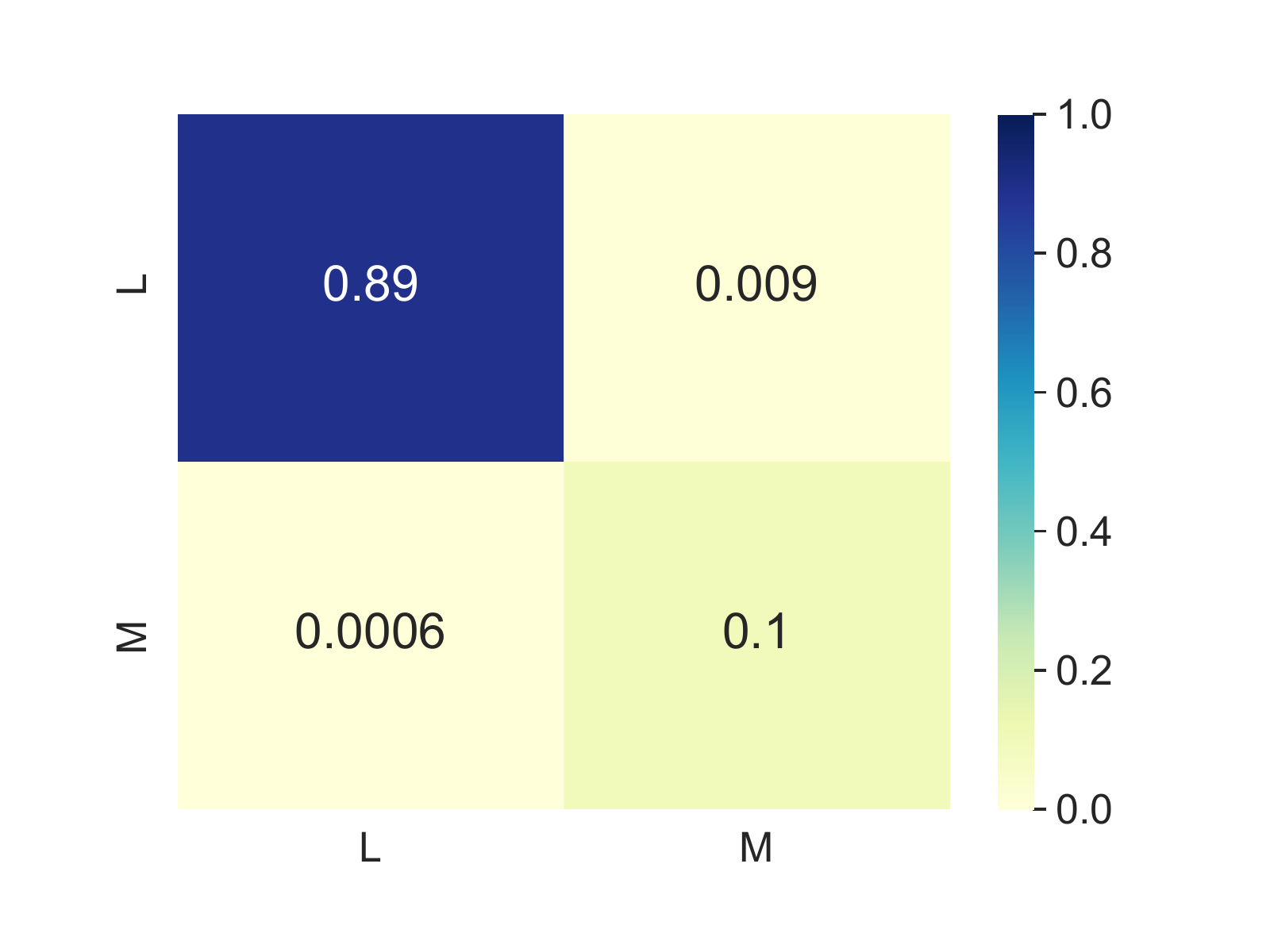}
    \caption{Game 2}
 \label{fig:acomacolam_g2}    
\end{subfigure}
\begin{subfigure}{0.3\textwidth}
    \centering
    \includegraphics[width=4cm]{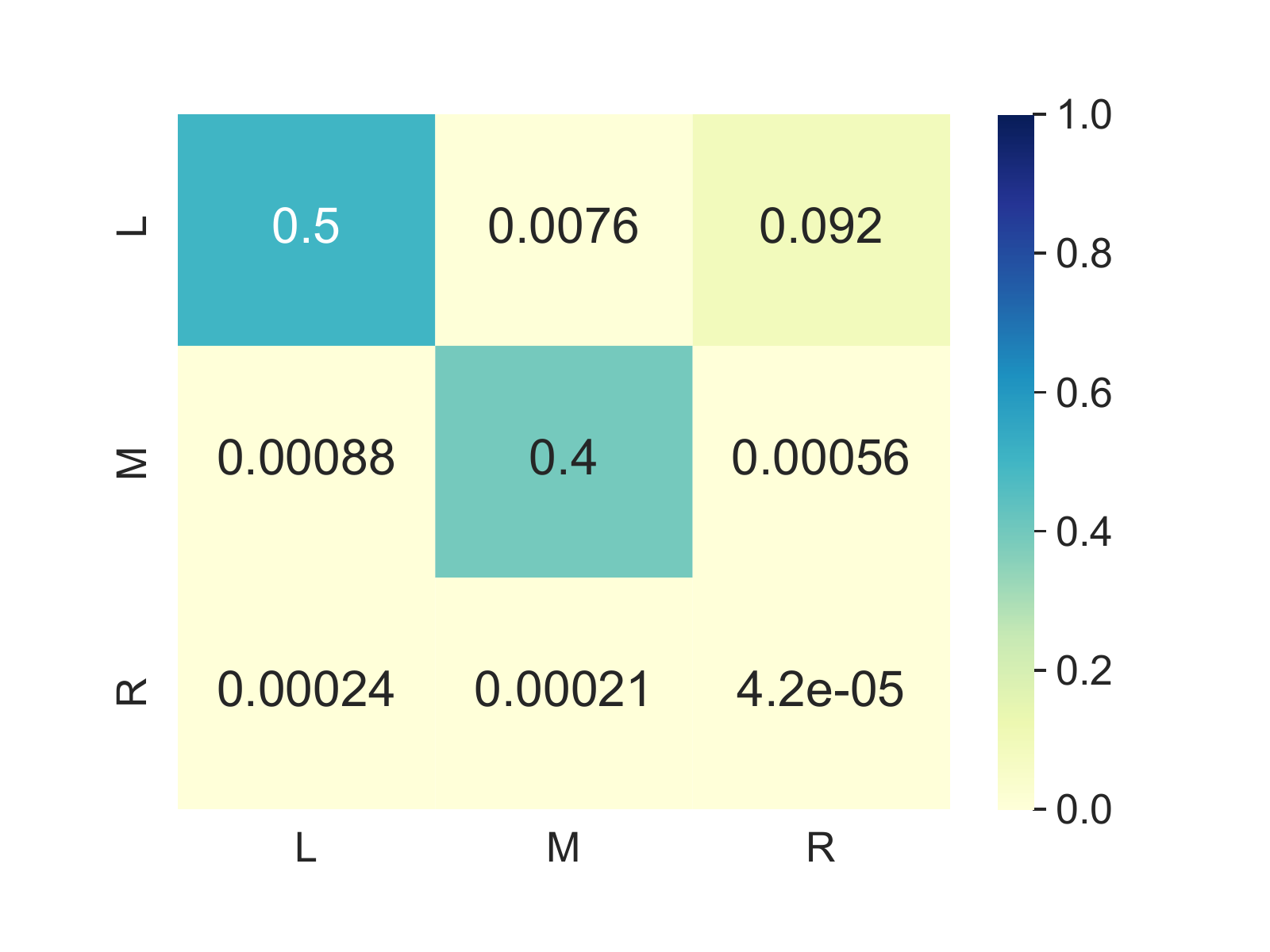}
    \caption{Game 3}
 \label{fig:acomacolam_g3}
\end{subfigure}\\
\begin{subfigure}{0.3\textwidth}
    \centering
    \includegraphics[width=4cm]{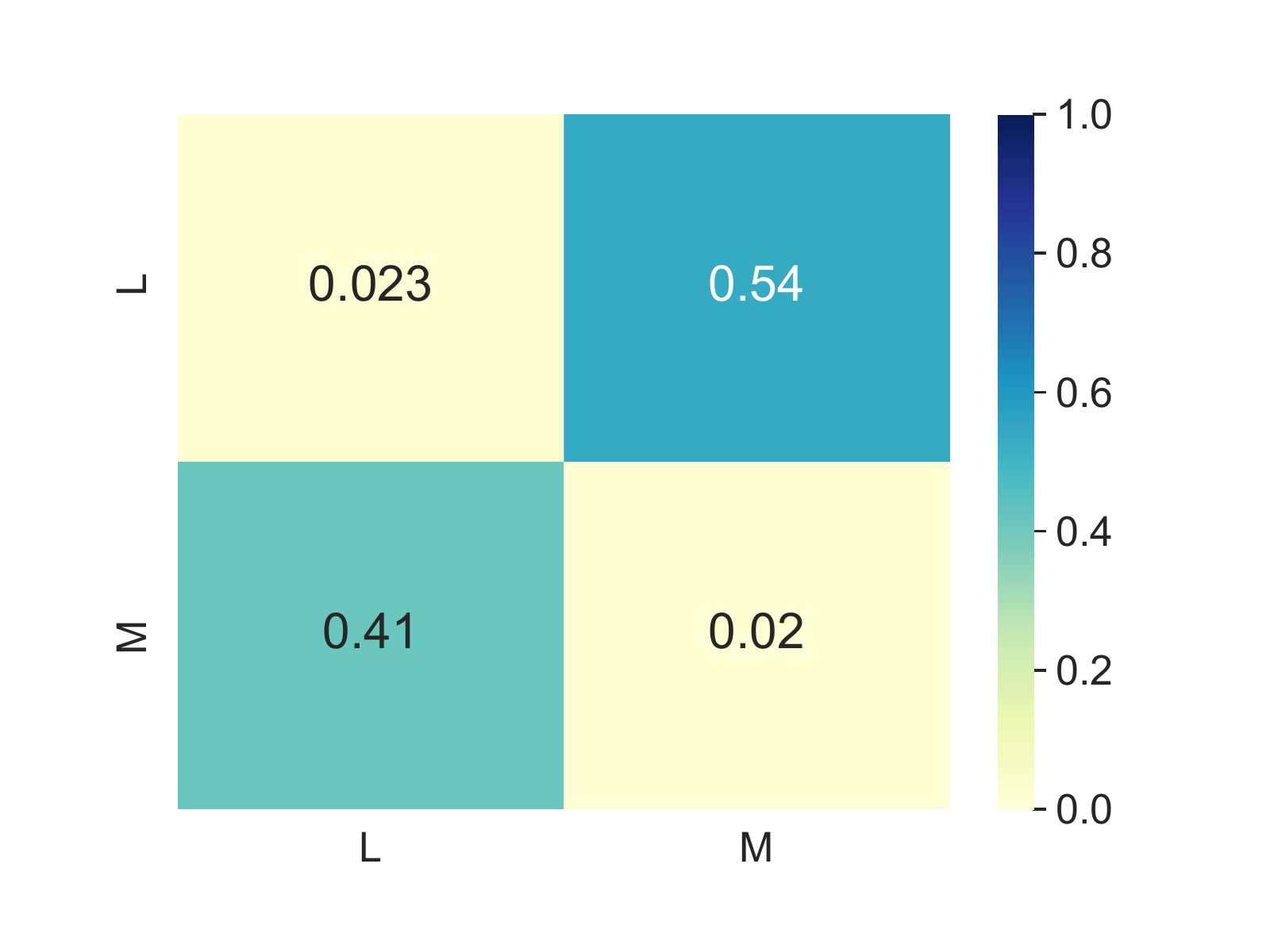}
    \caption{Game 4}
     \label{fig:acomacolam_g4}
\end{subfigure}
\begin{subfigure}{0.3\textwidth}
    \centering
    \includegraphics[width=4cm]{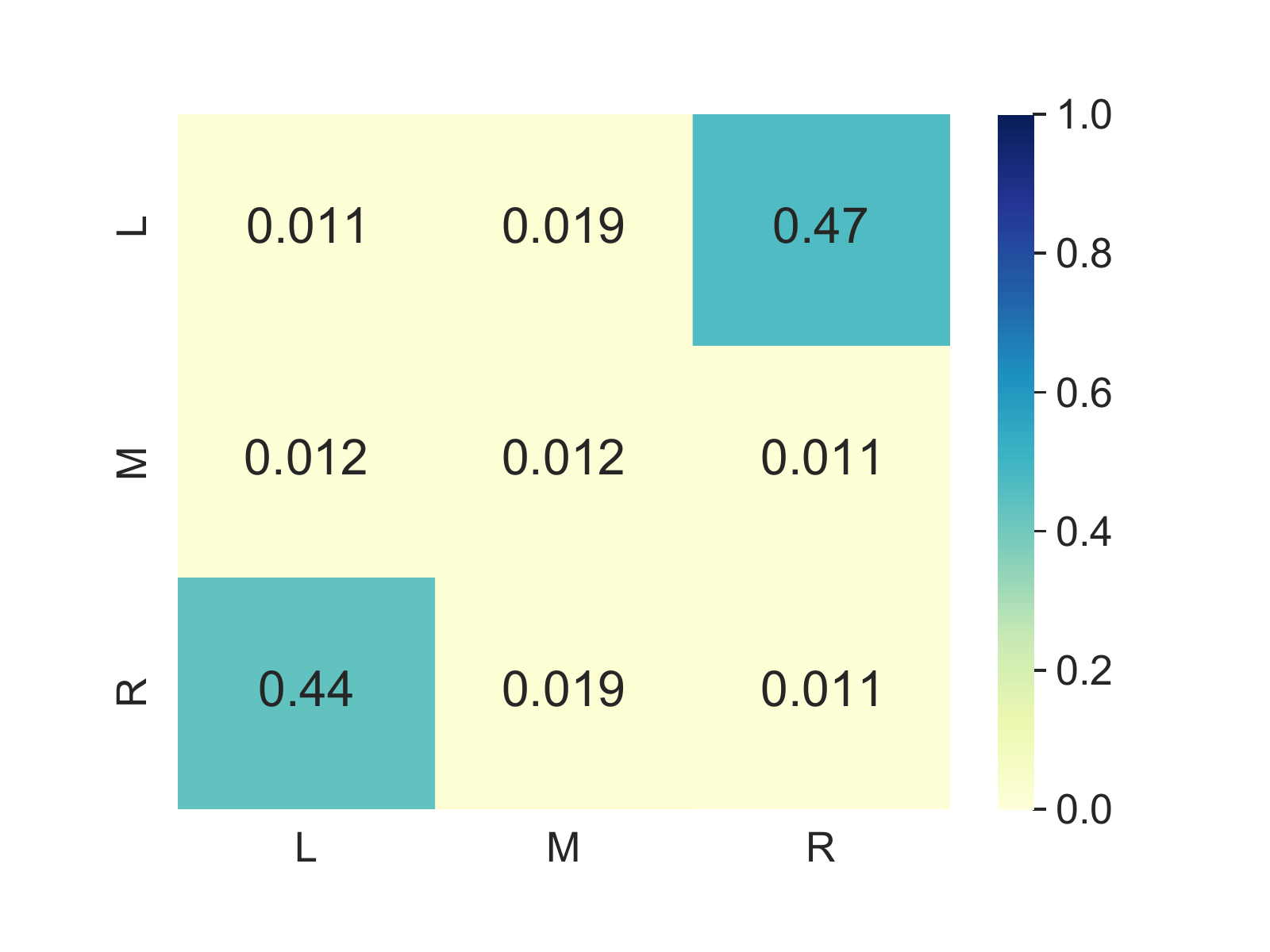}
    \caption{Game 5}
 \label{fig:acomacolam_g5}
\end{subfigure}
\caption{Empirical outcome distributions for ACOM vs. ACOLAM (lookahead value 1).}
\label{fig:acomacolam}
\end{figure*}

\begin{figure*}[h!]
\centering
\begin{subfigure}{0.3\textwidth}
    \centering
    \includegraphics[width=4cm]{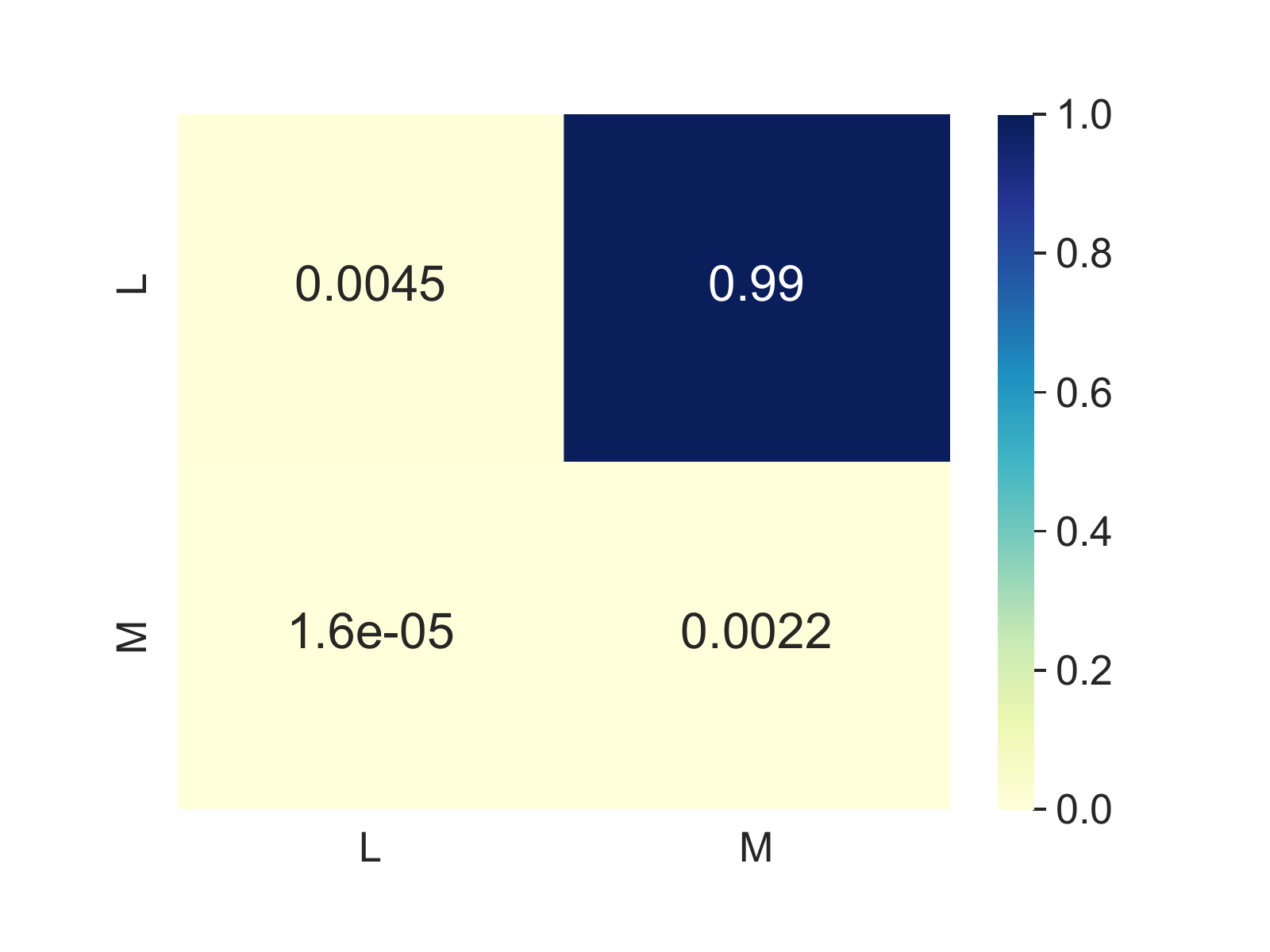}
    \caption{Game 1}
     \label{fig:acolamacom_g1}
\end{subfigure}
\begin{subfigure}{0.3\textwidth}
    \centering
    \includegraphics[width=4cm]{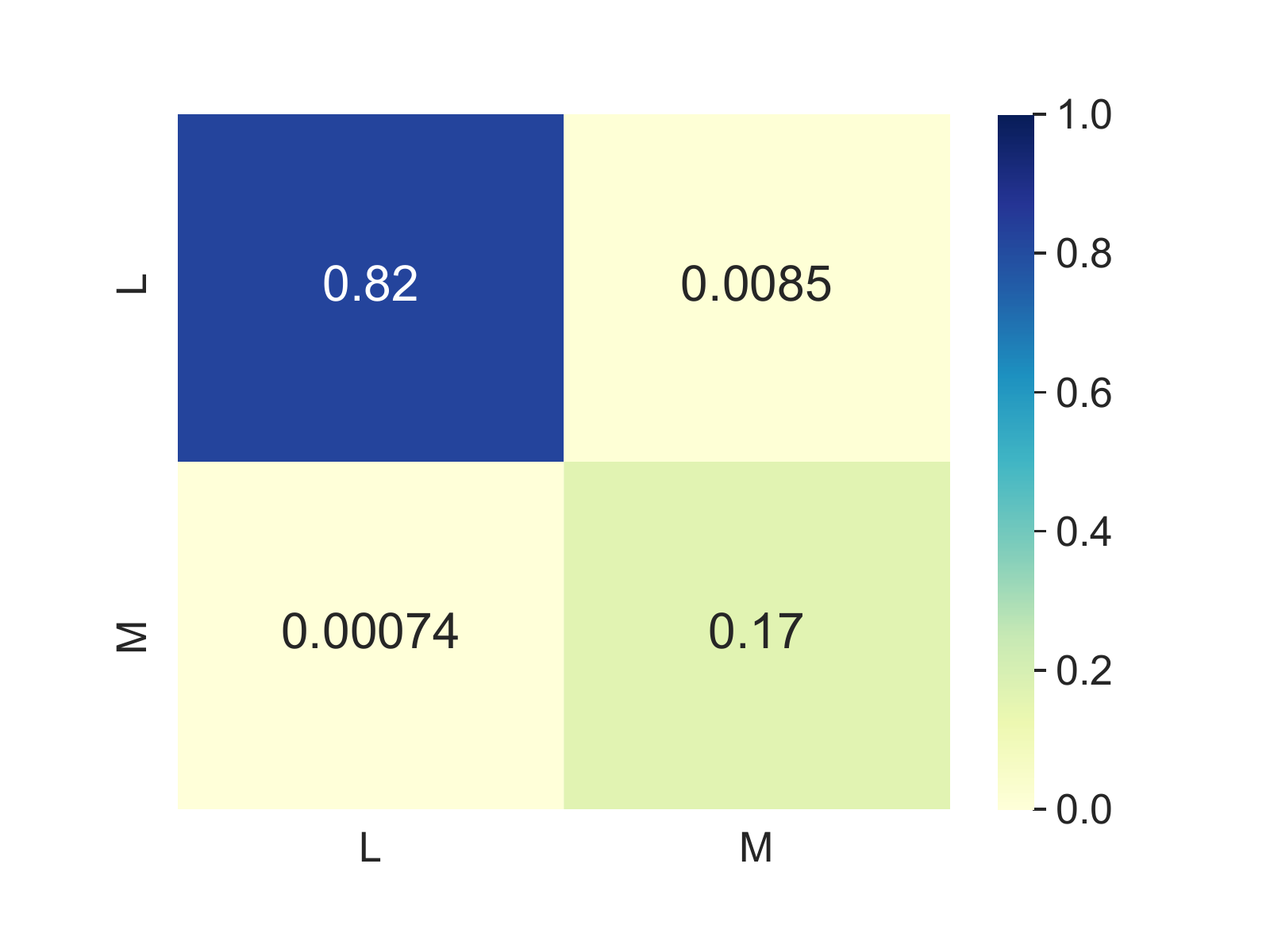}
    \caption{Game 2}
     \label{fig:acolamacom_g2}
\end{subfigure}
\begin{subfigure}{0.3\textwidth}
    \centering
    \includegraphics[width=4cm]{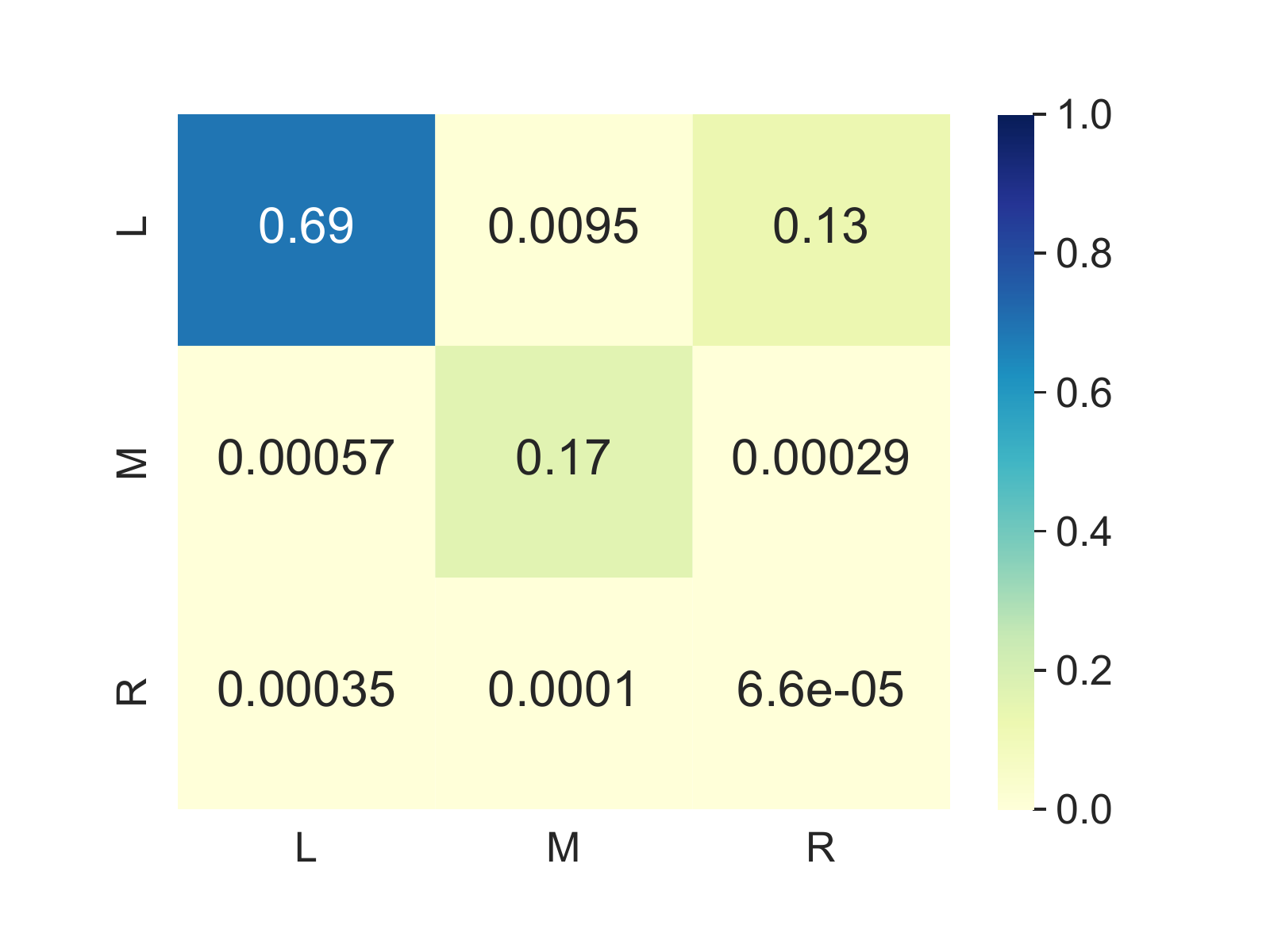}
    \caption{Game 3}
     \label{fig:acolamacom_g3}
\end{subfigure}\\
\begin{subfigure}{0.3\textwidth}
    \centering
    \includegraphics[width=4cm]{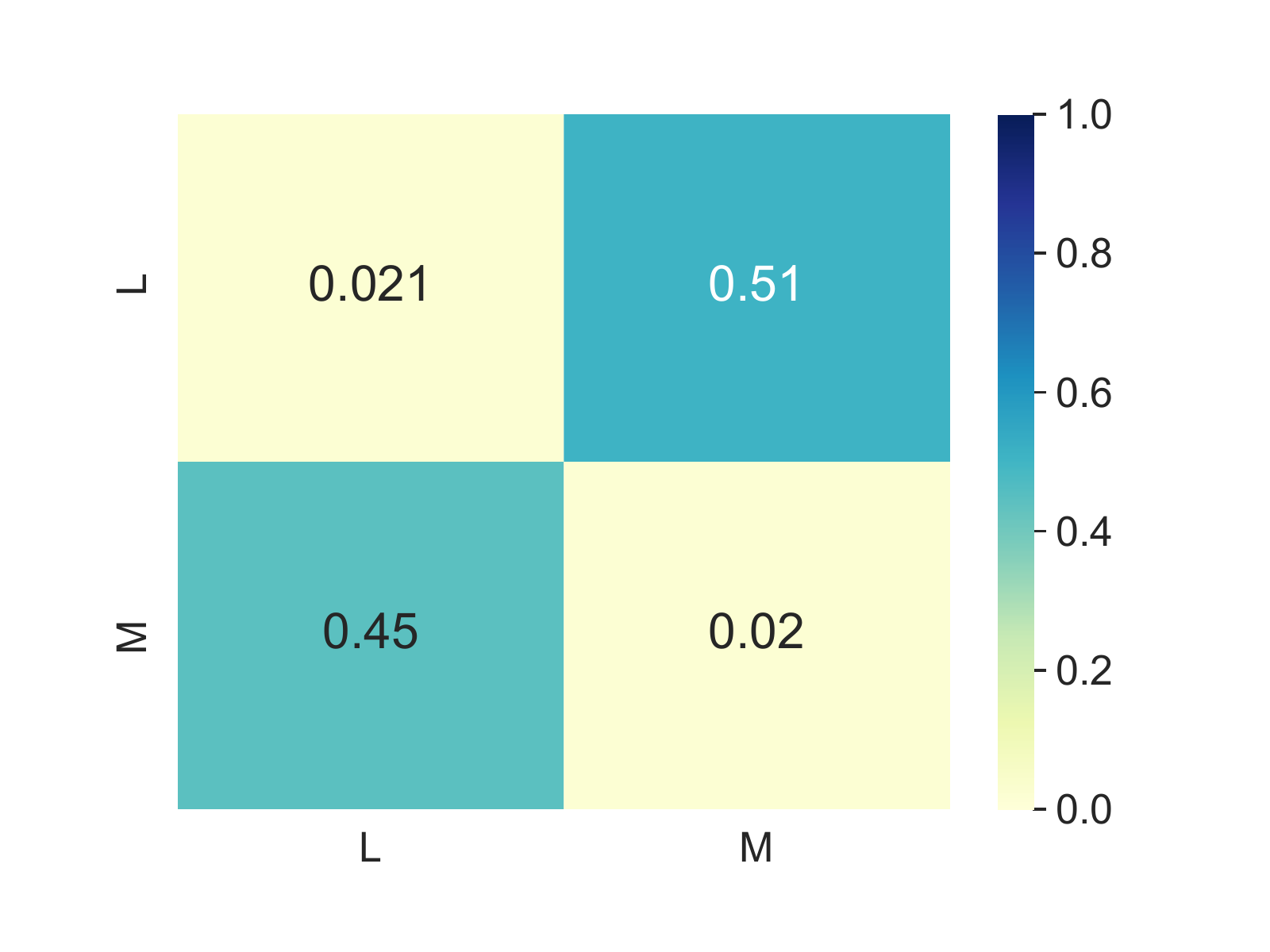}
    \caption{Game 4}
         \label{fig:acolamacom_g4}
\end{subfigure}
\begin{subfigure}{0.3\textwidth}
    \centering
    \includegraphics[width=4cm]{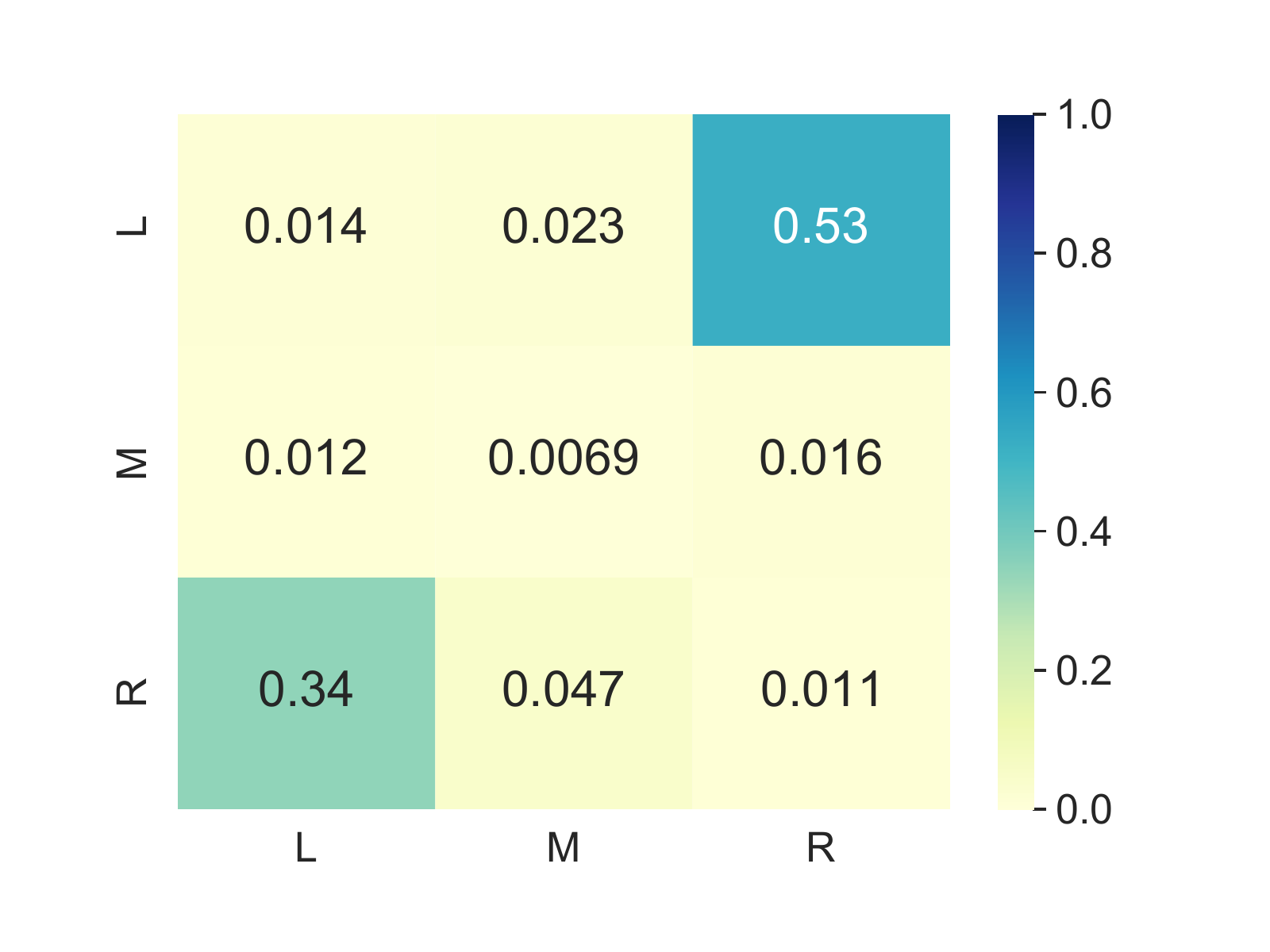}
    \caption{Game 5}
         \label{fig:acolamacom_g5}
\end{subfigure}
\caption{Empirical outcome distributions for ACOLAM vs. ACOM (lookahead value 1).}
\label{fig:col3lolam}
\end{figure*}

Looking at Game 1, where one NE is present (L,M), we see that the combination of ACOM and ACOLAM manages to reach this equilibrium quickly, with a probability of $\approx99\%$. In Game 2, it seems that agent 1, be it ACOM or ACOLAM, manages to steer the outcome towards its preferred NE (Figures~\ref{fig:acomacolam_g2} and \ref{fig:acolamacom_g2}). 

Only by looking at Game 3 (Figures~\ref{fig:acomacolam_g3} and \ref{fig:acolamacom_g3}), we can observe an asymmetry in the behaviour of the two approaches. Specifically, when agent 1 uses ACOM (Figure~\ref{fig:acomacolam_g3}), it does not manage to shift the outcome significantly in its favour anymore. Moreover, when agent 1 uses ACOLAM (Figure~\ref{fig:acolamacom_g3}), we can observe a more pronounced difference between the probabilities of (L,L) and (M,M) in its favour. This suggests that incorporating the idea of opponent learning awareness and modelling using a Gaussian process to estimate the local opponent learning step can be beneficial and improve upon the case of only using the current estimated opponent policy.

\subsubsection{LOLAM}

\begin{figure*}[h]
\centering
\begin{subfigure}{0.3\textwidth}
    \centering
    \includegraphics[width=4cm]{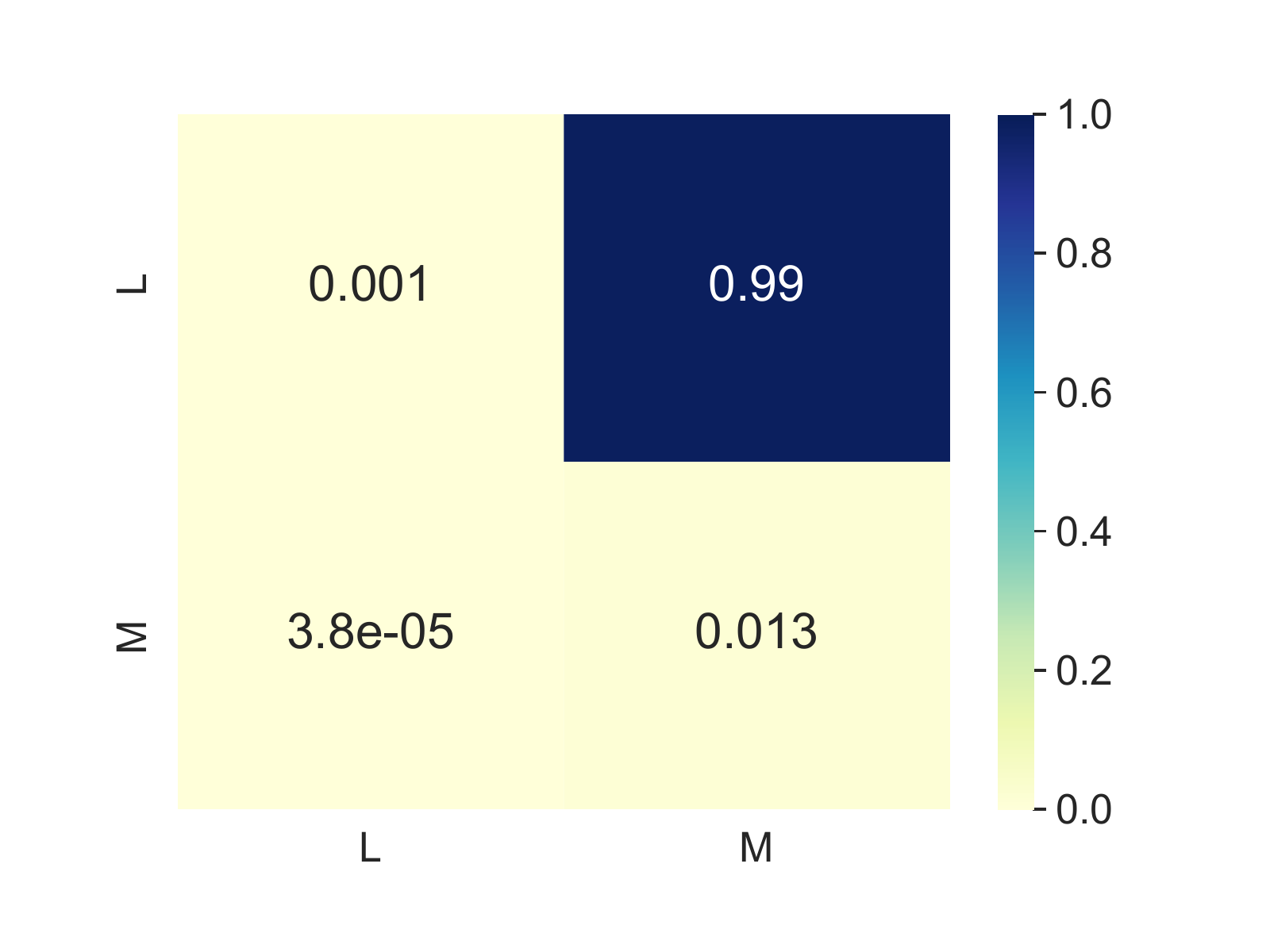}
    \caption{Game 1}
    \label{fig:lolam_g1}
\end{subfigure}
\begin{subfigure}{0.3\textwidth}
    \centering
    \includegraphics[width=4cm]{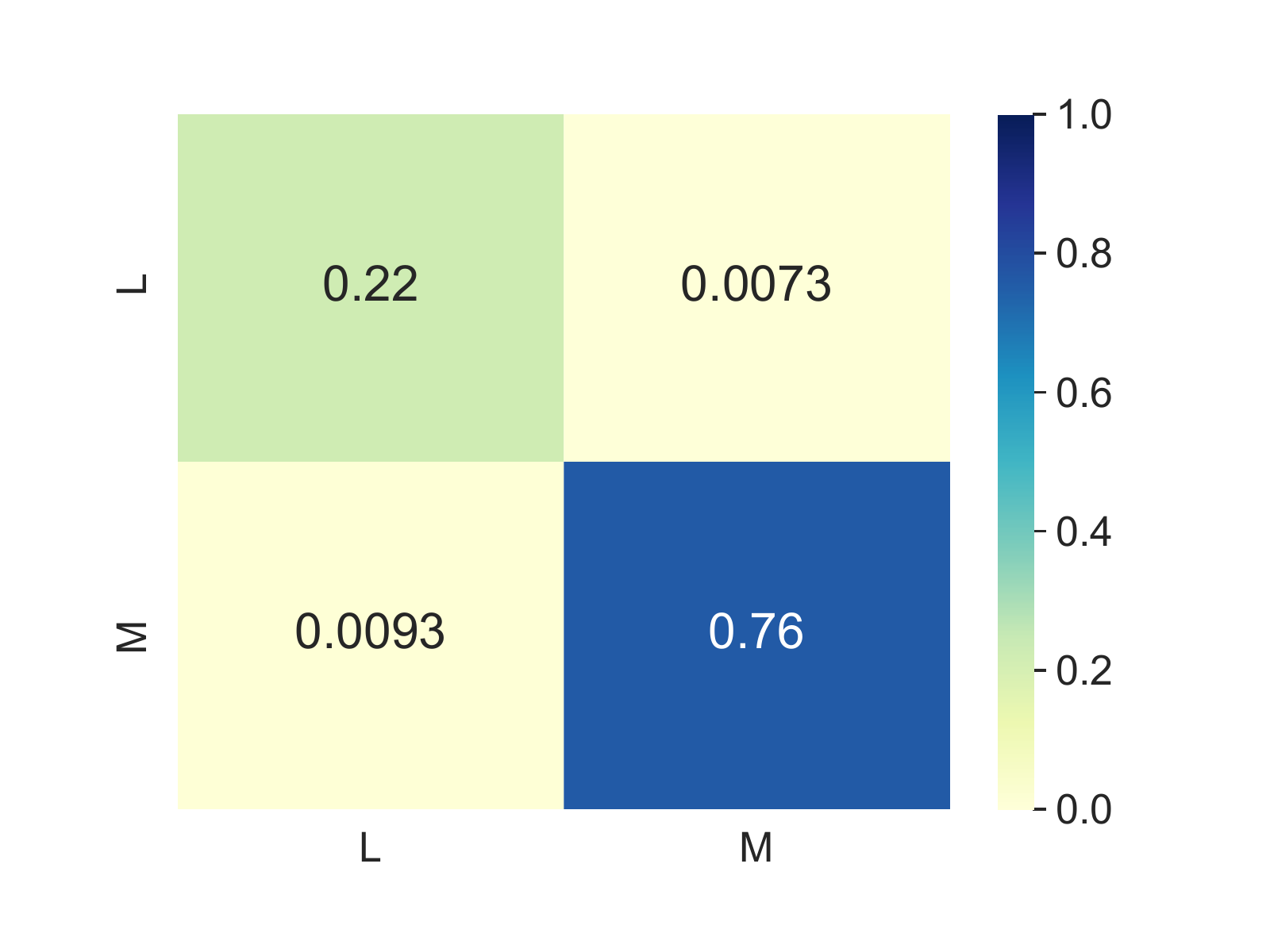}
    \caption{Game 2}
    \label{fig:lolam_g2}
\end{subfigure}
\begin{subfigure}{0.3\textwidth}
    \centering
    \includegraphics[width=4cm]{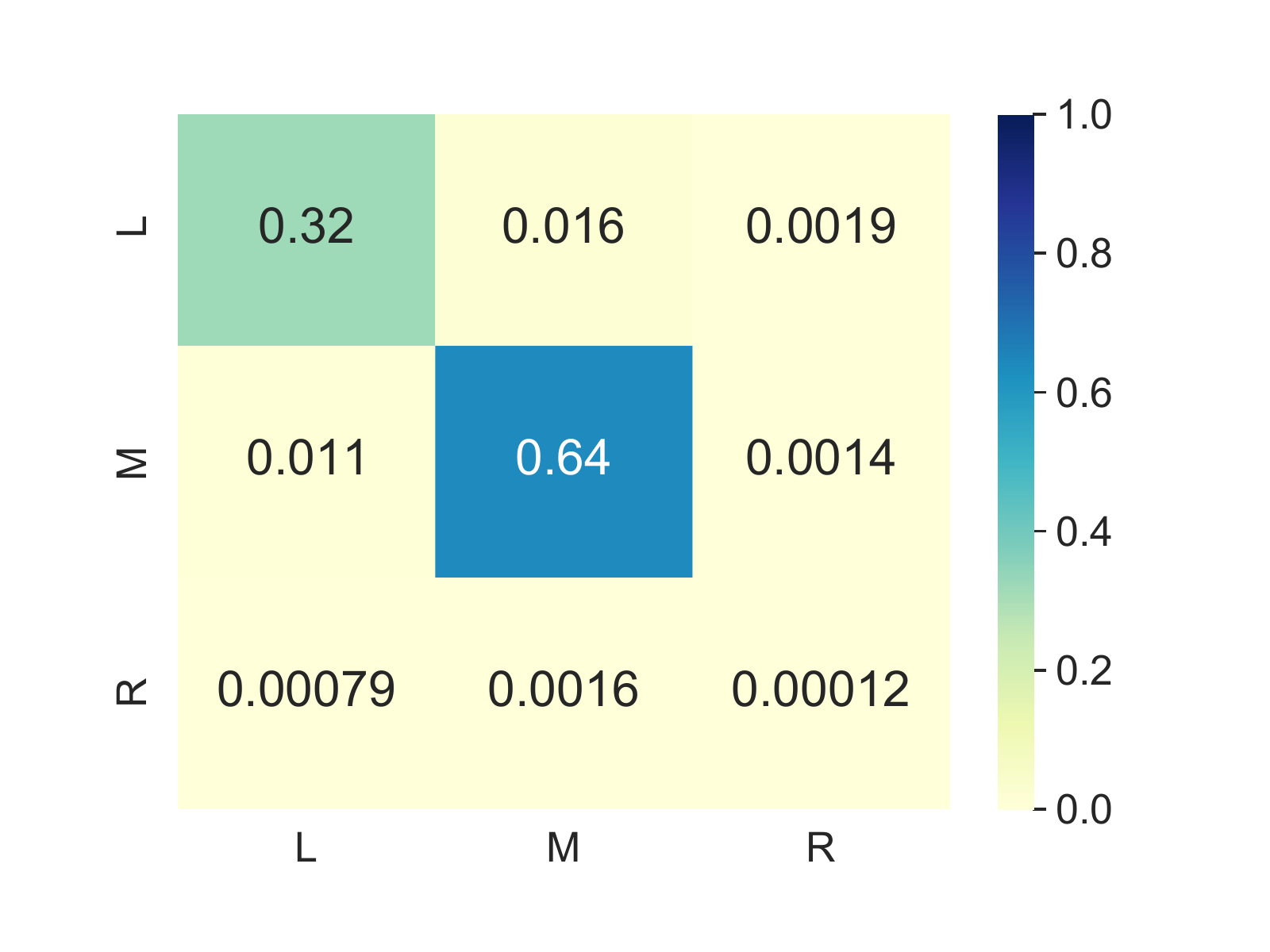}
    \caption{Game 3}
    \label{fig:lolam_g3}
\end{subfigure}\\
\begin{subfigure}{0.3\textwidth}
    \centering
    \includegraphics[width=4cm]{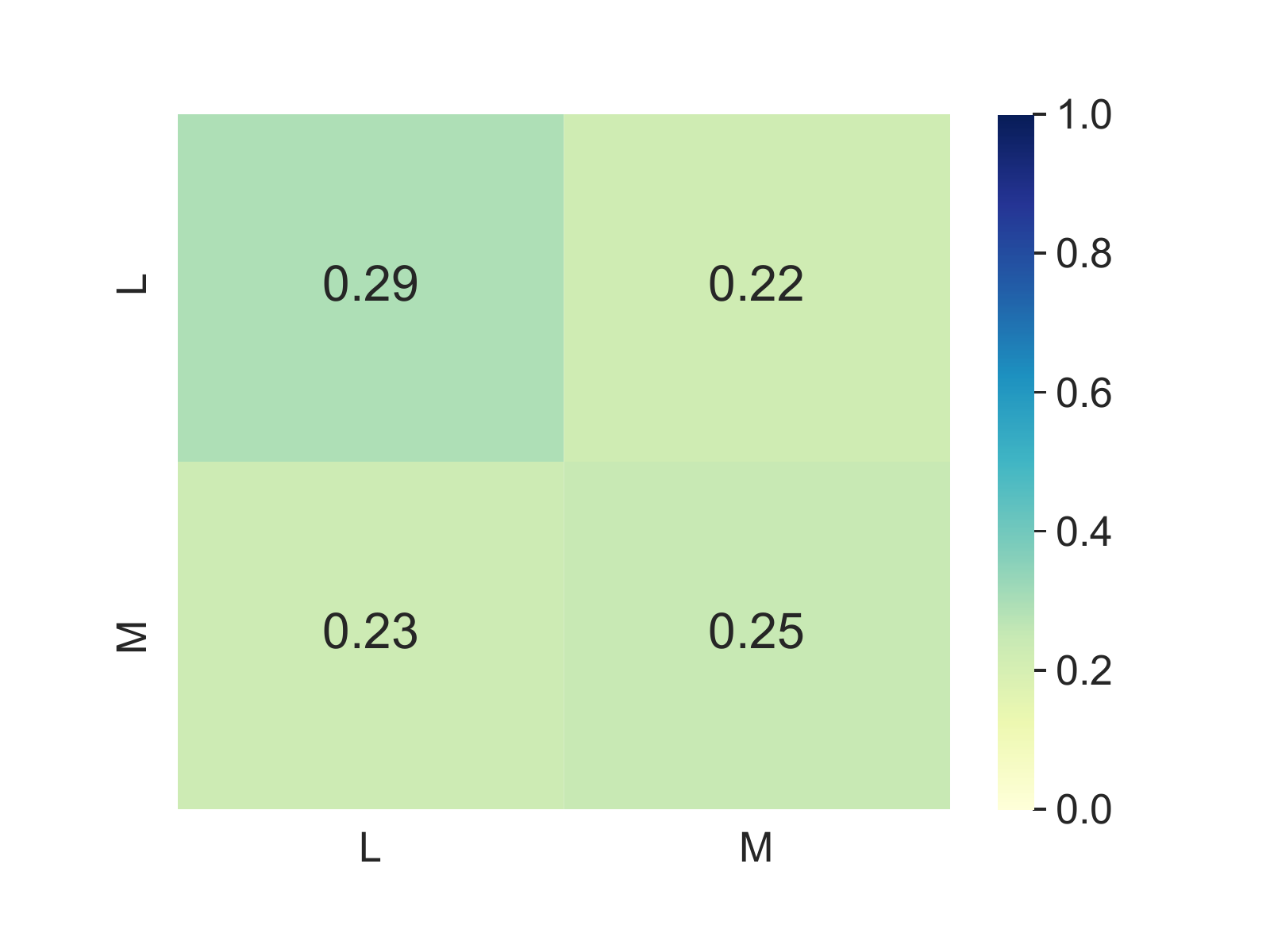}
    \caption{Game 4}
    \label{fig:lolam_g4}
\end{subfigure}
\begin{subfigure}{0.3\textwidth}
    \centering
    \includegraphics[width=4cm]{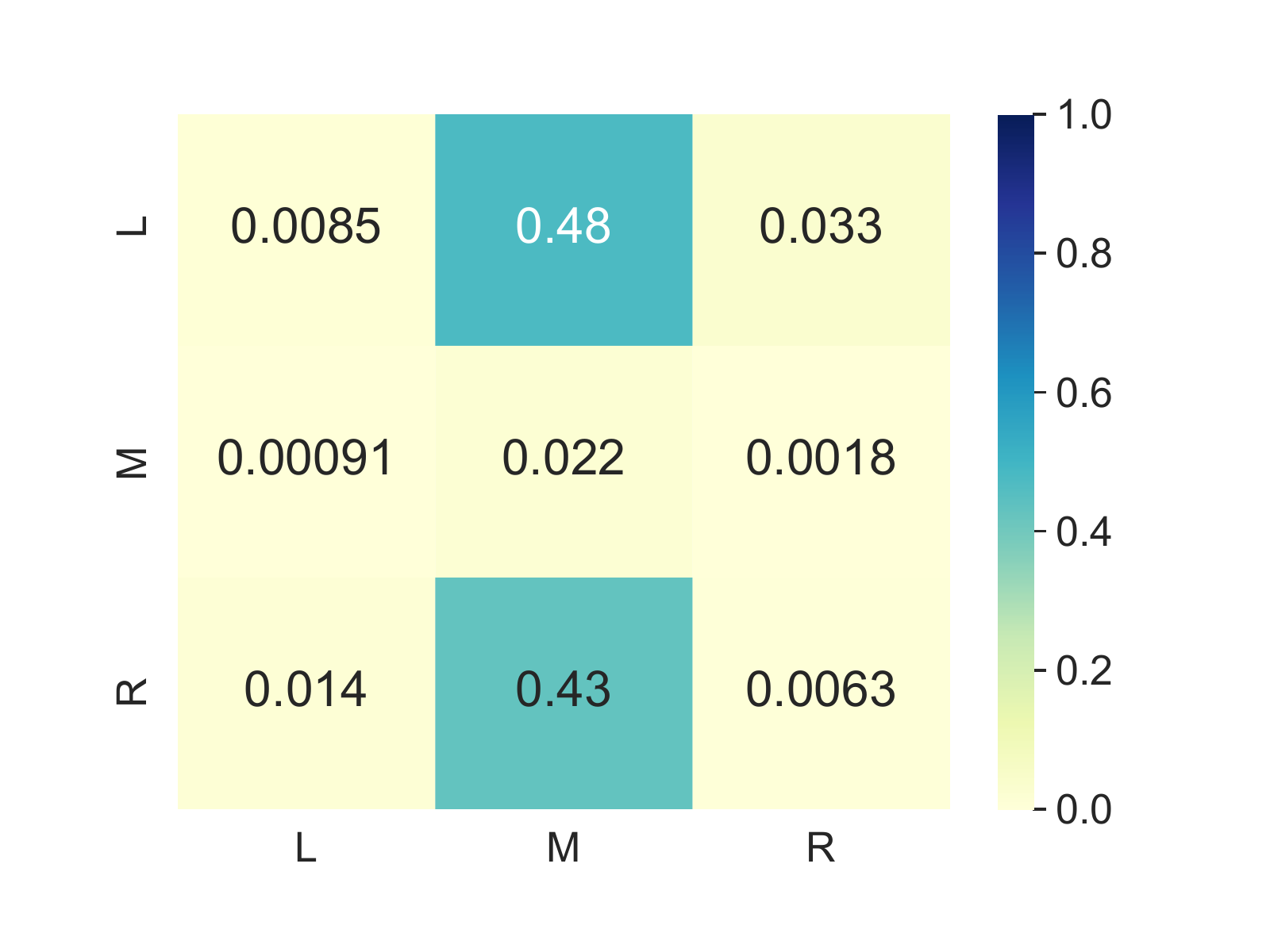}
    \caption{Game 5}
    \label{fig:lolam_g5}
\end{subfigure}
\caption{Empirical outcome distributions for LOLAM vs. LOLAM (lookahead values 1 and 3).}
\label{fig:lolamlolam}
\end{figure*}

The final approach we investigate here is LOLAM. We first look at how LOLAM vs. LOLAM  (Figure~\ref{fig:lolamlolam}) performs, especially in comparison to the full information setting (Section~\ref{sec:fullinfo}, Figures \ref{fig:lolagms}--\ref{fig:lolag5}).

Despite the fact that LOLAM agents do not have access to the same level of information compared to MO-LOLA, and make decisions based on a model of the opponent built from observations, they exhibit very similar behaviour to MO-LOLA in the unrealistic full information setting. In Games 1--3 (Figures \ref{fig:lolam_g1}--\ref{fig:lolam_g3}) the LOLAM agents have no trouble reaching the NE under SER and also learn to avoid the dominated (R,R) outcome of Game 3. We also note that we do not observe any correlation or relationship between the lookahead values used by the agents and the final empirical outcome distribution they converge towards. This means that a higher lookahead value does not translate to an agent being able to shift the outcome in its favour more often.

We observe the same behaviour as before for the games without a NE, Games 4 and 5 (Figures \ref{fig:lolam_g4} and \ref{fig:lolam_g5}), where the LOLAM agents' policies converge to meaningful outcomes previously identified as correlated equilibria for the games \citep{radulescu2019equilibria}.

These results further validate the use of Gaussian Process as estimators for the opponents' learning step in multi-objective settings, and alleviate the problem of not knowing the utility function of the opponent.

\begin{figure*}[h!]
\centering
\begin{subfigure}{0.3\textwidth}
    \centering
    \includegraphics[width=4cm]{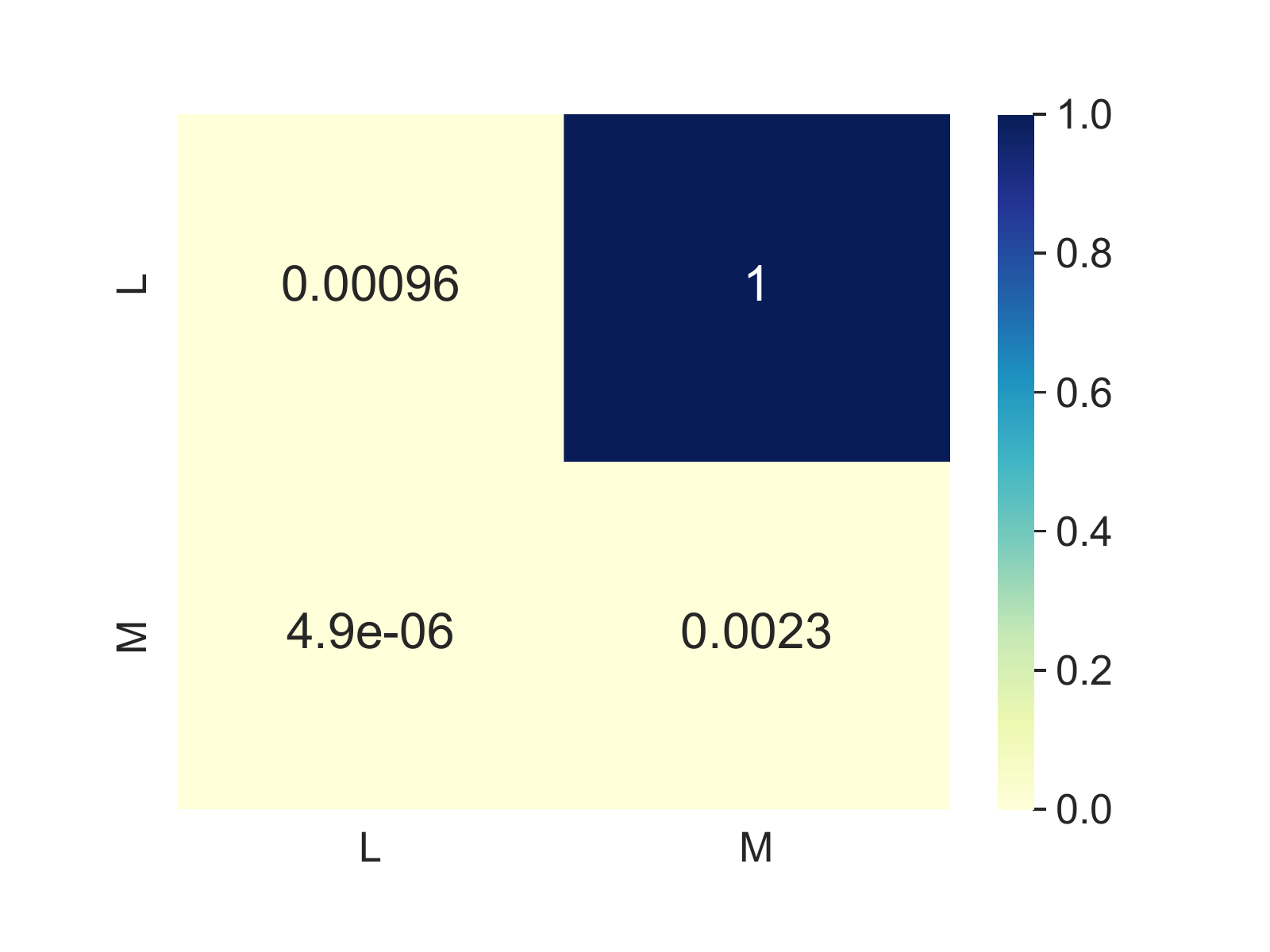}
    \caption{Game 1}
    \label{fig:acolamlolam_g1}
\end{subfigure}
\begin{subfigure}{0.3\textwidth}
    \centering
    \includegraphics[width=4cm]{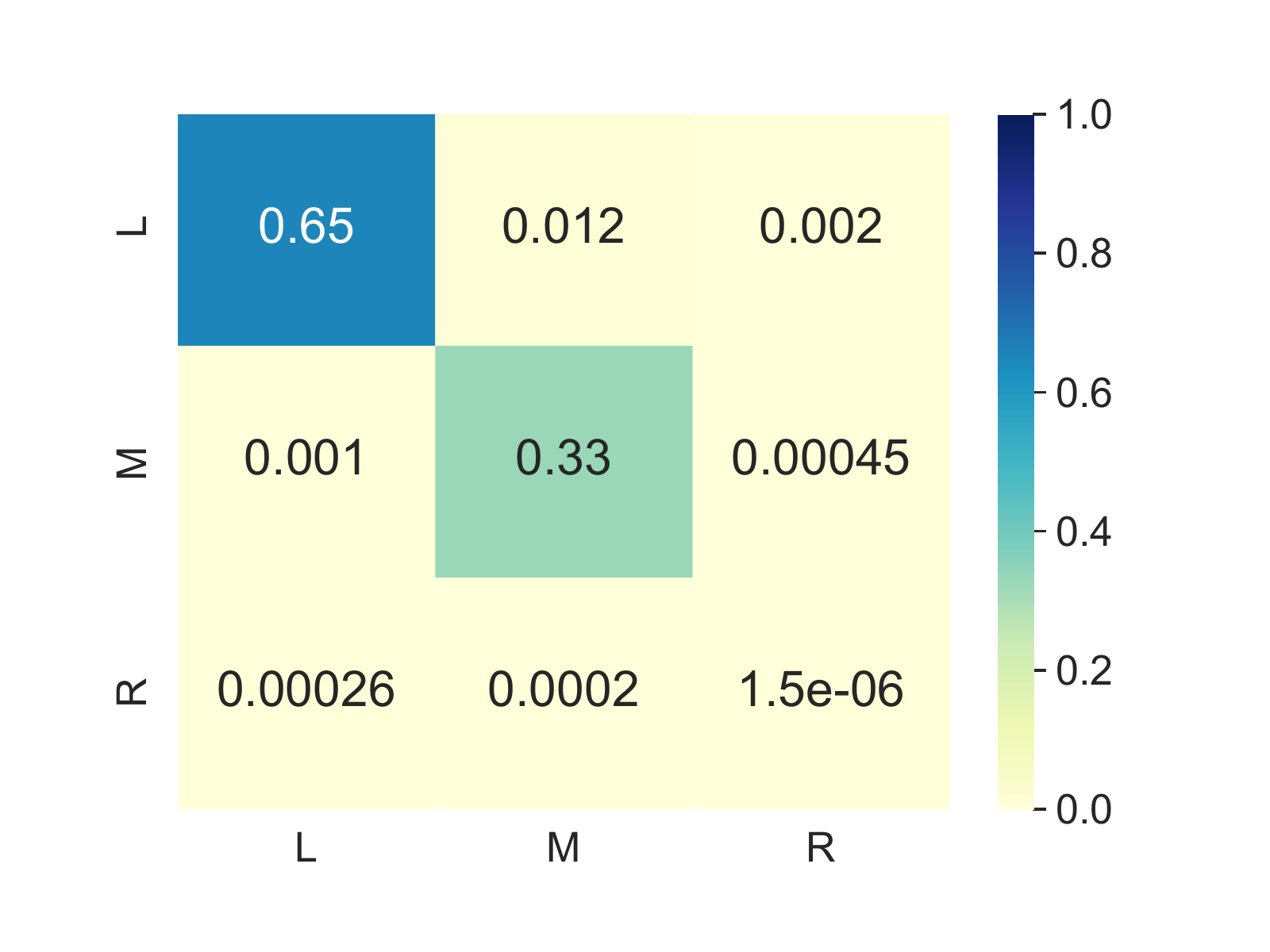}
    \caption{Game 3}
    \label{fig:acolamlolam_g3}
\end{subfigure}
\begin{subfigure}{0.3\textwidth}
    \centering
    \includegraphics[width=4cm]{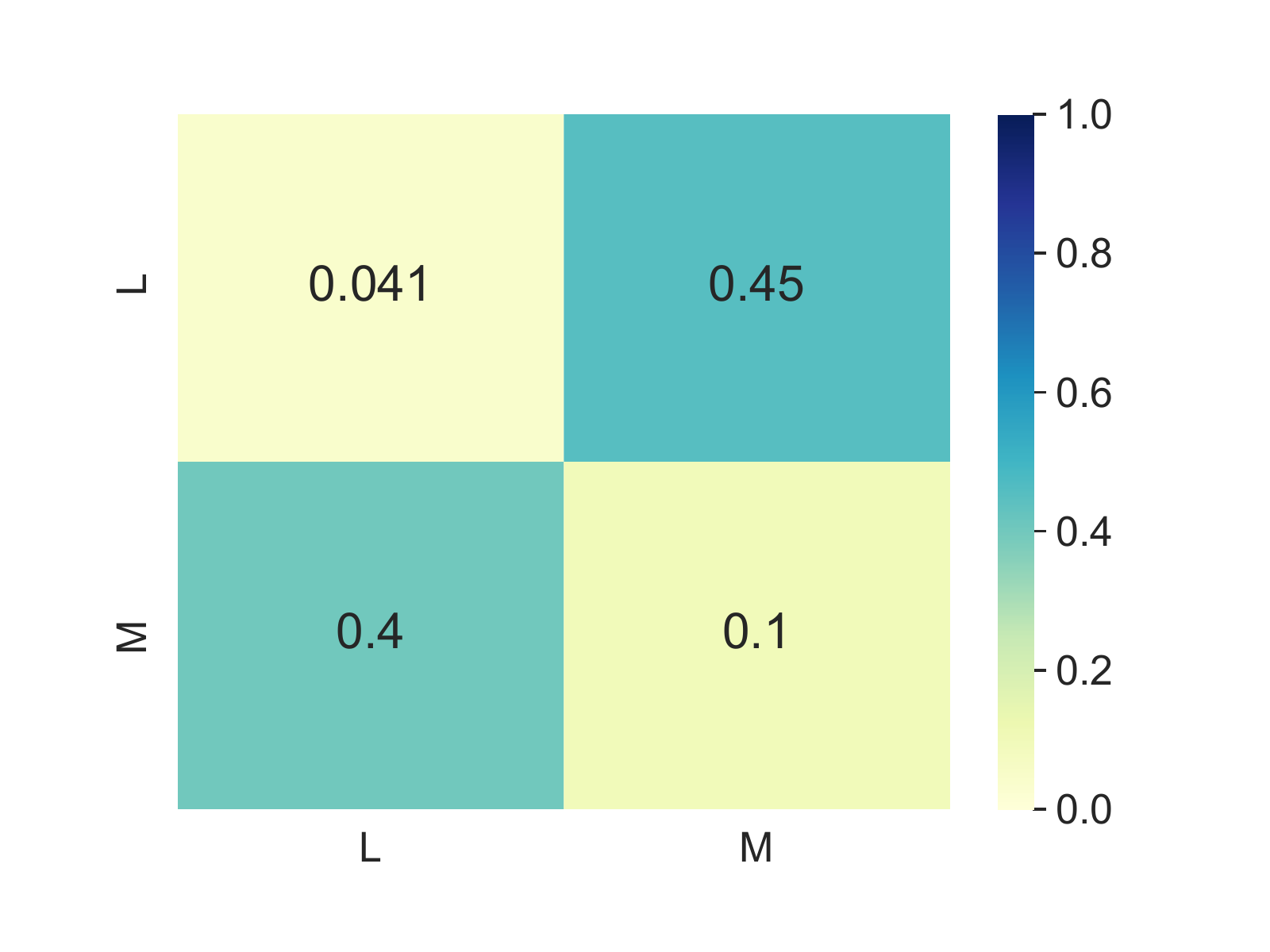}
    \caption{Game 4}
    \label{fig:acolamlolam_g4}
\end{subfigure}\\
\caption{Empirical outcome distributions for ACOLAM vs. LOLAM (lookahead values 2 and 2).}
\label{fig:acolamlolam}
\end{figure*}
\begin{figure*}[h!]
\centering
\begin{subfigure}{0.3\textwidth}
    \centering
    \includegraphics[width=4cm]{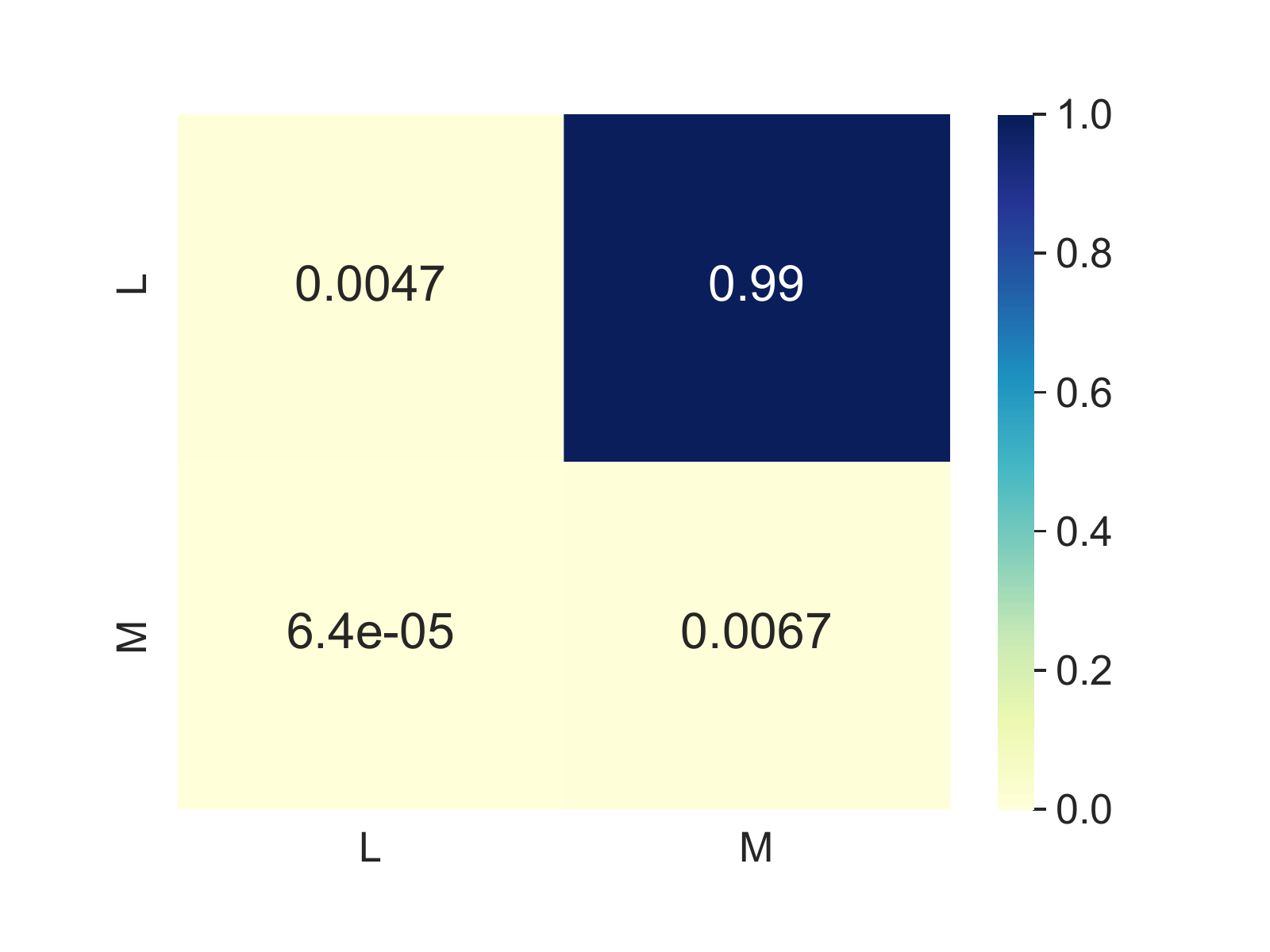}
    \caption{Game 1}
    \label{fig:lolamacolam_g1}
\end{subfigure}
\begin{subfigure}{0.3\textwidth}
    \centering
    \includegraphics[width=4cm]{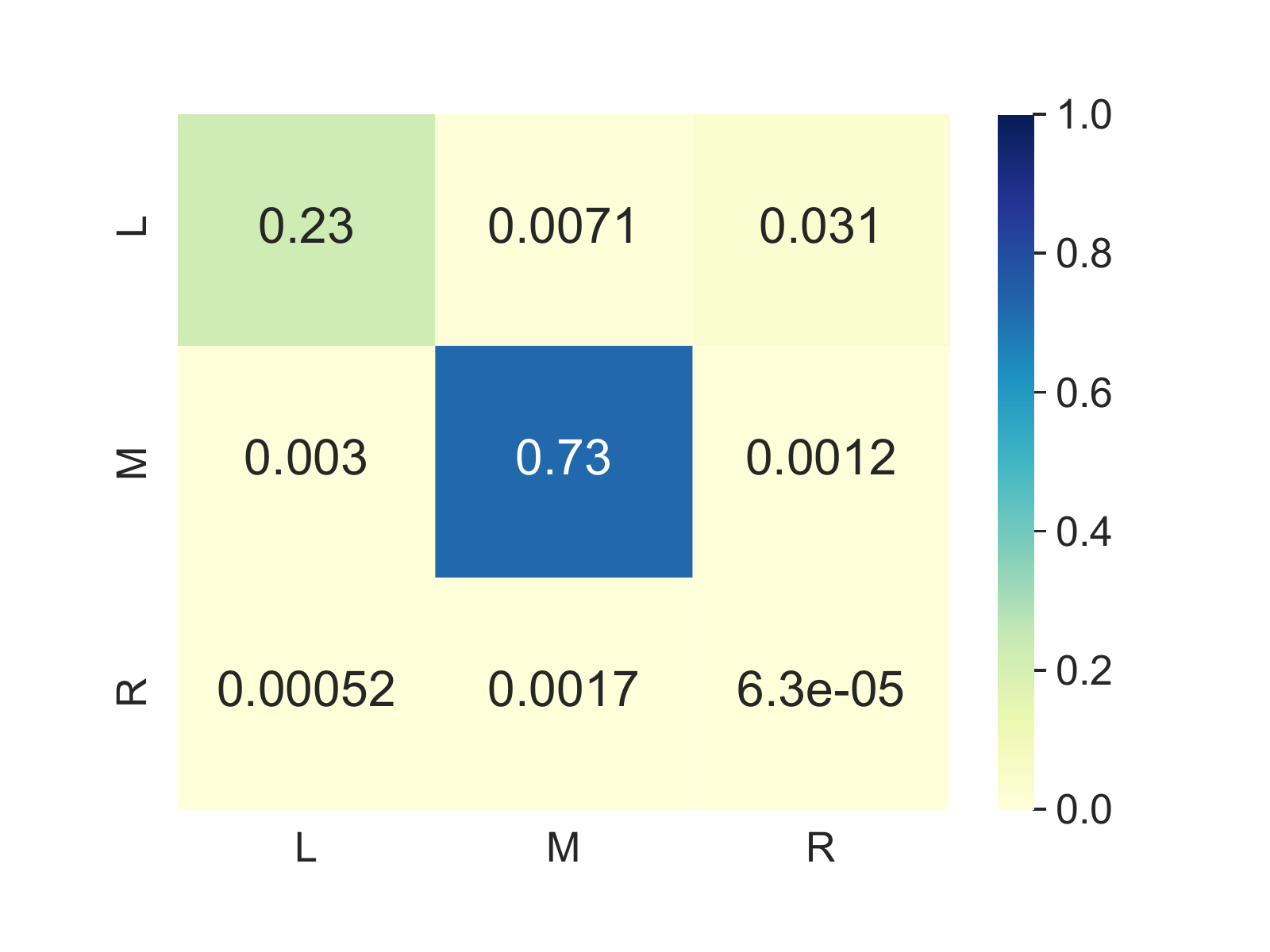}
    \caption{Game 3}
    \label{fig:lolamacolam_3}
\end{subfigure}
\begin{subfigure}{0.3\textwidth}
    \centering
    \includegraphics[width=4cm]{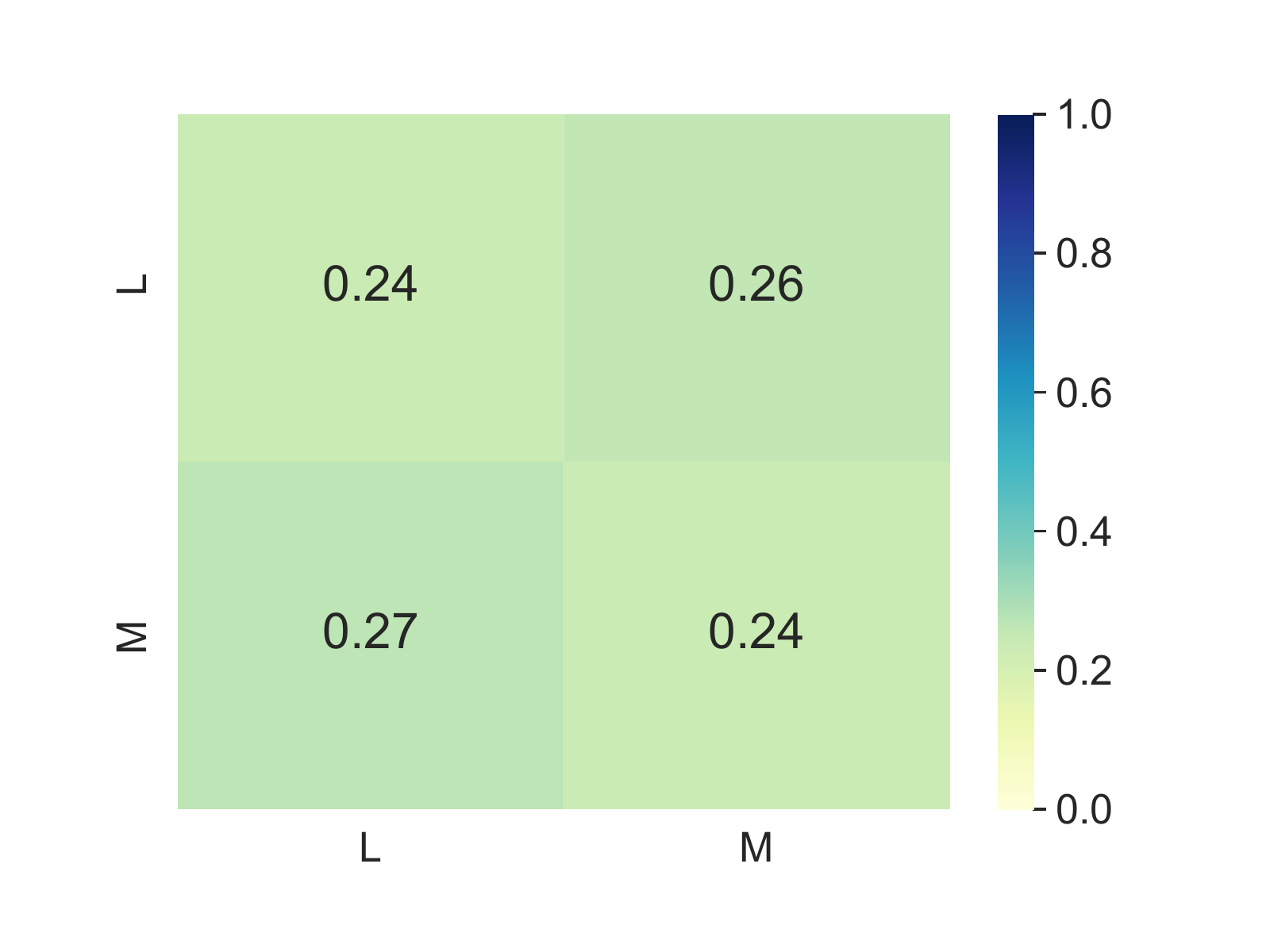}
    \caption{Game 4}
    \label{fig:lolamacolam_g4}
\end{subfigure}\\
\caption{Empirical outcome distributions for LOLAM vs. ACOLAM (lookahead values 1 and 1).}
\label{fig:lolamacolam}
\end{figure*}

\paragraph{ACOLAM vs. LOLAM and LOLAM vs. ACOLAM}
At this point we have an idea about the type of behaviour that each of our actor-critic and policy gradient approaches output in the studied MONFGs. The last remaining comparison we look at is between the two families of algorithms.

Before we discuss the obtained results, we first highlight a few differences between the approaches that might play a role in the observed learning dynamics. First of all, ACOM and ACOLAM learn a joint-action Q-table. Since we are in a deterministic setting, the learning rate $\alpha_{\boldsymbol{Q}}$ is set to 1. In comparison to the policy gradient approaches, this seems to allow AC-based agents to converge faster, as demonstrated in Figure~\ref{fig:acolamlolam}. Secondly, each method makes an assumption regarding the type of policy used by the other agent. AC-based agents assume a sofmax function when marginalising over the opponent's action, while PG-based agents will simulate a sigmoid function during the internal rollouts. This assumption does not invalidate the comparison, but it might affect the observed dynamics. It would be interesting to further investigate the effects of such assumptions, since in a competitive multi-agent setting, there is no guarantee for the type of opponent an agent can encounter, or whether the opponents will use the same learning approach.

Both combination of the ACOLAM and LOLAM methods manage to reach the pure NE (L,M) in Game 1 (Figures~\ref{fig:acolamlolam_g1} and \ref{fig:lolamacolam_g1}). For Game 3, the ACOLAM agent is able to shift the outcome in its favour more often. 

In Game 4, the dynamics and final interaction results seem to be dictated by the type of approach employed by agent 1. If agent 1 is an ACOLAM learner, we notice the same type of output as for the previous AC-based approaches, where agent 2 has an advantage. If agent 1 is a LOLAM learner, then the agents converge to equal probabilities over their actions and thus to a correlated equilibrium for this game under SER. 

\begin{figure*}[h!]
\centering
\begin{subfigure}{0.3\textwidth}
    \centering
    \includegraphics[width=4cm]{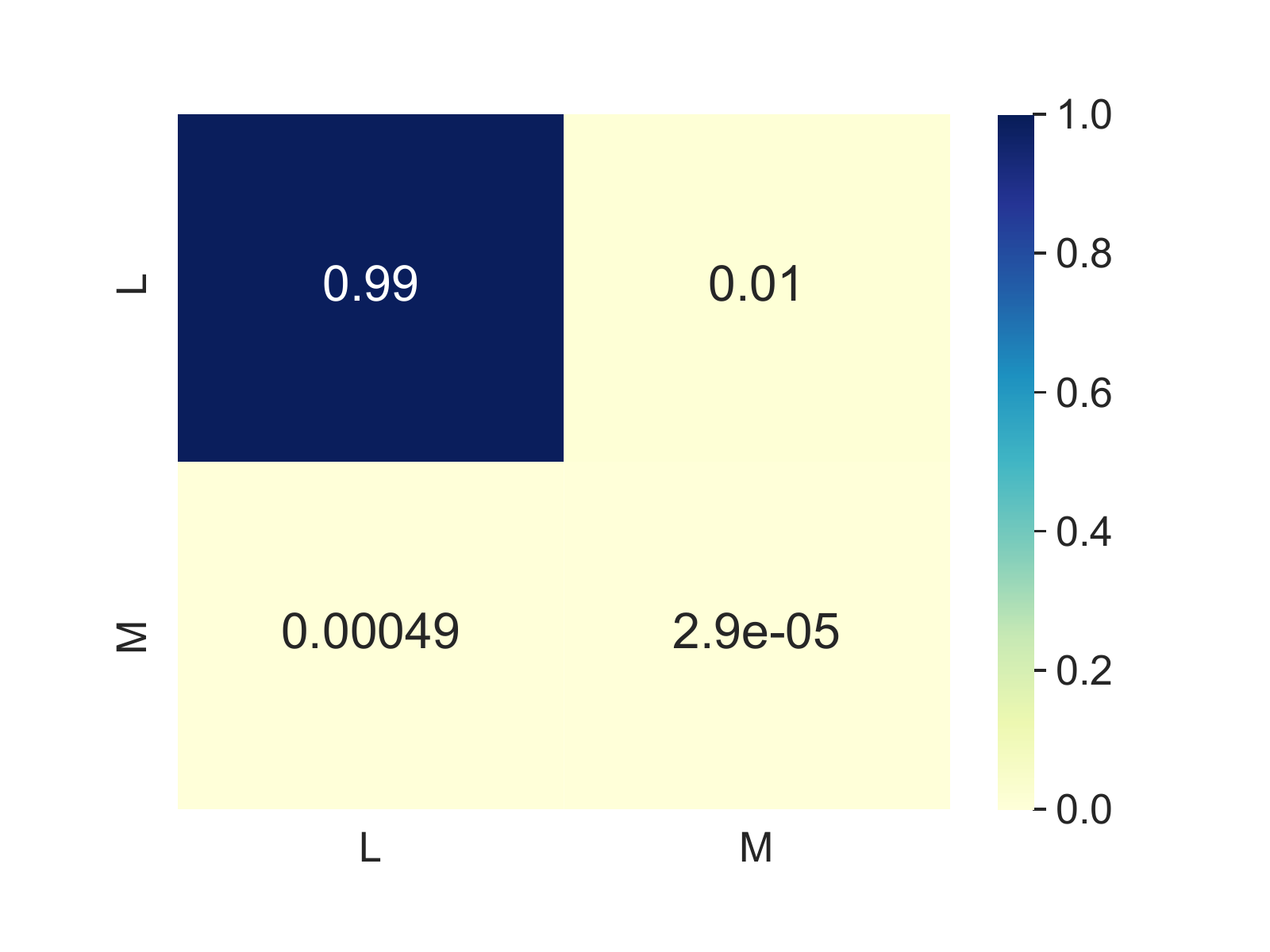}
    \caption{Empirical outcome distribution}
    \label{fig:acolamlolam_g2_states}
\end{subfigure}
\begin{subfigure}{0.3\textwidth}
    \centering
    \includegraphics[width=4cm]{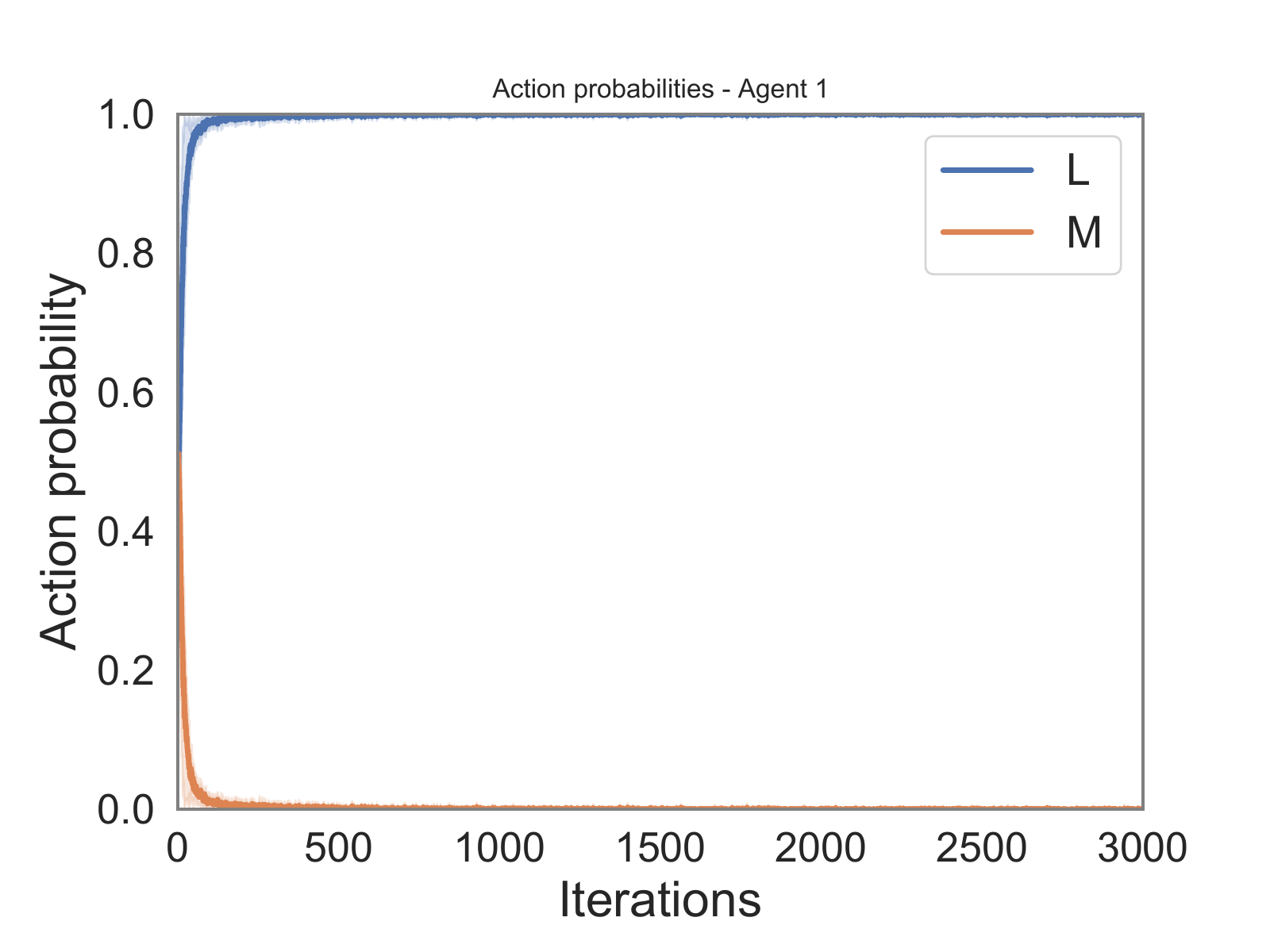}
    \caption{Action probabilities - agent 1}
    \label{fig:acolamlolam_g2_a1}
\end{subfigure}
\begin{subfigure}{0.3\textwidth}
    \centering
    \includegraphics[width=4cm]{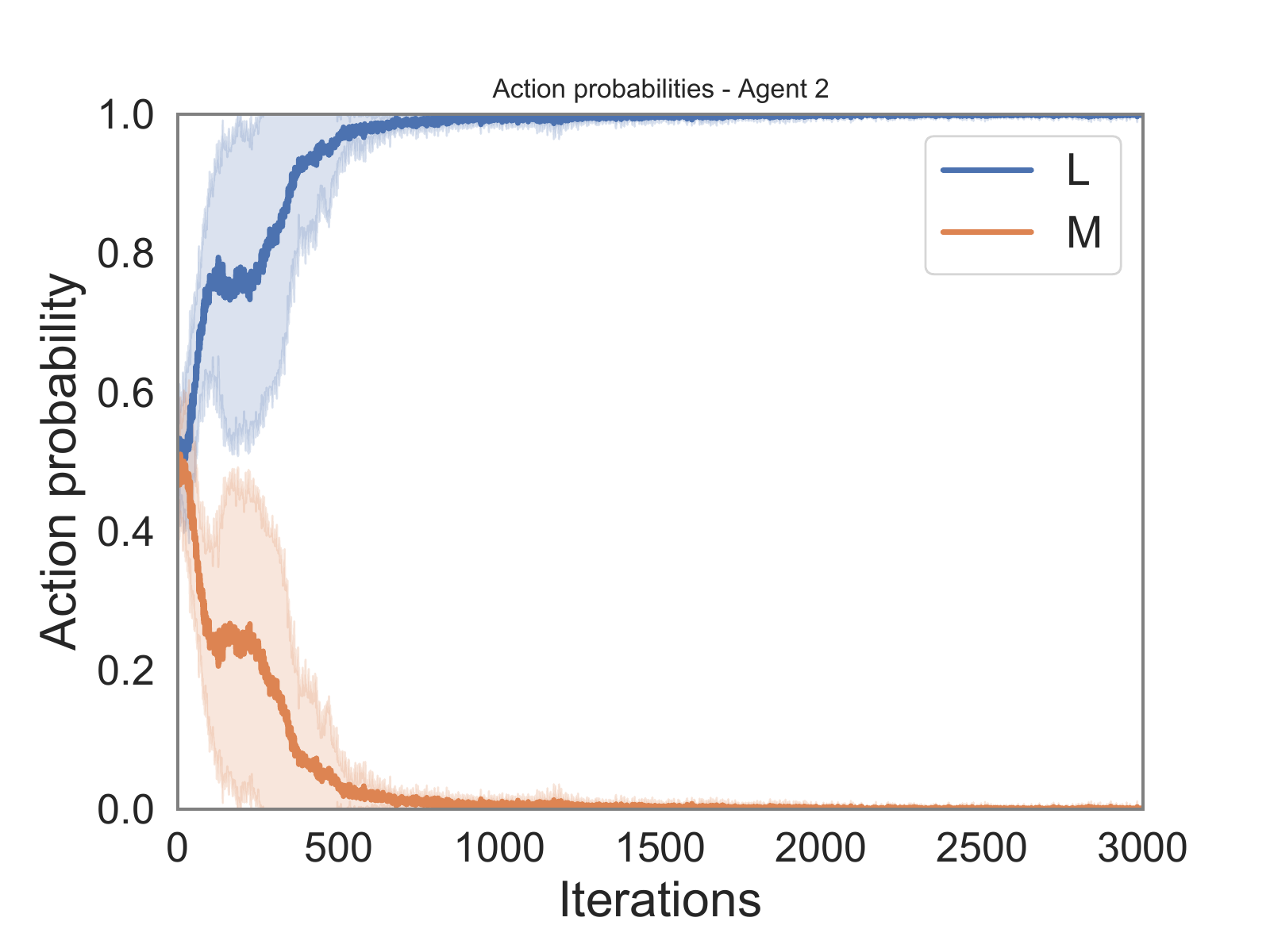}
    \caption{Action probabilities - agent 2}
    \label{fig:acolamlolam_g2_a2}
\end{subfigure}
\caption{Game 2 - ACOLAM vs. LOLAM (lookahead values 2 and 2).}
\label{fig:acolamlolam_g2}
\end{figure*}

\begin{figure*}[h!]
\centering
\begin{subfigure}{0.3\textwidth}
    \centering
    \includegraphics[width=4cm]{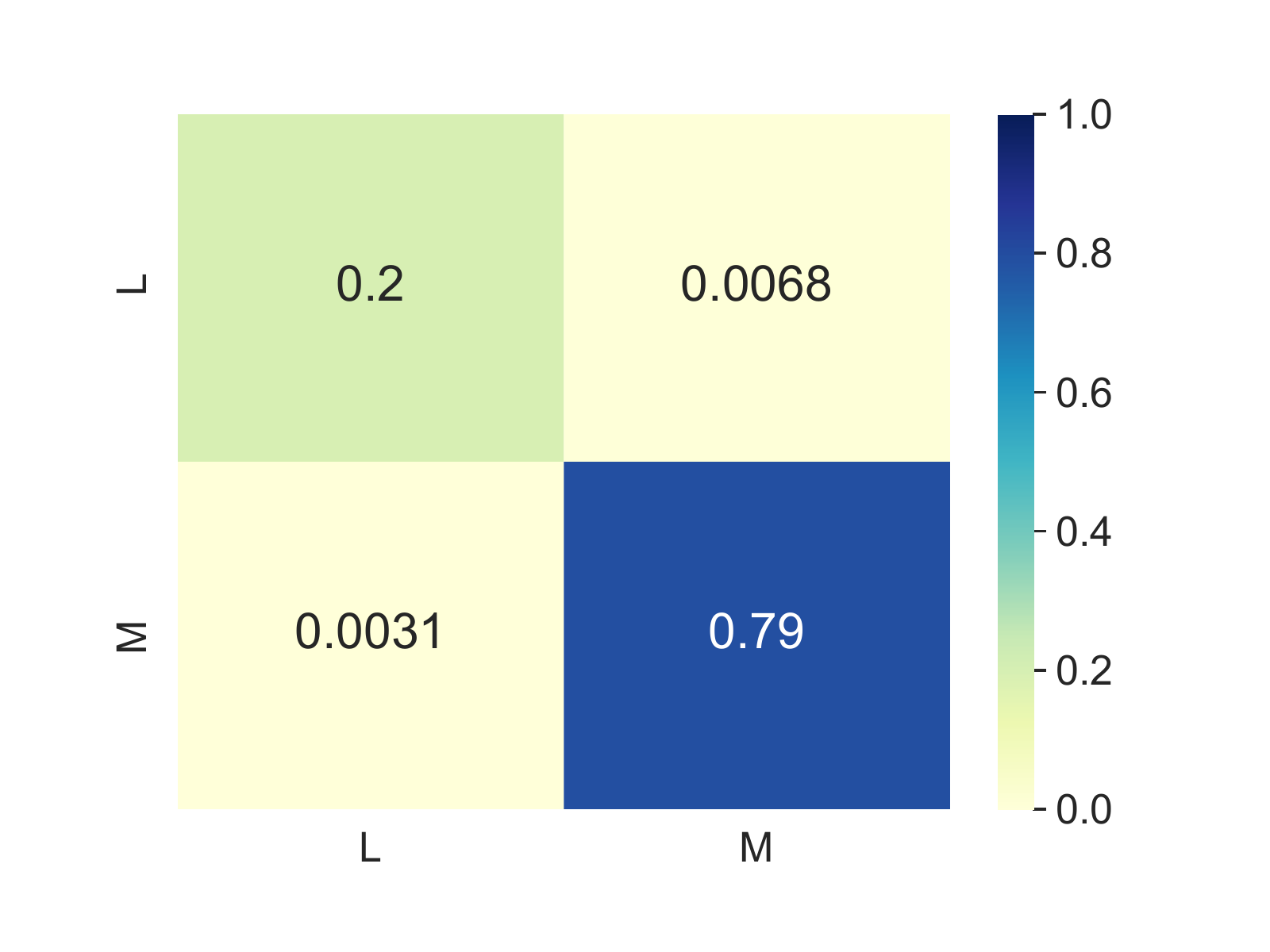}
    \caption{Joint action distribution}
     \label{fig:lolamacolam_g2_states}
\end{subfigure}
\begin{subfigure}{0.3\textwidth}
    \centering
    \includegraphics[width=4cm]{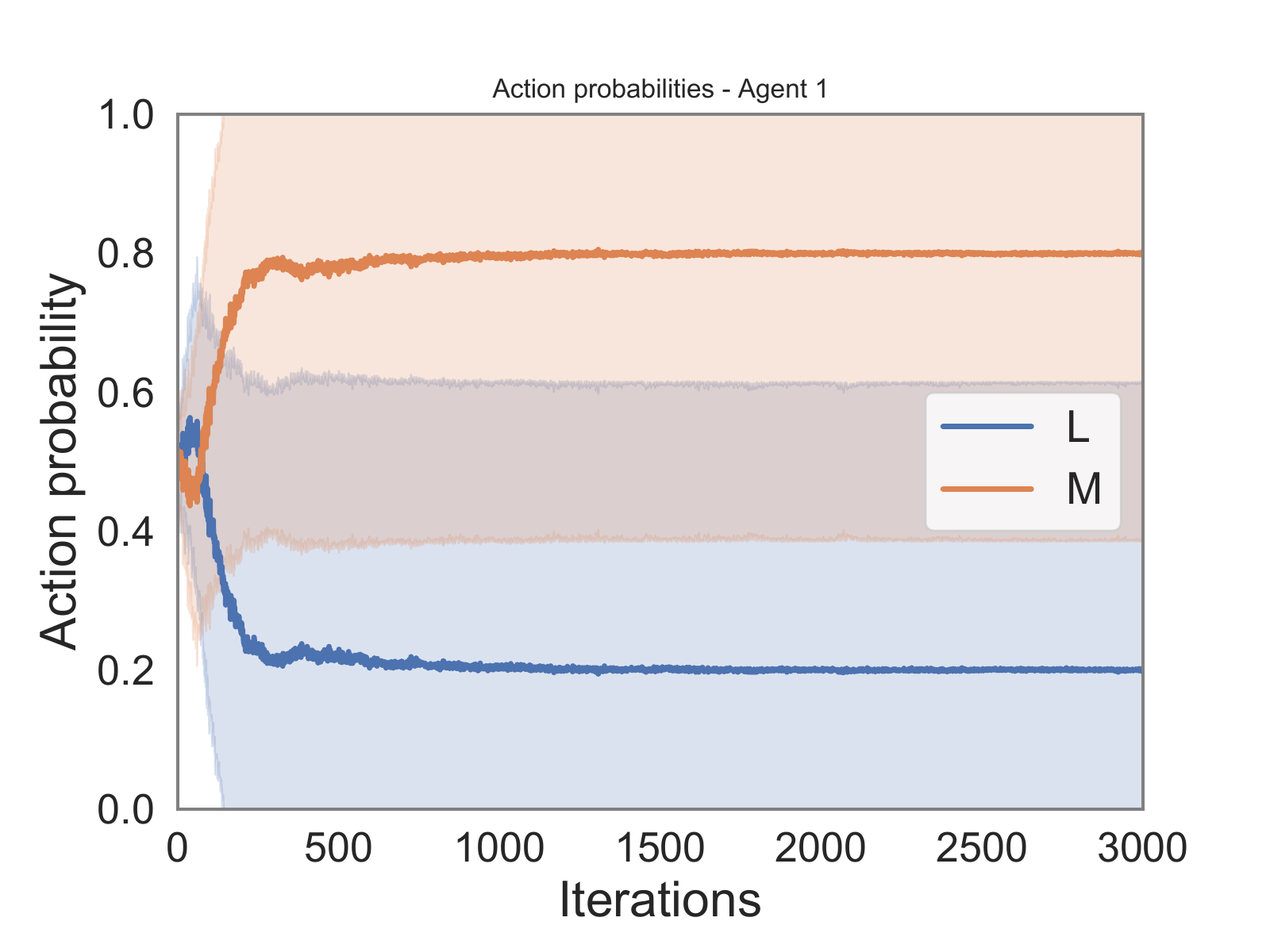}
    \caption{Action probabilities - agent 1}
    \label{fig:lolamacolam_g2_ag1}
\end{subfigure}
\begin{subfigure}{0.3\textwidth}
    \centering
    \includegraphics[width=4cm]{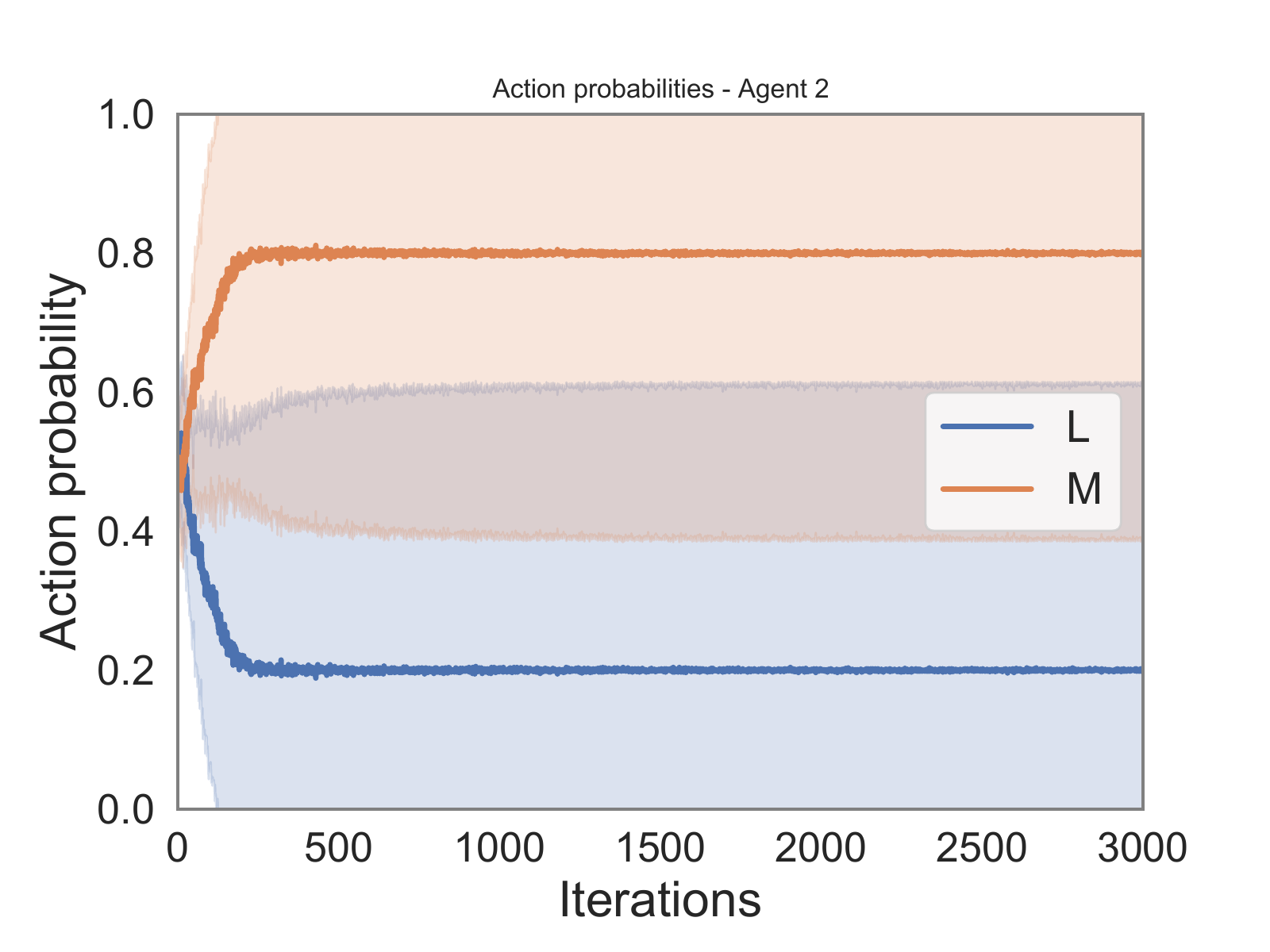}
    \caption{Action probabilities - agent 2}
     \label{fig:lolamacolam_g2_ag2}
\end{subfigure}
\caption{Game 2 - LOLAM vs. ACOLAM (lookahead values 1 and 1).}
\label{fig:lolamacolam_g2}
\end{figure*}

\begin{figure*}[h!]
\centering
\begin{subfigure}{0.3\textwidth}
    \centering
    \includegraphics[width=4cm]{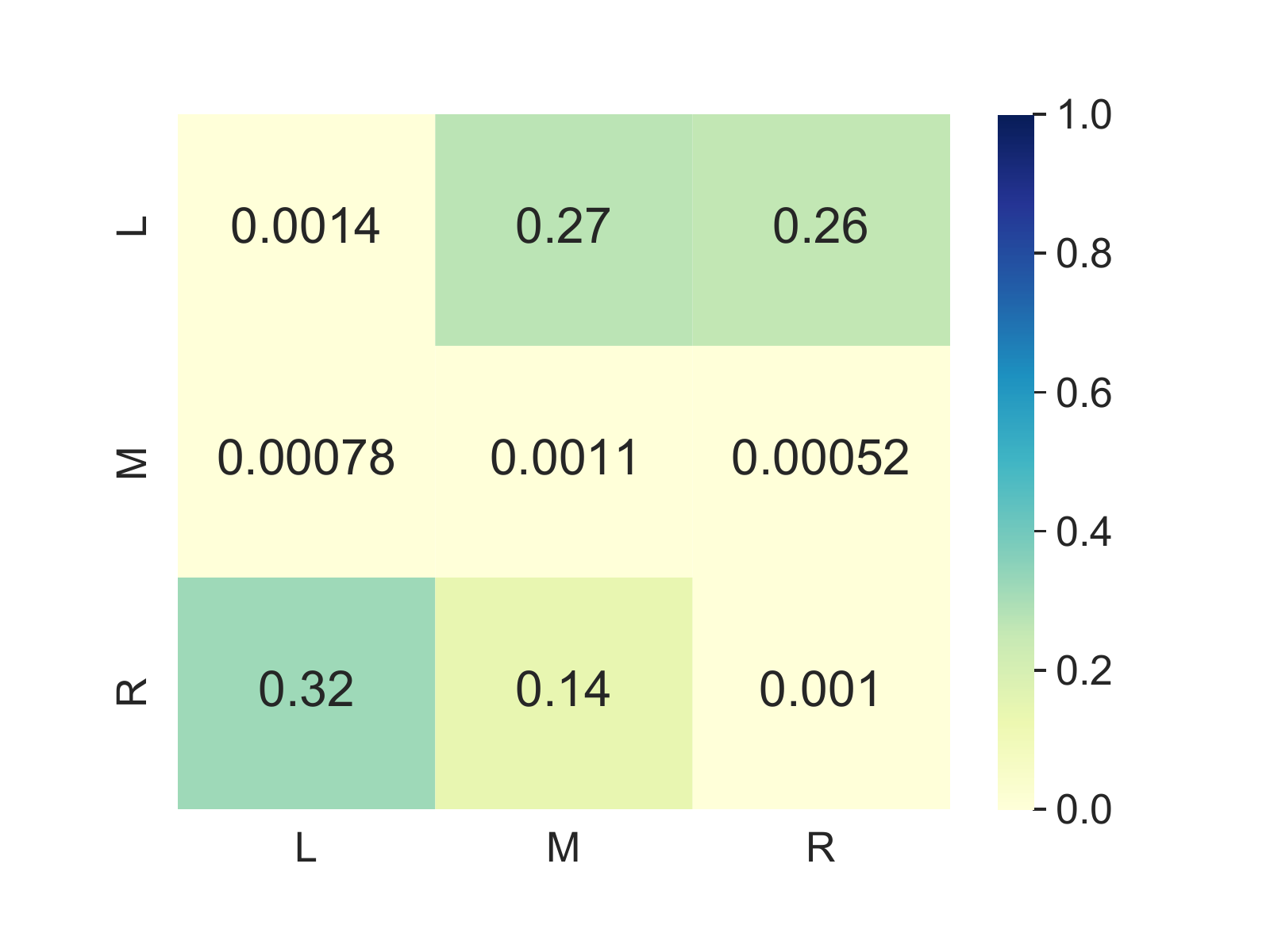}
    \caption{Joint action distribution}
    \label{fig:acolamlolam_g5_states}
\end{subfigure}
\begin{subfigure}{0.3\textwidth}
    \centering
    \includegraphics[width=4cm]{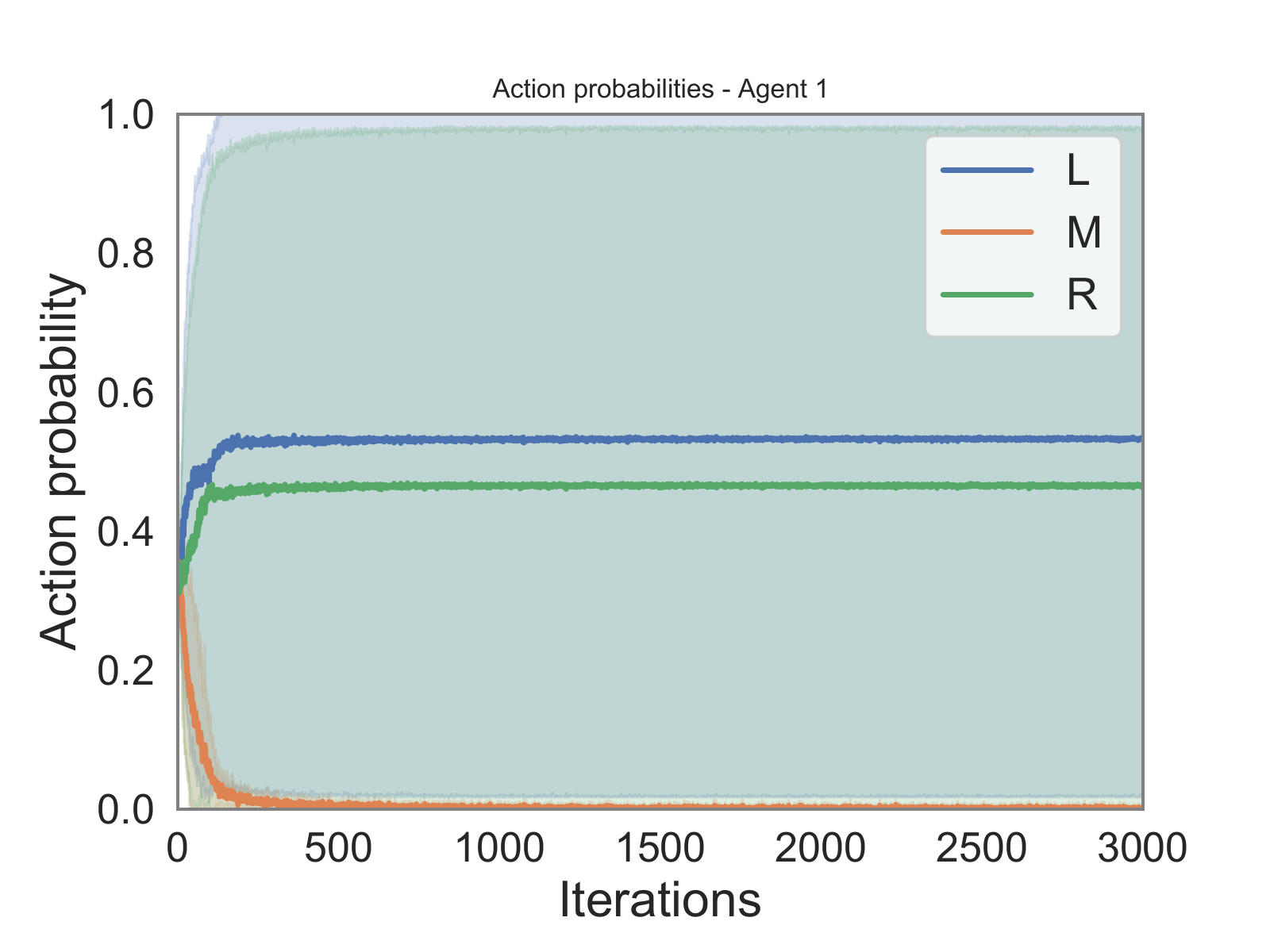}
    \caption{Action probabilities - agent 1}
    \label{fig:acolamlolam_g5_a1}
\end{subfigure}
\begin{subfigure}{0.3\textwidth}
    \centering
    \includegraphics[width=4cm]{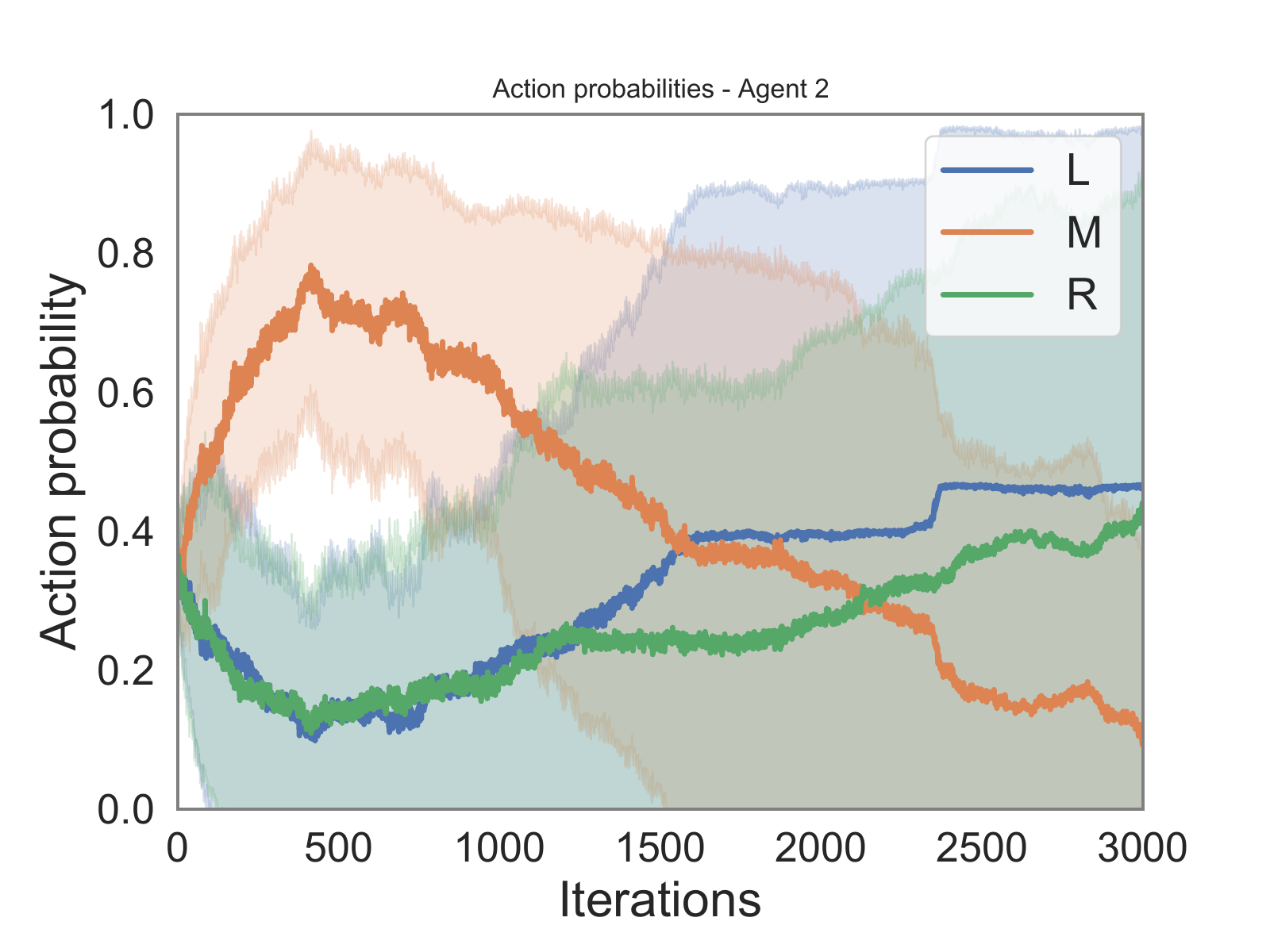}
    \caption{Action probabilities - agent 2}
    \label{fig:acolamlolam_g5_a2}
\end{subfigure}\\
\caption{Game 5 - ACOLAM vs. LOLAM (lookahead values 2 and 2).}
\label{fig:acolamlolam_g5}
\end{figure*}

\begin{figure*}[h]
\centering
\begin{subfigure}{0.3\textwidth}
    \centering
    \includegraphics[width=4cm]{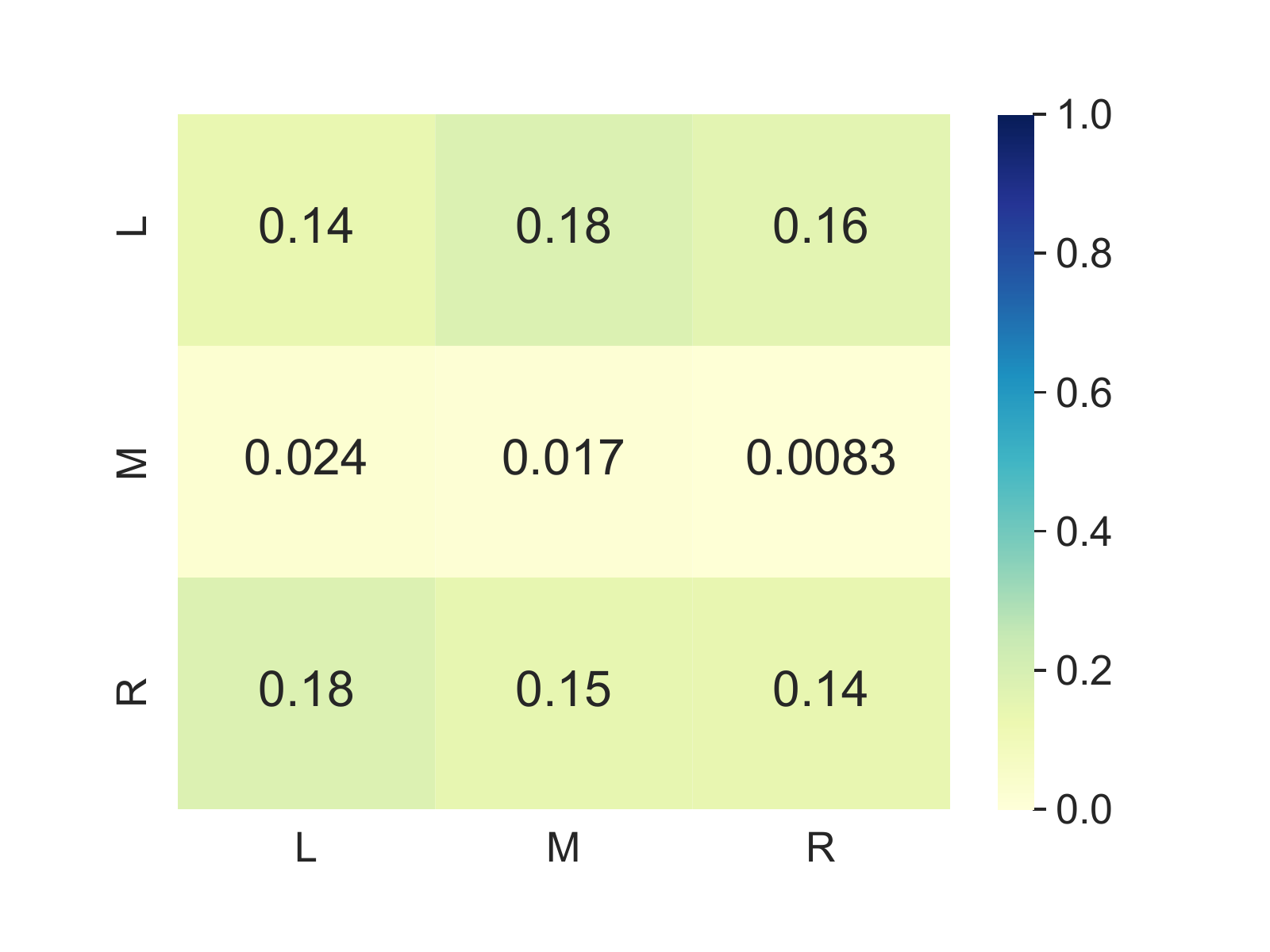}
    \caption{Joint action distribution}
    \label{fig:lolamacolam_g5_states}
\end{subfigure}
\begin{subfigure}{0.3\textwidth}
    \centering
    \includegraphics[width=4cm]{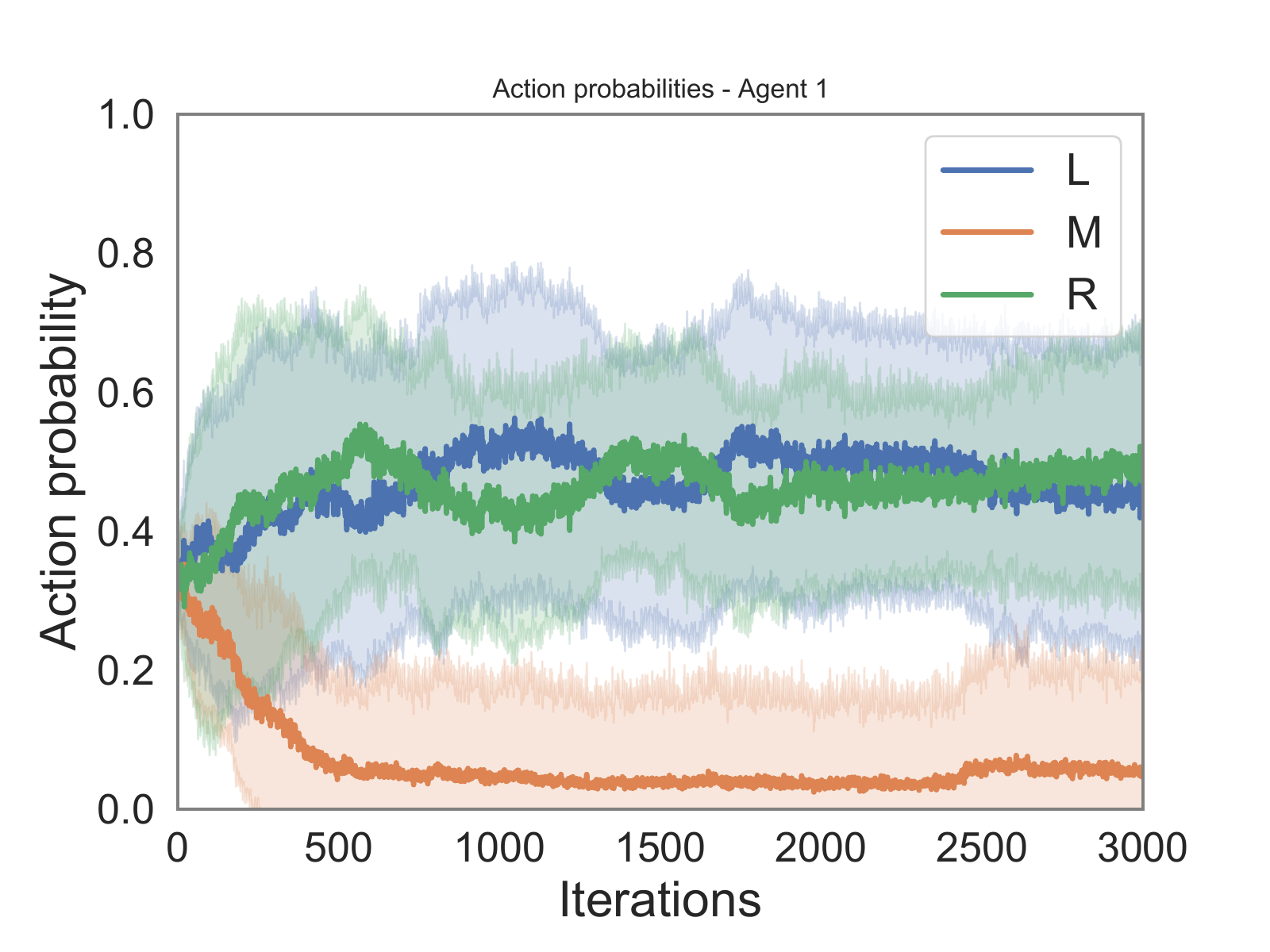}
    \caption{Action probabilities - agent 1}
    \label{fig:lolamacolam_g5_a1}
\end{subfigure}
\begin{subfigure}{0.3\textwidth}
    \centering
    \includegraphics[width=4cm]{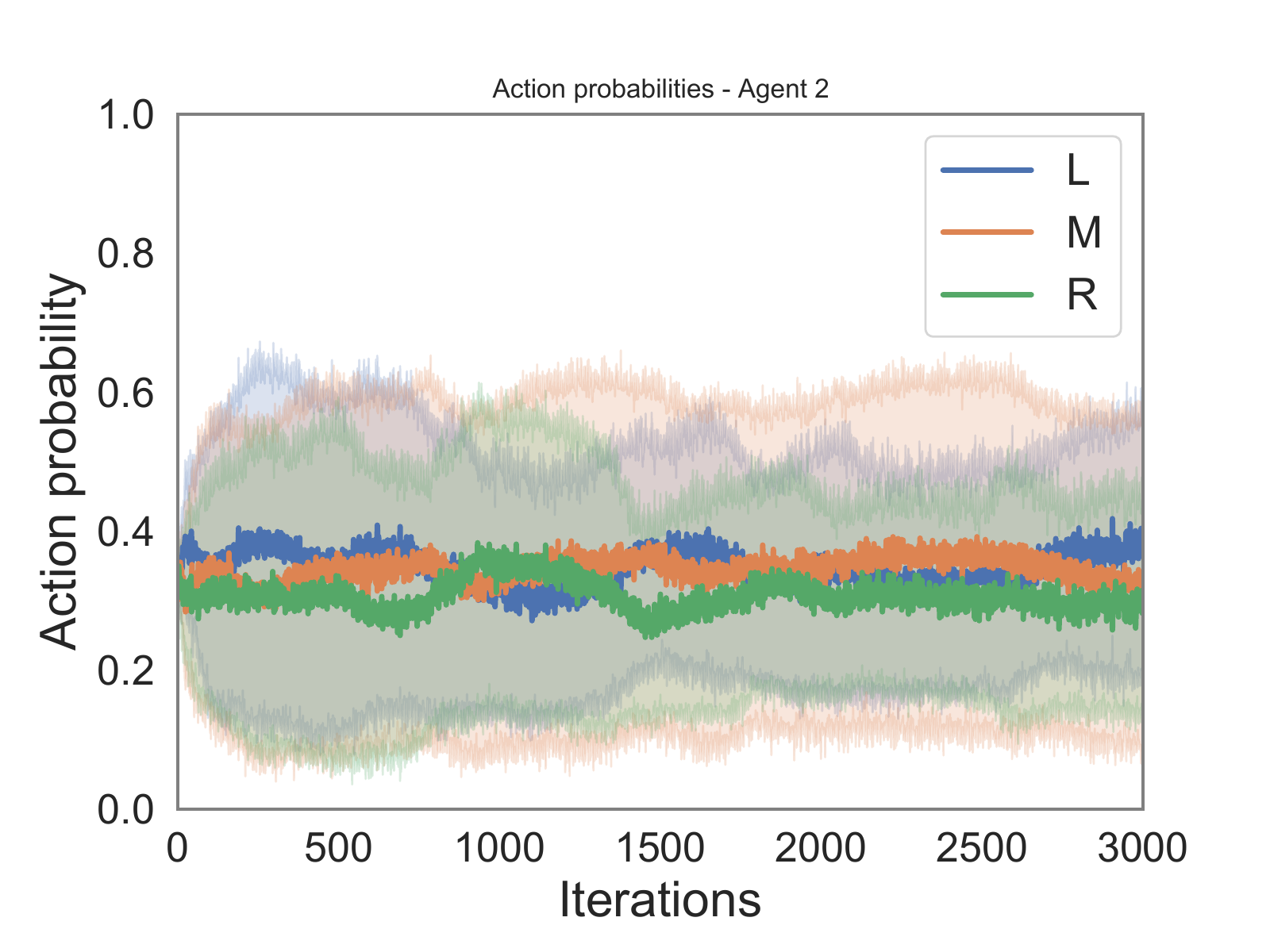}
    \caption{Action probabilities - agent 2}
    \label{fig:lolamacolam_g5_a2}
\end{subfigure}\\
\caption{Game 5 - LOLAM vs. ACOLAM (lookahead values 1 and 1).}
\label{fig:lolamacolam_g5}
\end{figure*}

The results for Game 2 highlight how the speed of convergence for the ACOLAM agent (Figures~\ref{fig:acolamlolam_g2} and \ref{fig:lolamacolam_g2}) can shift the outcome in that agent's favour. We can notice in Figure~\ref{fig:lolamacolam_g2_ag1} that initially the LOLAM agent starts playing action L, but immediately shifts to M to match the second agent.

Finally, the results of Game 5 present some interesting characteristics, depending on which agent uses which approach. When ACOLAM is used by agent 1, we observe the same evolution in the action probabilities as in the previous ACOLAM analyses (Figure~\ref{fig:acolamlolam_g5_a1}), i.e. a convergence to either fully action L or M. The initial action probability trajectory for LOLAM in this case is to play action M, as seen in the LOLAM vs. LOLAM analysis. However, since its opponent is immediately converging to either L or R, the LOLAM agent seems to start playing the opposing action R or L. This shifts the outcome from the correlated equilibrium actions (L,M)--(R,M) to (L,R)--(R,L), which is a better situation for agent 2. When LOLAM is used by agent 1, it converges, as seen in the LOLAM vs. LOLAM analysis, to a mixed strategy between actions L and R (Figure~\ref{fig:lolamacolam_g5_a1}). In this situation the ACOLAM agent maintains a uniform distribution over its actions as its policy (Figure~\ref{fig:lolamacolam_g5_a2}).

While we are not able to say that the LOLAM approach manages every time to obtain a more preferred outcome, we do notice that LOLAM is capable of finding interesting, middle ground solutions, in situations where no NE under SER exist. The outcomes that LOLAM converges to approximate correlated equilibria, without having received any prior correlation signal. We consider this to be a valuable and interesting finding for future analyses of MONFGs, as it opens up the idea that different solution concepts are attainable in such settings.

\section{Related Work}
\label{sec:related_work}
Here we present a brief overview of prior work on MONFGs and opponent modelling, with a specific focus on works which are closely related to the contributions of this paper. A comprehensive survey of opponent modelling techniques is presented by \citet{albrecht2018survey}.

Since the introduction of MONFGs in \citep{blackwell1956analog}, they have mostly been considered under linear utility functions \citep{borm1988pareto,lozovanu2005multiobjective} or when implicitly assuming the ESR criterion \citep{borm2003structure}.  \citet{radulescu2019equilibria} revisit MONFGs and explicitly distinguish between ESR and SER under non-linear utility functions. They demonstrate the effect of using different criteria on the set of Nash and correlated equilibria. We note, however, that their experimental framework only incorporates a simplistic Q-learning approach, using $\epsilon$-greedy as an action selection mechanism. This needed to be coupled with a non-linear optimisation solver to enable agents to determine at each step their optimal mixed strategies.

A straightforward method for opponent modelling in reinforcement learning is building a model of the other agents' policy. Opponent Modelling Q-learning \citep{uther1997adversarial} extends Q-learning in a similar manner to our approach for extending the critic in our actor-critic-based algorithms: it calculates the probability distribution of the opponent's actions from the observed behaviour, and then derives the best action for the agent by marginalising out the opponent's actions from the joint Q-table. More recently,
\citet{foerster2018lola} used maximum likelihood to infer the opponent's policy from the observed state-action trajectories, in their opponent modelling setting.

Opponent modelling has also been incorporated in RL methods based on neural function approximators by augmenting the model with a module that is able to predict the action of the other agent \citep{he2016opponent,knegt2018opponent}. Finally, goal prediction is another approach for opponent modelling, presented in Self Other-Modeling \citep{raileanu2018modeling}, where the agent uses his own policy to learn to predict the goal of the other agent.

In multi-objective settings, another choice for modelling other agents is to build a model of their utility functions. However, this task is not trivial, and an important idea is to use, anytime possible, knowledge regarding the structure of the utility space. For example, Chajewska and Koller \citep{chajewska2000utilities} build a probabilistic model for utilities elicited from a population of users and show how one can find a factorisation of the utility function.
\citet{chajewska2001learning} continue this line of work and show how one can use such a model as a probabilistic prior to obtain the sub-utility components for a specific agent (having a linear additive utility function) when passively observing his behaviour. 
They derive a set of linear constraints on the set of sub-utilities from the observed actions to obtain a posterior and sample a new utility function similar to the inverse reinforcement learning approach of \citep{ng2000algorithms}. 
More recently, several works have used Gaussian processes to model utility functions \citep{chu2005preference,guo2010gaussian,zintgraf2018ordered}.

Using an active learning approach, the decision making process -- based on partial utility information -- can be intertwined with a querying process in order to elicit additional utility information \citep{chajewska2000making,zintgraf2018ordered}. However, revealing information regarding one's preferences will not always be in the best interest of agents, especially in competitive settings. Hence, the agent would need to extract preference information from interactions, which is far from trivial.

\section{Conclusion and Future Work}
\label{sec:conclusion}
In this work, we presented the first study on the effects of opponent learning awareness and modelling in multi-objective multi-agent settings under the SER optimisation criterion. In contrast to much prior work on opponent modelling in multi-criteria problems, we considered opponents with non-linear utility functions. We adopted the MONFG model for our experimental evaluations. Novel formulations of actor-critic and policy gradient approaches for this setting were introduced, along with extensions that incorporate: (1) opponent modelling via policy reconstruction using action frequencies; and (2) modelling the opponent's learning step using a Gaussian process, when no information regarding the opponent's utility function is available. 

Empirical results in {five} different MONFGS ({three} with Nash equilibria, and {two} without under the SER criterion) demonstrated that opponent learning awareness and modelling can significantly alter the learning dynamics of a MONFG. In cases where NE are present, opponent learning awareness and modelling can confer significant benefits for agents that implement it. When there are no NE, we showed that our policy gradient approach, LOLAM, allows agents to compromise and still converge to a middle ground solution, which corresponded to an approximate correlated equilibrium for the games, without having received any correlation signal. This brings forth the idea that different solution concepts are attainable in such settings.

This study has a number of limitations, leaving much scope for future research to build upon the present work. As we adopted the MONFG model, our analysis considered stateless decision making problems only; therefore this line of work should be extended to sequential settings such as multi-objective stochastic games (MOSGs) \citep{Mannion2018Reward,Radulescu2020Survey}. Furthermore, our experimental evaluations were limited to games with two agents only, so there is much work to be done on opponent modelling in larger MOMAS. In many real world settings (e.g. online games such as MMORPGs, or political negotiations between multiple states), the utility functions of agents in the environment often have varying degrees of alignment to one another. Therefore an agent that can effectively model opponent utility from interactions could make predictions about the intentions (i.e. cooperative vs. competitive) of other agents, based on the degree of alignment of an estimated opponent utility function with her own private utility function.

As multi-objective multi-agent decision making is a relatively under-explored area of MAS research, many significant and interesting open questions remain within the field. The choice of optimisation criterion (ESR versus SER) can have drastic effects on the set of equilibria in MOMAS. We already made the surprising observation that opponent learning awareness and modelling can allow agents to find compromise solutions under SER, when there are no NEs. We want to further investigate this phenomenon. Larger MOMAS may contain agents that choose different optimisation criteria or different learning mechanisms. This could add further complications when determining the conditions for a stable outcome to be reached. While we have proposed several new MONFGs in this paper, in future work it would be worthwhile to develop a larger set of standardised benchmarks that could be used to evaluate the performance of algorithms in a variety of multi-objective multi-agent decision making settings, e.g., cooperative and competitive games, negotiations, and sequential settings.

\bibliographystyle{plainnatCustom}
\bibliography{bibliography}

\end{document}